\documentclass[aps,pra,amsmath,twocolumn,showpacs,superscriptaddress]{revtex4-2}
\usepackage[colorlinks=true,linkcolor=blue,citecolor=blue,urlcolor=blue,filecolor=blue]{hyperref}
\usepackage[english]{babel}
\usepackage{graphicx}
\usepackage{subfigure}
\usepackage{dcolumn}
\usepackage{bm}
\usepackage{units}
\usepackage{color}
\usepackage{xcolor}
\usepackage{longtable}
\usepackage{array}
\usepackage{multirow}
\usepackage{physics}
\usepackage{sidecap}
\usepackage{CJKutf8}
\usepackage{siunitx}
\usepackage{tikz}

\definecolor{lime}{HTML}{A6CE39}
\DeclareRobustCommand{\orcidicon}{
	\begin{tikzpicture}
	\draw[lime, fill=lime] (0,0) 
	circle [radius=0.16] 
	node[white] {{\fontfamily{qag}\selectfont \tiny ID}};
	\draw[white, fill=white] (-0.0625,0.095) 
	circle [radius=0.007];
	\end{tikzpicture}
	\hspace{-0.3mm}
}

\foreach \x in {A, ..., Z}{\expandafter\xdef\csname orcid\x\endcsname{\noexpand\href{https://orcid.org/\csname orcidauthor\x\endcsname}
			{\noexpand\orcidicon}}
}

\begin{document}

\title{Laser cooling for quantum gases}
\author{Florian Schreck$\orcidA{}$}
\email[]{LC4QG@strontiumBEC.com}
\affiliation{Van der Waals-Zeeman Institute, Institute of Physics, University of Amsterdam, Science Park 904, 1098XH Amsterdam, The Netherlands}
\affiliation{QuSoft, Science Park 123, 1098 XG Amsterdam, The Netherlands}
\author{Klaasjan van Druten$\orcidB{}$}
\email[]{LC4QG@strontiumBEC.com}
\affiliation{Van der Waals-Zeeman Institute, Institute of Physics, University of Amsterdam, Science Park 904, 1098XH Amsterdam, The Netherlands}

\maketitle
\onecolumngrid
{\bf Laser cooling exploits the physics of light scattering to cool atomic and molecular gases to close to absolute zero. It is the crucial initial step for essentially all atomic gas experiments in which Bose–Einstein condensation and, more generally, quantum degeneracy is reached. The ongoing development of laser-cooling methods has allowed more elements to be brought to quantum degeneracy, with each additional atomic species offering its own experimental opportunities. Improved methods are opening new avenues, for example, reaching Bose-Einstein condensation purely through laser cooling as well as the realization of continuous Bose-Einstein condensation. Here we review these recent innovations in laser cooling and provide an outlook on methods that may enable new ways of creating quantum gases.}
\vspace*{2mm}
\twocolumngrid
\section{Introduction}

Reaching ever lower temperatures has been a frontier of physics for over a century, as the reduction of thermal excitations renders the quantum behaviour of matter more apparent and controllable, enables low-energy quantum phases to arise and reduces decoherence. Laser cooling allows us to cool atomic and molecular gases from room temperature to the microkelvin regime, where particles move with velocities of as little as centimetres per second \cite{Chu89,Met99}. Combined with a trapping mechanism, thousands to billions of particles can be accumulated in a small volume. These laser-cooled gas clouds are exploited for precision measurement, enabling clocks that would only go wrong by one second over the lifetime of the Universe \cite{Lud15}, atom interferometers that can accurately measure accelerations \cite{Cro09}, molecular spectroscopy to search for new physics \cite{Jon06,Saf18}, quantum memories \cite{Hes16}, cold electron beams \cite{Spe15} and more \cite{Che99,Eik00}. They are also a crucial stepping stone in nearly all degenerate quantum gas experiments (the exception being experiments based on buffer-gas cooling \cite{Fri98,Dor09}) thereby enabling the exploration of intriguing few- and many-body quantum physics \cite{Blu12,Blo12}. The importance of this research direction has been recognized by several Nobel Prizes, in particular those for laser cooling and trapping in 1997 \cite{Phi98,Chu98,Coh98} and for Bose-Einstein condensation of atomic gases in 2001 \cite{Cor02,Ket02}.
\begin{figure}
\centerline{\includegraphics[width=0.95\columnwidth]{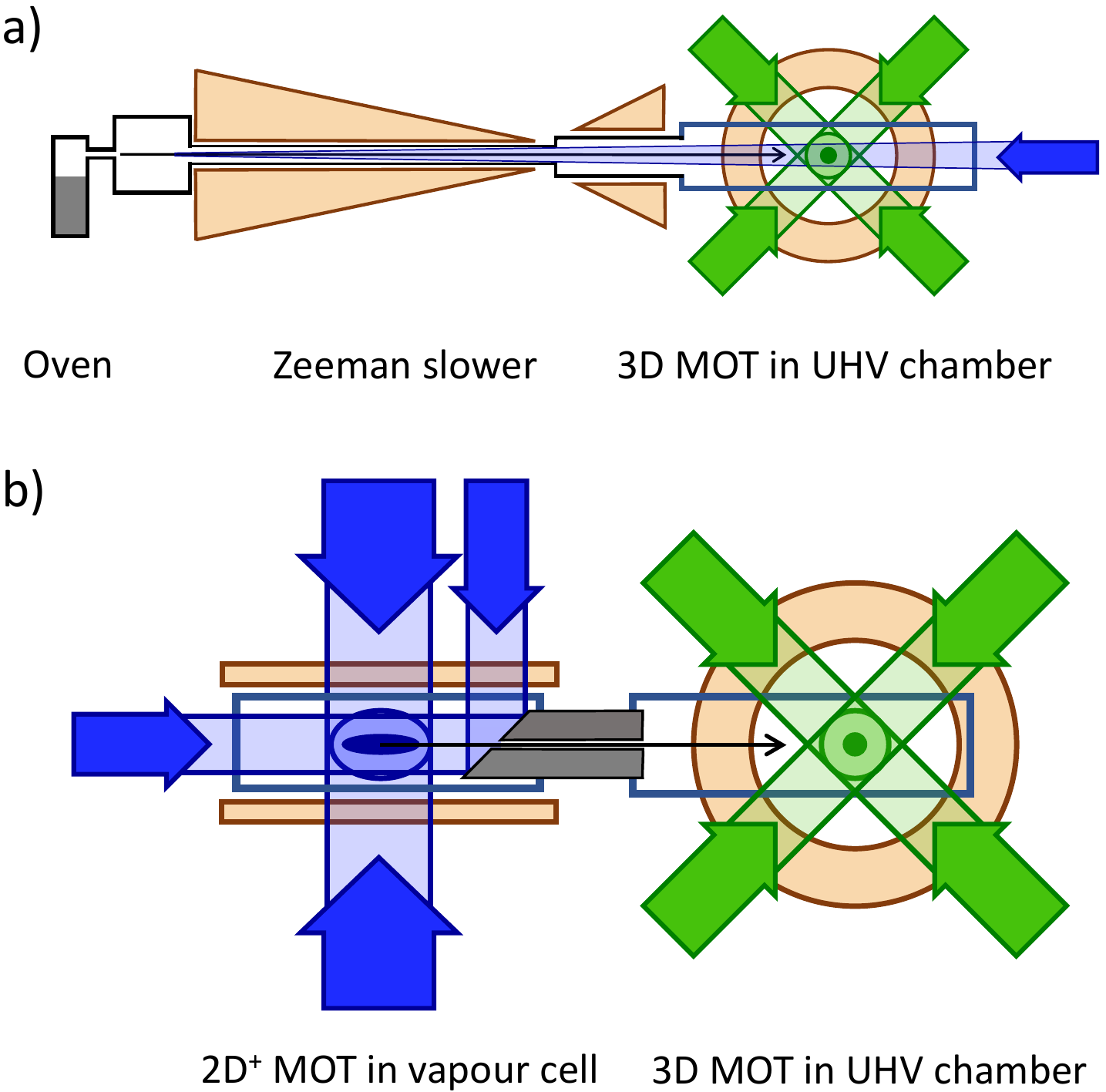}}
\caption{\label{fig:TypicalLaserCoolingConfigurations}
\textbf{Typical experimental set-ups for ultracold gas creation.} A 3D magneto-optical trap (MOT) of ultracold atoms is created in an ultrahigh-vacuum (UHV) chamber. Laser beams used to slow hot thermal atoms are shown in blue, 3D MOT beams in green and magnetic field coils in brown. Vacuum pumps are omitted. a) The MOT is loaded from an atomic beam that is effusing from an oven and Zeeman slowed by a counter-propagating laser beam. b) The 3D MOT is loaded from a 2D$^+$ MOT located in a vapour cell. One of the axial 2D$^+$ MOT beams is reflected from a mirror located around a differential pumping tube. Atoms on axis with this tube experience a light pressure imbalance and are pushed towards the 3D MOT. To create a quantum-degenerate gas in either set-up, atoms are transferred from the 3D MOT into an optical dipole trap or magnetic trap, where they are evaporatively cooled.}
\end{figure}

\section{Basic laser-cooling scheme}
\label{SecII}

To reach quantum degeneracy, the most important quantity to consider is the phase-space density (PSD) of an atomic gas. This represents the number of particles per cubic de Broglie wavelength, and has to reach a value of 1 or higher for a Bose-Einstein condensate (BEC) or a degenerate Fermi gas to form. With an experimental set-up such as that shown in Fig.\,\ref{fig:TypicalLaserCoolingConfigurations}a) as the baseline of our discussion, a series of cooling and compression techniques are applied to achieve this goal. In brief, a beam of fast atoms effuses under vacuum from an oven and is slowed in a Zeeman slower by the scattering of photons from a counter-propagating laser beam. The slowed atoms are captured in a magneto-optical trap (MOT), where they are further cooled and compressed by light scattering until this cooling method reaches its limits. To go beyond this, the atoms are transferred into a magnetic or optical dipole trap and cooled by evaporation until quantum degeneracy is reached.

Let us now take a more detailed look at these slowing and cooling stages. The atomic beam is slowed by scattering photons from a counter-propagating laser beam via the absorption of the momentum of the photons, leading to a deceleration of $\sim 1$\,cm/s in each scattering event. The optically excited atoms emit photons in random directions, thereby recoiling in the opposite direction, which averages out over many scattering events. The Doppler frequency shift of the laser beam reduces while the atoms are slowing down. To keep the atoms in resonance the Zeeman effect can be used by applying a magnetic field that changes along the atoms’ trajectory \cite{Chu89,Met99}. Such a Zeeman slower, first demonstrated in 1982 \cite{Phi82} creates an atomic beam with a velocity of a few tens of m/s that arrives in a central ultrahigh vacuum (UHV) chamber in which the quantum gas will be created.

Doppler cooling, first proposed in the 1970s \cite{Han75,Win79}, is used to cool the atoms further \cite{Chu89,Met99}. Three pairs of counter-propagating laser beams of a few cm diameter are sent along three orthogonal axes onto the centre of the UHV chamber. The laser beams are tuned to a frequency below the atomic transition frequency, such that atoms that travel against them with a few 10\,m/s are Doppler shifted onto resonance. Atoms will therefore mostly scatter photons from beams against which they are propagating and less from the other beams. This results in a friction force irrespective of the direction in which the atoms travel, which will bring the atoms close to standstill. It is as if the atoms are moving through a substance with high viscosity, similar to a finger moving through honey, which led to the name molasses cooling for this technique. The atoms will not quite reach a standstill, because once they are close they will randomly scatter photons from all six beams. Thus the atoms will undergo a random walk in momentum space, which is counterbalanced by the friction force. This leads to a gas sample with a finite temperature, the Doppler temperature $T_D$, which under optimal conditions \cite{Chu89,Met99} turns out to be simply related to the linewidth $\Gamma$ of the atomic transition used for the cooling $T_D=\hbar\Gamma/2k_B$, with $\hbar$ and $k_B$ the reduced Planck constant and the Boltzmann constant, respectively.

\begin{figure*}
\includegraphics[width=0.8\textwidth]{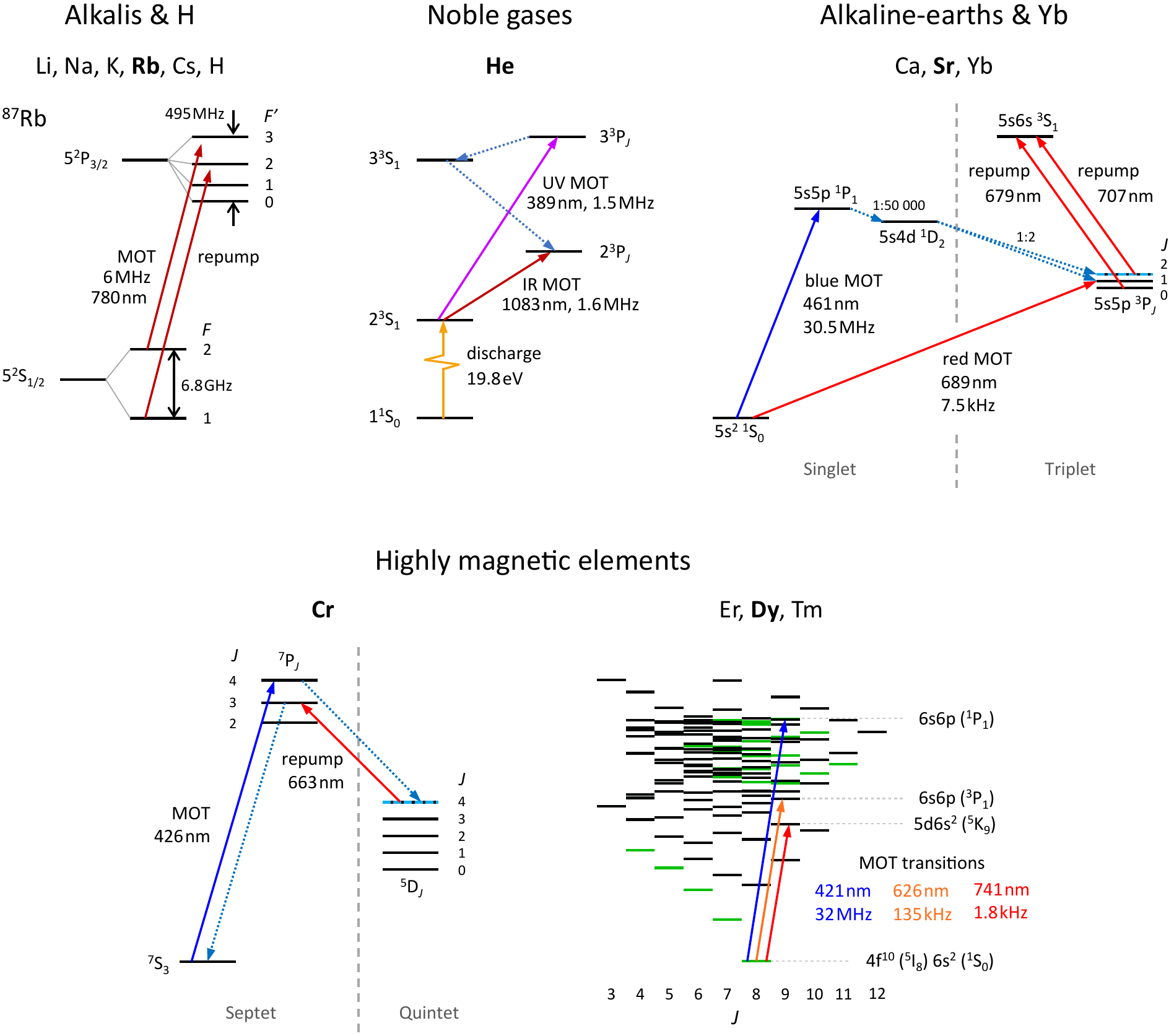}
\caption{\label{fig:TermSchemes}
\textbf{Partial electronic term schemes of elements cooled to quantum degeneracy.} Elements with similar term schemes are grouped and represented by the element marked in bold. The term scheme determines availability and performance of the various laser-cooling schemes for each element. Some MOT transitions are marked. $J$ and $F$ indicate the quantum numbers of electronic and total angular momentum, respectively. Operation of a MOT can lead to decay into dark states, which can be emptied using repump transitions. Atoms in magnetic metastable dark states of Sr ($^3$P$_2$) and Cr ($^5$D$_4$) can be magnetically trapped and accumulated. The Dy states marked in green have even parity, the others odd. The term schemes are adapted from Refs.~\cite{Ste03} (Rb), \cite{Vas12} (He), \cite{Ste14} (Sr), \cite{Sch03a} (Cr) and \cite{You10} (Dy).}
\end{figure*}

The molasses can easily be extended into a trap for atoms by adding a three-dimensional (3D) quadrupole magnetic field centered on the molasses region. The Zeeman effect creates a spatially dependent shift of the laser cooling transition frequency, such that atoms somewhat away from the quadrupole centre are in resonance with the laser cooling beams and scatter more photons. The polarization of the laser cooling beams is chosen such that an atom displaced from the quadrupole centre scatters most photons from the beam pushing it back to the centre. Inside such a MOT, densities exceeding $10^{11}$\,cm$^{-3}$ can be reached, as the first MOT experiment demonstrated  \cite{Raa87}, and  billions of atoms can be accumulated within seconds. In a MOT, the atom number is limited by atom loss processes, such as photo-association of atom pairs into molecules by laser-cooling photons \cite{Jon06}. The density is limited by repulsion between atoms that arises when an atom absorbs a photon emitted by another atom.

Several alkali elements can easily be cooled to a lower temperature by an additional cooling mechanism, Sisyphus cooling \cite{Chu89,Met99}. Sisyphus cooling of alkalis exploits that the ground-state hyperfine levels experience periodic energy shifts in the laser-cooling field through the AC Stark effect. A moving atom that is initially in a potential minimum will climb up a potential hill, converting kinetic into potential energy. When the atom is close to the potential maximum it has a high likelihood of being optically pumped by the laser-cooling light into a hyperfine level that has a potential minimum nearby. The atom will then climb another hill, slowing down, until it is optically pumped again. Clearly the name Sisyphus cooling is well chosen. Optical molasses cools atoms to a velocity of a few cm/s, corresponding to the microkelvin regime.

To reach quantum degeneracy, further cooling stages are employed, in particular evaporative cooling \cite{KET96b}. To avoid the negative side-effects of laser-cooling photons, the first step after the MOT (and Sisyphys cooling) is to store the atoms in a trap that does not rely on near-resonant light, a magnetic trap \cite{Per13} or an optical dipole trap. The latter uses far-off-resonant light and the dispersive response of the atoms to it (via the induced electric dipole moment of the atoms in the oscillating electric field of the light), to create a near-conservative trapping potential \cite{Gri00}. The gas is further cooled by evaporation, during which the hottest atoms are removed from the gas and the remaining atoms thermalize by elastic collisions at a lower temperature. While the gas cools it shrinks in volume if confined in a suitable trap, which can lead to density increase. The evaporation is forced by continuously lowering the trap depth as the cloud cools, maintaining a roughly constant depth-to-temperature ratio \cite{KET96b}. Forced evaporative cooling typically takes seconds to a few tens of seconds. During this time it is necessary to reduce all heating and atom loss processes sufficiently to avoid depletion of the gas before quantum degeneracy is reached. In particular this usually means performing evaporative cooling in an ultrahigh vacuum, where few residual room-temperature gas molecules exist that could knock ultracold atoms out of the trap. Successful evaporative cooling increases the PSD of the gas and eventually a Bose-Einstein condensate or a degenerate Fermi gas forms, depending on the quantum statistics of the isotope used \cite{Dal99,Gio08}.

\section{Improved laser cooling techniques for quantum gas creation}
\label{SecIII}
The laser-cooling techniques outlined so far are sufficient to provide a sample of atoms suitable for reaching the initial conditions for evaporative cooling to BEC. However even the first BEC experiments included additional techniques to more quickly accumulate more atoms in a MOT or to improve the PSD of the laser-cooled cloud before evaporation. We will now discuss laser-cooling innovations that are used to obtain quantum gases of each element. The discussion is ordered along increasing term scheme complexity of the elements, see Fig.\,\ref{fig:TermSchemes}.

\subsection{Alkali quantum gases}
Alkali atoms, especially rubidium (Rb) and sodium (Na), are easy to cool to quantum degeneracy because of their convenient energy structure and scattering properties. Nevertheless, as detailed below, several innovations have been introduced to improve the atom source (2D MOT), to make the set-up more compact (atom chip and mirror MOT), and to increase the laser cooling performance (dark SPOT, narrow-line, grey molasses and Raman cooling).

{\it Improved atom sources} --- The first BEC experiment \cite{And95} captured Rb atoms from a near-room-temperature Rb vapour by ``simply'' operating a MOT in a UHV vapour cell. The Rb pressure at the beginning of the MOT loading phase was increased to enhance the MOT atom number, making it necessary to wait for the Rb pressure to decrease before evaporation to BEC could be started.

The MOT loading time can be dramatically reduced and much larger MOT atom numbers can be reached by adding a separate source of pre-cooled atoms, see Fig.~\ref{fig:TypicalLaserCoolingConfigurations}. A Zeeman slower was employed as such a source in the first Na BEC experiment \cite{Dav95} and this demonstrated much larger condensates and faster cycle times. Another source design (which was used in the second Rb BEC machine at JILA \cite{Mya97}) uses a MOT in a separate source chamber, and periodically accelerates the accumulated atoms by a resonant laser beam, pushing the atoms through a differential pumping tube into a UHV chamber, where they are further accumulated in a second MOT. In later machines the first MOT was given an elongated cylindrical symmetry using a 2D quadrupole magnetic field. A high-flux beam of cold atoms was continuously sent to the UHV MOT, either by replacing the axial MOT beams by a push beam or by creating a shadow in the centre of one of the axial MOT beams to imbalance their light pressure and create an escape path for the atoms, see 2D$^+$ MOT in Fig.\,\ref{fig:TypicalLaserCoolingConfigurations}b). The 2D MOT can be loaded either from a thermal vapour or from an oven beam source \cite{Wey97,Die98}. Somewhat related, the atom flux into the MOT in machines based on an oven and Zeeman slower can be improved by transversally molasses-cooling the atoms escaping the oven.

A range of BEC experiments target simplified laser cooling architectures. A UHV-chamber MOT is not needed in machines that transfer atoms from a vapour cell MOT into a magnetic trap that moves the atoms from the vapour cell to a UHV chamber \cite{Gre01, Lew03}. Compact BEC machines with fast cycle time are enabled by the use of magnetic microtraps (`atom chips') \cite{For07}, often employing a mirror MOT. Other methods to simplify the MOT beam geometry, such as a pyramid MOT \cite{Lee96} and a grating MOT \cite{Van10} have also been demonstrated. The compactness of atom chip machines has even enabled BEC in space \cite{Bec18,Ave20}.

{\it Rb and Na BEC} --- Alkali elements have two hyperfine ground states that are separated by several hundred times the linewidth of the laser cooling transition. Single-frequency MOT beams can only excite atoms from one hyperfine state into an optically excited state manifold, from which they sometimes decay into the other hyperfine state. Atoms in this `dark' state are out of resonance with the MOT light and would fly out of the trap if no further action is taken. The simplest approach to keep the atoms is to optically repump them back into the original hyperfine state using an additional laser beam that addresses the dark state and illuminates the whole MOT. A slight variation of this approach can be used to reduce the light-dependent processes limiting atom density and number in a MOT. By applying the repumping light only in the outer regions of the MOT, atoms in the MOT centre are pumped into the dark state and can accumulate in free flight at high density and in high number. Before they fly out of the MOT they will enter the region with repumping light, become sensitive to MOT light again and are pushed back into the MOT centre. This approach can be implemented by using two repump laser beams crossing at roughly 90$^\circ$ that have a dark spot shadowed into their centre, which inspired the name `dark SPOT' (SPontaneous-force Optical Trap) for this type of MOT \cite{Ket93, And94}. This dark SPOT makes it easier to reach the threshold at which evaporative cooling starts to work and allows to create condensates with more atoms.

{\it Li and K quantum gases} --- The alkalis potassium (K) and lithium (Li) stand out because they have naturally occurring and stable fermionic isotopes that were used to create the first degenerate atomic Fermi gases \cite{DeM99,Tru01}. Bosonic Li was used for one of the first condensates \cite{Bra95,BRA97b}. These original experiments did not require additional laser cooling techniques. However, advanced laser-cooling techniques have improved the performance of K and Li quantum gas machines, in particular narrow-line laser cooling and grey-molasses cooling. Potassium and lithium are more difficult to laser-cool than the elements discussed so far, because the hyperfine splitting of these lighter elements is small, which compromises the performance of Sisyphus cooling. Improved cooling performance has been reached by using transitions with narrow linewidth \cite{Dua11,McK11}, because the Doppler temperature is proportional to that linewidth. Even better performance has been obtained by using a grey molasses \cite{Sal13}. Grey molasses exploits bright and dark eigenstates of the atomic ground state manifold in presence of cooling light. The bright states are periodically light-shifted, which allows Sisyphus-like cooling. Moving atoms are transferred from the dark to the bright state at the potential minimum and back again near the maximum, thereby losing kinetic energy. The coupling between dark and bright states is velocity-dependent, which leads to accumulation of slow atoms in dark states, where photon scattering is suppressed. This aspect of grey molasses cooling is similar to velocity-selective coherent population trapping, which can even reach momentum spreads below the level of a single-photon recoil \cite{Asp88}.

\begin{figure}
\centerline{\includegraphics[width=0.95\columnwidth]{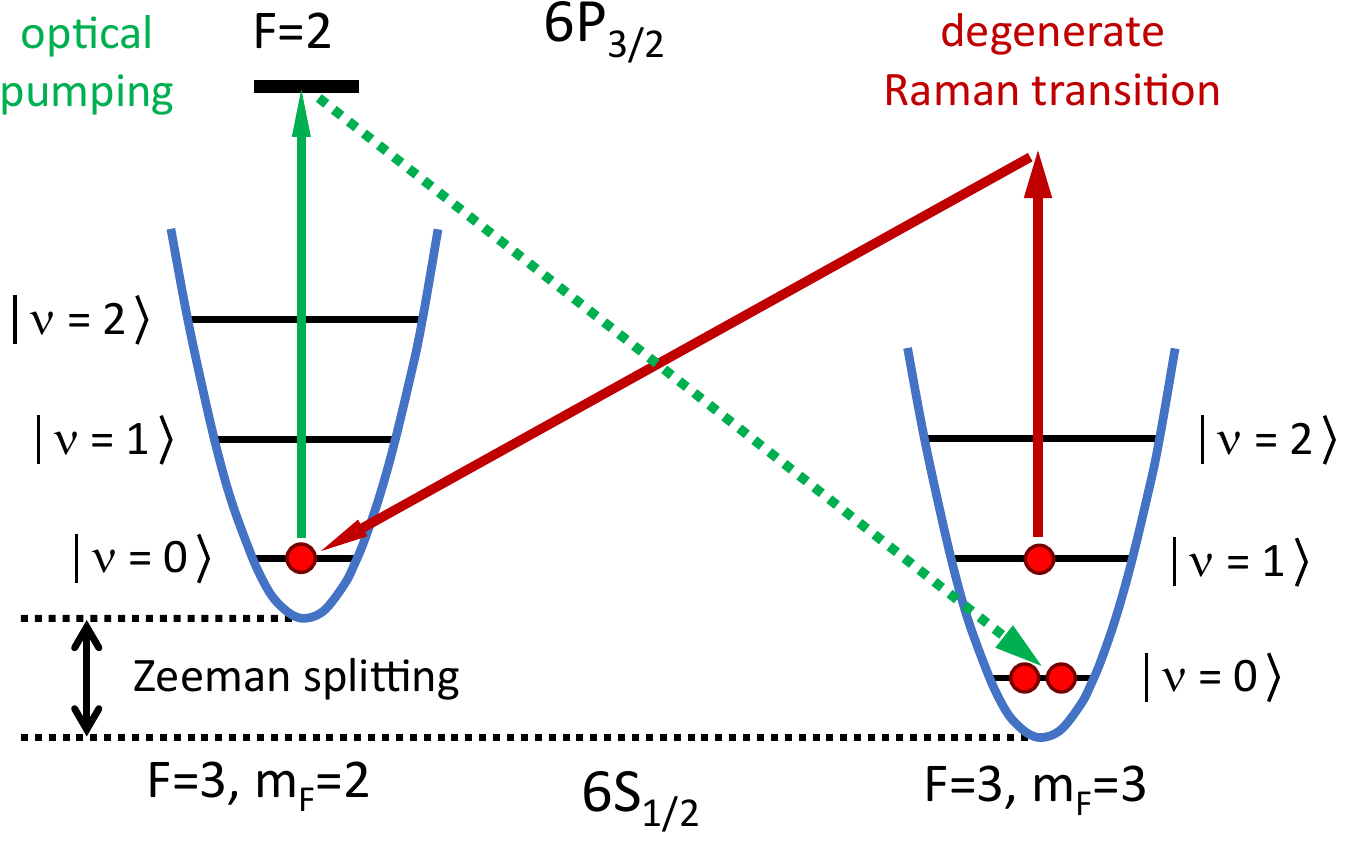}}
\caption{\label{fig:RamanSidebandCooling}
\textbf{Raman sideband cooling of Cs.} Relevant energy levels and optical transitions involved in Raman sideband cooling, which was an essential laser-cooling stage to create the first Cs condensate. Figure adapted from \cite{Vul98}.}
\end{figure}

{\it Cs BEC} --- The achievement of BEC with caesium (Cs) \cite{Web03} relied on a powerful laser cooling technique, Raman sideband cooling \cite{Ham98,Vul98,Han00,Ker00,Tre01}. This technique can be applied after the MOT stage, with atoms trapped in an 3D optical lattice. In Ref.~\cite{Web03} the lattice is an optical dipole trap formed by a standing wave of four interfering laser beams, resulting in periodic lattice wells, which are sparsely populated by atoms. Initially most atoms are trapped in excited vibrational levels $\ket{\nu}$ of their lattice well. Raman sideband cooling transfers the atoms into the lowest vibrational level, reducing the entropy of the gas. To start a cooling cycle the atoms are optically pumped into the $\ket{F=3, m_F=3}$ state ($F$ and $m_F$ are the quantum numbers of total and projected angular momentum, respectively) and a magnetic field is applied that splits the $m_F$ states by the same energy as the low-lying vibrational levels of a lattice well, see Fig.\,\ref{fig:RamanSidebandCooling}. The lattice laser light then couples $\ket{F=3, m_F=3}\ket{\nu}$ to $\ket{F=3, m_F=2}\ket{\nu-1}$ and $\ket{F=3, m_F=1}\ket{\nu-2}$ by two-photon Raman transitions that do not change the energy (for simplicity the second coupling is not shown in the figure). A repump laser couples $\ket{F=3, m_F=1,2}$ atoms to an excited-state manifold from which atoms can decay to $\ket{F=3, m_F=3}$ by spontaneous emission. This absorption-emission cycle is unlikely to change the vibrational level because the lattice wells are so tight that internal and external atomic degrees of freedom are nearly decoupled. This regime is reached when the quantum-mechanical ground state of the well is smaller than $\lambda/2\pi$, and is called the Lamb-Dicke regime \cite{Dic53,Phi98}. The end result of one cooling cycle is that the vibrational level is lowered by one or two units. After several cooling cycles most atoms will reach the harmonic oscillator ground state. To subsequently perform evaporative cooling the gas is released into a large volume optical dipole trap by ramping off the optical lattice. The PSD reached at this stage is at least one order of magnitude higher than what could be reached from a Cs MOT alone, which was crucial to obtain the first Cs condensate.

{\it H BEC} --- Hydrogen (H), which has essentially the same electronic structure as the alkalis, was Bose condensed without laser cooling, via cryogenic buffer-gas precooling followed by evaporative cooling \cite{Fri98}. Laser cooling of H \cite{Set93}, and even of anti-H \cite{Bak21}, 
was demonstrated, but requires light near the Lyman-$\alpha$ wavelength (121 nm), which is very challenging to produce.

\subsection{Beyond alkalis}

The first degenerate quantum gases were created with alkali atoms because these elements possess a simple energy level structure and laser-cooling transitions at convenient wavelengths. Since then, other elements have been cooled to quantum degeneracy, tremendously enriching what can be explored with quantum gases. In addition, the more complex level structures of these elements have often allowed to improve laser cooling, in some cases with vastly enhanced results.

{\it He$^*$ quantum gases} --- The first gas of multi-valence-electron atoms to be cooled to quantum degeneracy consisted of metastable bosonic helium-4 ($^4$He$^*$), later followed by fermionic $^3$He$^*$ \cite{Vas12}. These quantum gases stand out because the highly excited (18\,eV) metastable state makes it possible to detect individual atoms when they impact a microchannel plate detector. The metastable state is also the ground state of the two laser-cooling transitions, one in the infrared (wavelength 1083\,nm) and one in the ultraviolet (389\,nm). The highest MOT PSDs were achieved by accumulating a large sample of atoms in an infrared MOT and subsequently executing a short MOT stage on the ultraviolet transition. This combined the advantage of a high atom capture velocity of the infrared MOT with the relatively low rate of light-assisted collisions of the ultraviolet MOT.

{\it Yb, Ca, Sr quantum gases} --- Alkalis and metastable He have only one valence electron that is used for laser cooling. Elements with two valence electrons offer new laser cooling opportunities, which were used to create quantum gases of the lanthanide ytterbium (Yb) \cite{Tak03b} and the alkaline-earths calcium (Ca) \cite{Kra09} and strontium (Sr) \cite{Ste09, Mar09}. The two valence electrons lead to a richer energy-level structure as they can form a spin singlet or triplet. The ground state is a singlet. Optical transitions between it and electronically excited singlet states can have a broad linewidth, whereas intercombination transitions to triplet states are narrow. These narrow transitions and the associated long-lived excited states make these elements uniquely valuable for precision measurement \cite{Lud15,Yu11} and quantum simulation \cite{Gor10,Coo19}. Furthermore, the $^1$S$_0$-$^3$P$_1$ intercombination transitions are well suited for narrow-line laser cooling. Their linewidths (182\,kHz for Yb, 7.5\,kHz for Sr and 370\,Hz for Ca) enable MOTs of a few $\mu$K for Yb \cite{Kuw99} and Ca \cite{Bin01} and below 0.5\,$\mu$K for Sr \cite{Kat99}. These temperatures are even well below what can be reached in alkalis by Sisyphus cooling (a few tens of microkelvin). The density of Sr narrow-line MOTs can reach $10^{12}$\,atoms/cm$^3$, ten times denser than alkali MOTs. Combined with the lower temperatures, this can lead to MOT PSDs of up to 10$^{-2}$ \cite{Kat99}. This upper limit arises from the necessity of MOTs in earth-based laboratories to levitate atoms against gravity, which requires a minimum photon scattering rate (and therefore a minimum light intensity) which in turn leads to a maximum density because of multiple scattering and photo-associative loss of atoms. The PSD can be further increased by levitating the atoms against gravity in an optical dipole trap, which allows one to reduce the cooling laser intensity, leading to PSDs of up to 0.1 for Sr \cite{Ido00}. Recently, Sisyphus cooling has been implemented with Sr \cite{Coo18,Che19,Cov19}, exploiting the difference in AC polarizability of the $^1$S$_0$ and $^3$P$_1$ states and the relatively long lifetime of the $^3$P$_1$ state (22\,$\mu$s). Yb and Sr also have fermionic isotopes that have been cooled to quantum degeneracy \cite{Fuk07,Des10}. These isotopes possess hyperfine structure, necessitating more involved laser-cooling schemes \cite{Muk03,Nor18SWAP,Mun18,Bar18}.

There is a price to pay in laser system complexity to benefit from narrow-line laser cooling. The photon scattering rate on the narrow transitions is insufficient to slow most atoms effusing from the oven sources typically used for these experiments. Luckily these elements possess a broad linewidth transition suitable for this task ($^1$S$_0$-$^1$P$_1$), which requires construction of a second cooling laser system. Ca possesses an additional difficulty: the 370-Hz-wide laser-cooling transition does not provide enough photon scattering rate to support the atoms against gravity, even at high MOT light intensity. To obtain a sufficient scattering rate, the transition is quenched by laser-coupling the $^3$P$_1$ state to a higher-lying singlet state, which quickly decays to the ground state \cite{Bin01}. Finally, the broad $^1$S$_0$-$^1$P$_1$ laser cooling transition of Sr is not completely closed and atoms can decay through an intermediate state into the metastable $^3$P$_2$ state. So as not to lose these atoms, repumping can be used, requiring one or two additional laser sources.

Instead of being a nuisance, the decay to $^3$P$_2$ can also be exploited to easily create mixtures of Sr isotopes \cite{Ste14}, to access advantageous second-stage MOT transitions \cite{Gru02,Hob20}, to enable continuous dipole trap loading \cite{Rie12}, or to accumulate more atoms through a magnetic trapping stage. The latter was key in obtaining the first Sr quantum gases, which consisted of the low abundance (0.6\%) $^{84}$Sr isotope because of its good elastic scattering properties.

{\it Cr, Dy, Er, Tm quantum gases} --- The most complex atoms cooled to quantum degeneracy to date are chromium (Cr) \cite{Gri05} and the lanthanides dysprosium (Dy) \cite{Lu11}, erbium (Er) \cite{Aik12} and thulium (Tm) \cite{Dav20}. They have many unpaired electrons in their ground state and as a result have large electronic spin and magnetic dipole moment, up to 10 Bohr magneton for Dy. This leads to very interesting dipolar quantum many-body physics \cite{Lah09,Bar12} and is valuable for the creation of artificial gauge fields \cite{Bur16,Cha20}. Cr was the first of these elements to be condensed. The initial laser-cooling stage is similar to the one of Sr: atoms leak from a broad line MOT cycle into a dark, metastable state and are magnetically trapped and accumulated \cite{Sch03a}. After repumping to the ground state the atoms are Doppler-cooled while remaining magnetically trapped \cite{Sch03b}. This in-trap cooling delivers good starting conditions for evaporative cooling. In Dy and Er the large electronic spin leads to an extremely complicated electronic level scheme with many metastable states into which optically excited atoms could in principle decay and be lost. It came as a surprise that it is still possible to operate a MOT, even without any repumper \cite{McC06}. These elements each possess narrow-line cooling transitions, enabling cooling performance close to Sr \cite{Ber08,Lu12,Ilz18}. The large magnetic moment of these elements also enables demagnetization cooling, as demonstrated with Cr \cite{Fat06}. An initially spin-polarized gas depolarizes by inelastic dipolar collisions, which leads to the conversion of kinetic energy into magnetic work in presence of a magnetic field that Zeeman splits the spin states. Optical pumping is used to repolarize the sample, allowing continuous demagnetization cooling.

Other elements have been captured in MOTs, but not yet cooled to quantum degeneracy, e.g. Mg \cite{Sen94},  Ba \cite{De09}, Ra \cite{Gue07}, Fr \cite{Sim96}, Ne*, Ar*, Kr*, and Xe* \cite{Vas12}, Fe, Cd \cite{Bri07}, Hg \cite{Hac08}, Ho \cite{Mia14} and Ag \cite{Uhl00}. A recent proposal further extends the range of transition-metal atoms to be laser cooled \cite{Eus20}. These elements form natural targets to create quantum gases with novel properties, enabling new explorations of quantum physics and opportunities for precision measurement.

\begin{figure*}
\includegraphics[width=0.8\textwidth]{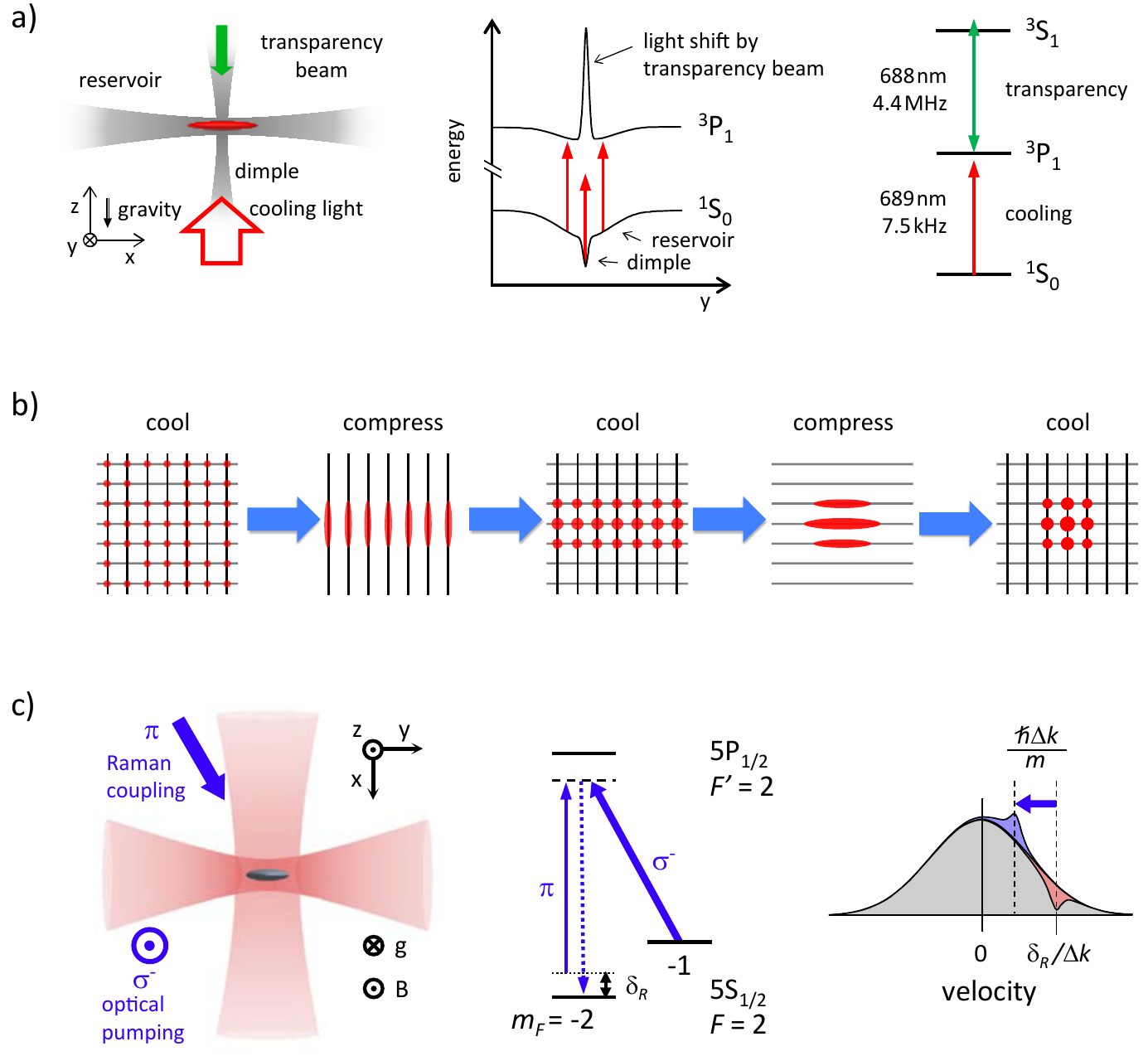}
\caption{\label{fig:LaserCoolingToBEC}
\textbf{Schemes of three experiments achieving BEC by laser cooling.} a) Strontium atoms are confined in a crossed optical dipole trap creating a large reservoir and a small and deep dimple region. The atoms in the reservoir are laser-cooled on the narrow $^1$S$_0$-$^3$P$_1$ transition, whereas the atoms in the dimple region are protected from laser-cooling photons by AC Stark shifting the $^3$P$_1$ out of resonance. A condensate forms in the dimple, while being in thermal contact with the laser-cooled reservoir atoms through elastic collisions \cite{Ste13}. b) Rubidium atoms are confined in a 2D optical lattice. They are Raman sideband-cooled in stages that alternate with compression stages during which one lattice is ramped off till a condensate forms \cite{Hu17}. c) Raman cooling of a Rb cloud in a 3D dipole trap leads to BEC \cite{Urv19}. 
The direction of gravity and magnetic field are indicated by $g$ and $B$ respectively. The Raman lasers have a two photon detuning of $\delta_R$ and address atoms with velocity $\delta_R/\Delta k$ in the direction of the two photon momentum $\Delta k$. A Raman transition reduces the velocity by $\hbar \Delta k/m$. Panels adapted from: a) \cite{Ste13}, b) \cite{Hu17}, c) \cite{Urv19}.}
\end{figure*}

\section{New laser cooling routes to BEC}
\label{SecIV}

{\it Laser cooling to BEC} --- In all experiments described above, laser cooling was followed by evaporative cooling to remove enough entropy from the gas to achieve condensation. This raises the question whether it is possible to create atomic Bose-Einstein condensates by removing the entropy by laser cooling {\em alone}, and thus render the condensates' existence compatible with simultaneous laser cooling. This was achieved in three experiments \cite{Ste13, Hu17, Urv19}, see Fig.\,\ref{fig:LaserCoolingToBEC}. A key challenge of those experiments is to sufficiently suppress the negative effects of laser-cooling photons on quantum gases, such as heating and loss of atoms. 

The first experiment cools a Sr cloud in a `reservoir' dipole trap on the narrow $^1$S$_0$-$^3$P$_1$ laser cooling transition to 1\,$\mu$K and a PSD of 0.1. Atoms in the central part of the cloud are protected from laser-cooling photons by shifting the $^3$P$_1$ state out of resonance using an AC Stark shift provided by a `transparency beam'. The optical dipole potential in this small region is lowered using an additional `dimple' dipole laser beam \cite{Pin97, Sta98}. Atoms collect in this deeper dimple well, while remaining thermalized with the surrounding laser-cooled cloud through elastic collisions. This increases the density and PSD of the gas in the dimple and a condensate is formed. This scheme was recently adopted to quickly create cold and dense samples of fermionic Sr in a 3D optical lattice clock \cite{Son20}. 

The more recent experiments \cite{Hu17,Urv19} do not separate the cooling and BEC spatial regions and Raman-cool Rb samples to quantum degeneracy. Careful choice of parameters minimizes the negative effects of spontaneously emitted photons. The first of those experiments holds the Rb sample in a 2D optical lattice. To increase the PSD, Raman sideband cooling is interrupted by periods during which one of the two lattices is ramped off, allowing the gas to shrink in that dimension and thermalize by elastic collisions. After three such cycles in alternate directions an array of 1D condensates is formed during Raman cooling, which can be merged into 2D condensates by ramping one lattice off. The second experiment even achieves BEC directly in an optical dipole trap by Raman cooling, see Fig.\,\ref{fig:LaserCoolingToBEC}c). The Raman transition selects $\ket{5S_{1/2}, F=2, m_F=-2}$ atoms in the high-velocity tail of the thermal distribution and couples them to $\ket{5S_{1/2}, F=2, m_F=-1}$, from which they are optically pumped back to the original internal state, reducing their velocity in the process. After five stages of Raman cooling under carefully chosen conditions a condensate is created.

{\it Continuous BEC} --- All BEC experiments discussed so far create condensates by executing a sequence of cooling steps one after the other in time. After the condensate is produced, it typically decays through unavoidable loss processes such as molecule formation. This limits what can be done with Bose-Einstein condensates. For example, it becomes impossible to extract a continuous-wave atom laser. Such a phase-coherent, uninterrupted beam of quantum-degenerate atoms has promising applications for precision measurement through atom interferometry \cite{Rob13}. An important step towards this goal was maintaining a trap filled with Bose-condensed atoms by periodically adding fresh condensates \cite{Chi02}. The remaining challenge was to overcome the problem that each refilling event was accompanied by motional excitations, atom number jumps and phase-slips disrupting the coherence. In order to coherently maintain a stationary condensate, a solution is to use a Bose-stimulated amplification process \cite{Mie98,Rob08} adding new atoms from a continuous source of near-degenerate gas. Such a system was recently realised \cite{Che20}, relying on the laser-cooling to BEC technique of Fig.\,\ref{fig:LaserCoolingToBEC}a). To provide the continuous source, all laser-cooling stages have to be executed simultaneously and rendered compatible with BEC. Sr atoms are streamed from an oven through a spatial sequence of laser-cooling stages, first using broad, then narrow laser-cooling transitions. The atoms finally accumulate in a reservoir dipole trap where the gas is laser-cooled to high PSD. Bose-stimulated elastic collisions within the thermal gas provide gain to a condensate located in a dimple. The only step missing to create the long-sought continuous-wave atom laser is the addition of a coherent atom outcoupling mechanism \cite{Rob13}.

Looking ahead, two developments involving laser cooling are rapidly progressing that, among many other things, may lead to a new way of producing BEC by laser cooling, or to condensates of ground-state molecules.

{\it Micro BEC} --- The first development is the approach to build up a condensate one-by-one from individual atoms. This approach would start by capturing individual atoms from a MOT in tightly confining dipole traps, called optical tweezers \cite{KaufmanNi2021}.
The capture is probabilistic, and light-assisted collisions cause each tweezer trap to be loaded with either one or no atom. The occupancy can be detected by fluorescence imaging without losing the atoms \cite{Ott16}. The atoms can then be sorted from the outer tweezers into unoccupied tweezers in the middle of the array, e.g. using a movable tweezer \cite{Bar16} controlled by a computer algorithm. After the innermost tweezer array reaches unity filling, the atoms are cooled to the vibrational ground state of the tweezers, e.g. using Raman sideband cooling. The final step to obtain a condensate will be to merge the tweezers into one larger trap, such that the atoms undergo a Mott insulator-to-superfluid transition \cite{Ols02}. In this way, condensates with a small and well defined atom number should be within reach.

{\it Molecular quantum gases} --- The second development is the extension of laser cooling to molecules \cite{Tar18,Fit21}. Molecules have vibrational and rotational structure. In many cases optical excitation is followed by decay into a large number of ro-vibrational levels, which makes it essentially impossible to scatter the thousands of photons needed to cool a room-temperature molecule to the ultracold regime. However quite recently it was found that some molecules, such as SrF, possess nearly closed optical transitions because they have a valence electron that is barely involved in the molecular bond. The use of only a few repump lasers in addition to the cooling laser allows to scatter enough photons. Since this discovery, many species of molecules have been laser-cooled \cite{Tar18,Fit21}. Slowers \cite{Bar12b} and MOTs \cite{Hum13,Bar14,Nor16} have been built, molecules have been trapped in magnetic \cite{Wil18,McC18} and optical dipole traps and even individually in optical tweezers \cite{And19}. An important enabling technology for molecules is the so-called radiofrequency MOT \cite{Hum13,Nor16,Fit21}. In such a MOT, the magnetic field and the polarization of the light beams are switched synchronously at a rate similar to that at which optical pumping occurs. Another technique is the dual frequency (dc) MOT \cite{Fit21} in which a second frequency with opposite circular polarization is added. In both schemes the molecules are repumped out of dark states so that the cooling process remains effective.  So far the highest PSD reached ($3.3\times 10^{-8}$, with $5\times10^4$ YO molecules \cite{Shi20}) is still orders of magnitude below quantum degeneracy. However it is likely that further improvements will lead to molecular samples that allow evaporative (or sympathetic) cooling to BEC. This requires a large enough sample in a magnetic or dipole trap with an elastic collision rate that is substantially higher than the loss rate. The latter is not trivial because the complex inner structure of molecules easily leads to strong inelastic losses during collisions. Fortunately it is possible to suppress such inelastic collisions by choosing suitable conditions, including the choice of molecular state, trap and applied electromagnetic fields \cite{Que12,Lem13}. Some of these methods have already been shown to work \cite{DeM19b,Son20cco,Mat20} with gases of ultracold, deeply bound molecules that were obtained not by direct laser cooling, but by associating pairs of ultracold atoms into weakly bound molecules \cite{Koh06} and then optically transferring those molecules into the desired bound state \cite{Boh17}. Even quantum gases of KRb molecules in their rotational and vibrational ground state have recently been created in this way \cite{DeM19a,DeM19b}. This strategy is limited to quantum gases of molecules consisting of elements that can be laser-cooled. Direct laser cooling of molecules is applicable to a different and large class of molecular species. Together these two strategies open great opportunities for quantum-state-controlled chemistry, quantum simulation and fundamental physics \cite{Que12,Lem13,Bar12,Boh17,Saf18}.

\section{Outlook}

Many more laser-cooling techniques have been and are constantly being developed, some of which will undoubtedly play a role in future quantum gas experiments. Examples of these methods are one-way wall cooling \cite{Sch10}, cavity cooling \cite{Vul00,Mau04,Hos17,Del20}, bichromatic force cooling \cite{Gri90,Sod97,Cor15}, or EIT cooling \cite{Sch01}. Furthermore, laser cooling can be applied beyond atoms and molecules, for example to cool ions \cite{Esc03}, micromechanical oscillators \cite{Asp14,Ros18b} or beads in dipole traps \cite{Del20}. More than 45 years after laser cooling of atoms was proposed, and more than 25 years after a BEC of an atomic gas was realized, the field of laser cooling continues to give rise to many exciting new developments.

\begin{acknowledgments}
We thank Robert Spreeuw, Ben van Linden van den Heuvell and Shayne Bennetts for helpful comments on the manuscript. We are grateful for funding from the NWO through Vici grant No 680-47-619 and grant No 680.92.18.05 (QuSim 2.0 programme) and from the European Union’s Horizon 2020 research and innovation programme under grant agreement No 820404 (iqClock project) and No 860579 (MoSaiQC project).
\end{acknowledgments}


\begin{thebibliography}{155}%
\makeatletter
\providecommand \@ifxundefined [1]{%
 \@ifx{#1\undefined}
}%
\providecommand \@ifnum [1]{%
 \ifnum #1\expandafter \@firstoftwo
 \else \expandafter \@secondoftwo
 \fi
}%
\providecommand \@ifx [1]{%
 \ifx #1\expandafter \@firstoftwo
 \else \expandafter \@secondoftwo
 \fi
}%
\providecommand \natexlab [1]{#1}%
\providecommand \enquote  [1]{``#1''}%
\providecommand \bibnamefont  [1]{#1}%
\providecommand \bibfnamefont [1]{#1}%
\providecommand \citenamefont [1]{#1}%
\providecommand \href@noop [0]{\@secondoftwo}%
\providecommand \href [0]{\begingroup \@sanitize@url \@href}%
\providecommand \@href[1]{\@@startlink{#1}\@@href}%
\providecommand \@@href[1]{\endgroup#1\@@endlink}%
\providecommand \@sanitize@url [0]{\catcode `\\12\catcode `\$12\catcode
  `\&12\catcode `\#12\catcode `\^12\catcode `\_12\catcode `\%12\relax}%
\providecommand \@@startlink[1]{}%
\providecommand \@@endlink[0]{}%
\providecommand \url  [0]{\begingroup\@sanitize@url \@url }%
\providecommand \@url [1]{\endgroup\@href {#1}{\urlprefix }}%
\providecommand \urlprefix  [0]{URL }%
\providecommand \Eprint [0]{\href }%
\providecommand \doibase [0]{https://doi.org/}%
\providecommand \selectlanguage [0]{\@gobble}%
\providecommand \bibinfo  [0]{\@secondoftwo}%
\providecommand \bibfield  [0]{\@secondoftwo}%
\providecommand \translation [1]{[#1]}%
\providecommand \BibitemOpen [0]{}%
\providecommand \bibitemStop [0]{}%
\providecommand \bibitemNoStop [0]{.\EOS\space}%
\providecommand \EOS [0]{\spacefactor3000\relax}%
\providecommand \BibitemShut  [1]{\csname bibitem#1\endcsname}%
\let\auto@bib@innerbib\@empty
%
\bibitem [{\citenamefont {Chu}\ and\ \citenamefont {{Wieman
  (Eds.)}}(1989)}]{Chu89}%
  \BibitemOpen
  \bibfield  {author} {\bibinfo {author} {\bibfnamefont {S.}~\bibnamefont
  {Chu}}\ and\ \bibinfo {author} {\bibfnamefont {C.}~\bibnamefont {{Wieman
  (Eds.)}}},\ }\bibfield  {title} {\bibinfo {title} {Feature issue on laser
  cooling and trapping of atoms},\ }\href
  {https://doi.org/10.1364/JOSAB.6.002020} {\bibfield  {journal} {\bibinfo
  {journal} {J. Opt. Soc. Am. B}\ }\textbf {\bibinfo {volume} {6}},\ \bibinfo
  {pages} {2020} (\bibinfo {year} {1989})}\BibitemShut {NoStop}%
\bibitem [{\citenamefont {Metcalf}\ and\ \citenamefont {van~der
  Straten}(1999)}]{Met99}%
  \BibitemOpen
  \bibfield  {author} {\bibinfo {author} {\bibfnamefont {H.~J.}\ \bibnamefont
  {Metcalf}}\ and\ \bibinfo {author} {\bibfnamefont {P.}~\bibnamefont {van~der
  Straten}},\ }\href@noop {} {\emph {\bibinfo {title} {Laser Cooling and
  Trapping of Neutral Atoms}}}\ (\bibinfo  {publisher} {Springer},\ \bibinfo
  {address} {New York},\ \bibinfo {year} {1999})\BibitemShut {NoStop}%
\bibitem [{\citenamefont {Ludlow}\ \emph {et~al.}(2015)\citenamefont {Ludlow},
  \citenamefont {Boyd}, \citenamefont {Ye}, \citenamefont {Peik},\ and\
  \citenamefont {Schmidt}}]{Lud15}%
  \BibitemOpen
  \bibfield  {author} {\bibinfo {author} {\bibfnamefont {A.~D.}\ \bibnamefont
  {Ludlow}}, \bibinfo {author} {\bibfnamefont {M.~M.}\ \bibnamefont {Boyd}},
  \bibinfo {author} {\bibfnamefont {J.}~\bibnamefont {Ye}}, \bibinfo {author}
  {\bibfnamefont {E.}~\bibnamefont {Peik}},\ and\ \bibinfo {author}
  {\bibfnamefont {P.~O.}\ \bibnamefont {Schmidt}},\ }\bibfield  {title}
  {\bibinfo {title} {Optical atomic clocks},\ }\href
  {https://doi.org/10.1103/RevModPhys.87.637} {\bibfield  {journal} {\bibinfo
  {journal} {Rev. Mod. Phys.}\ }\textbf {\bibinfo {volume} {87}},\ \bibinfo
  {pages} {637} (\bibinfo {year} {2015})}\BibitemShut {NoStop}%
\bibitem [{\citenamefont {Cronin}\ \emph {et~al.}(2009)\citenamefont {Cronin},
  \citenamefont {Schmiedmayer},\ and\ \citenamefont {Pritchard}}]{Cro09}%
  \BibitemOpen
  \bibfield  {author} {\bibinfo {author} {\bibfnamefont {A.~D.}\ \bibnamefont
  {Cronin}}, \bibinfo {author} {\bibfnamefont {J.}~\bibnamefont
  {Schmiedmayer}},\ and\ \bibinfo {author} {\bibfnamefont {D.~E.}\ \bibnamefont
  {Pritchard}},\ }\bibfield  {title} {\bibinfo {title} {Optics and
  interferometry with atoms and molecules},\ }\href
  {https://doi.org/10.1103/RevModPhys.81.1051} {\bibfield  {journal} {\bibinfo
  {journal} {Rev. Mod. Phys.}\ }\textbf {\bibinfo {volume} {81}},\ \bibinfo
  {pages} {1051} (\bibinfo {year} {2009})}\BibitemShut {NoStop}%
\bibitem [{\citenamefont {Jones}\ \emph {et~al.}(2006)\citenamefont {Jones},
  \citenamefont {Tiesinga}, \citenamefont {Lett},\ and\ \citenamefont
  {Julienne}}]{Jon06}%
  \BibitemOpen
  \bibfield  {author} {\bibinfo {author} {\bibfnamefont {K.~M.}\ \bibnamefont
  {Jones}}, \bibinfo {author} {\bibfnamefont {E.}~\bibnamefont {Tiesinga}},
  \bibinfo {author} {\bibfnamefont {P.~D.}\ \bibnamefont {Lett}},\ and\
  \bibinfo {author} {\bibfnamefont {P.~S.}\ \bibnamefont {Julienne}},\
  }\bibfield  {title} {\bibinfo {title} {Ultracold photoassociation
  spectroscopy: Long-range molecules and atomic scattering},\ }\href
  {https://doi.org/10.1103/RevModPhys.78.483} {\bibfield  {journal} {\bibinfo
  {journal} {Rev. Mod. Phys.}\ }\textbf {\bibinfo {volume} {78}},\ \bibinfo
  {pages} {483} (\bibinfo {year} {2006})}\BibitemShut {NoStop}%
\bibitem [{\citenamefont {Safronova}\ \emph {et~al.}(2018)\citenamefont
  {Safronova}, \citenamefont {Budker}, \citenamefont {DeMille}, \citenamefont
  {Kimball}, \citenamefont {Derevianko},\ and\ \citenamefont {Clark}}]{Saf18}%
  \BibitemOpen
  \bibfield  {author} {\bibinfo {author} {\bibfnamefont {M.~S.}\ \bibnamefont
  {Safronova}}, \bibinfo {author} {\bibfnamefont {D.}~\bibnamefont {Budker}},
  \bibinfo {author} {\bibfnamefont {D.}~\bibnamefont {DeMille}}, \bibinfo
  {author} {\bibfnamefont {D.~F.~J.}\ \bibnamefont {Kimball}}, \bibinfo
  {author} {\bibfnamefont {A.}~\bibnamefont {Derevianko}},\ and\ \bibinfo
  {author} {\bibfnamefont {C.~W.}\ \bibnamefont {Clark}},\ }\bibfield  {title}
  {\bibinfo {title} {Search for new physics with atoms and molecules},\ }\href
  {https://doi.org/10.1103/RevModPhys.90.025008} {\bibfield  {journal}
  {\bibinfo  {journal} {Rev. Mod. Phys.}\ }\textbf {\bibinfo {volume} {90}},\
  \bibinfo {pages} {025008} (\bibinfo {year} {2018})}\BibitemShut {NoStop}%
\bibitem [{\citenamefont {Heshami}\ \emph {et~al.}(2016)\citenamefont
  {Heshami}, \citenamefont {England}, \citenamefont {Humphreys}, \citenamefont
  {Bustard}, \citenamefont {Acosta}, \citenamefont {Nunn},\ and\ \citenamefont
  {Sussman}}]{Hes16}%
  \BibitemOpen
  \bibfield  {author} {\bibinfo {author} {\bibfnamefont {K.}~\bibnamefont
  {Heshami}}, \bibinfo {author} {\bibfnamefont {D.~G.}\ \bibnamefont
  {England}}, \bibinfo {author} {\bibfnamefont {P.~C.}\ \bibnamefont
  {Humphreys}}, \bibinfo {author} {\bibfnamefont {P.~J.}\ \bibnamefont
  {Bustard}}, \bibinfo {author} {\bibfnamefont {V.~M.}\ \bibnamefont {Acosta}},
  \bibinfo {author} {\bibfnamefont {J.}~\bibnamefont {Nunn}},\ and\ \bibinfo
  {author} {\bibfnamefont {B.~J.}\ \bibnamefont {Sussman}},\ }\bibfield
  {title} {\bibinfo {title} {Quantum memories: emerging applications and recent
  advances},\ }\href {https://doi.org/10.1080/09500340.2016.1148212} {\bibfield
   {journal} {\bibinfo  {journal} {Journal of Modern Optics}\ }\textbf
  {\bibinfo {volume} {63}},\ \bibinfo {pages} {2005} (\bibinfo {year}
  {2016})}\BibitemShut {NoStop}%
\bibitem [{\citenamefont {Speirs}\ \emph {et~al.}(2015)\citenamefont {Speirs},
  \citenamefont {Putkunz}, \citenamefont {McCulloch}, \citenamefont {Nugent},
  \citenamefont {Sparkes},\ and\ \citenamefont {Scholten}}]{Spe15}%
  \BibitemOpen
  \bibfield  {author} {\bibinfo {author} {\bibfnamefont {R.~W.}\ \bibnamefont
  {Speirs}}, \bibinfo {author} {\bibfnamefont {C.~T.}\ \bibnamefont {Putkunz}},
  \bibinfo {author} {\bibfnamefont {A.~J.}\ \bibnamefont {McCulloch}}, \bibinfo
  {author} {\bibfnamefont {K.~A.}\ \bibnamefont {Nugent}}, \bibinfo {author}
  {\bibfnamefont {B.~M.}\ \bibnamefont {Sparkes}},\ and\ \bibinfo {author}
  {\bibfnamefont {R.~E.}\ \bibnamefont {Scholten}},\ }\bibfield  {title}
  {\bibinfo {title} {Single-shot electron diffraction using a cold atom
  electron source},\ }\href {https://doi.org/10.1088/0953-4075/48/21/214002}
  {\bibfield  {journal} {\bibinfo  {journal} {J. Phys. B}\ }\textbf {\bibinfo
  {volume} {48}},\ \bibinfo {pages} {214002} (\bibinfo {year}
  {2015})}\BibitemShut {NoStop}%
\bibitem [{\citenamefont {Chen}\ \emph {et~al.}(1999)\citenamefont {Chen},
  \citenamefont {Li}, \citenamefont {Bailey}, \citenamefont
  {O{\textquoteright}Connor}, \citenamefont {Young},\ and\ \citenamefont
  {Lu}}]{Che99}%
  \BibitemOpen
  \bibfield  {author} {\bibinfo {author} {\bibfnamefont {C.~Y.}\ \bibnamefont
  {Chen}}, \bibinfo {author} {\bibfnamefont {Y.~M.}\ \bibnamefont {Li}},
  \bibinfo {author} {\bibfnamefont {K.}~\bibnamefont {Bailey}}, \bibinfo
  {author} {\bibfnamefont {T.~P.}\ \bibnamefont {O{\textquoteright}Connor}},
  \bibinfo {author} {\bibfnamefont {L.}~\bibnamefont {Young}},\ and\ \bibinfo
  {author} {\bibfnamefont {Z.-T.}\ \bibnamefont {Lu}},\ }\bibfield  {title}
  {\bibinfo {title} {Ultrasensitive isotope trace analyses with a
  magneto-optical trap},\ }\href
  {https://doi.org/10.1126/science.286.5442.1139} {\bibfield  {journal}
  {\bibinfo  {journal} {Science}\ }\textbf {\bibinfo {volume} {286}},\ \bibinfo
  {pages} {1139} (\bibinfo {year} {1999})}\BibitemShut {NoStop}%
\bibitem [{\citenamefont {Eike}\ \emph {et~al.}(2000)\citenamefont {Eike},
  \citenamefont {Luger}, \citenamefont {Manek-Hönninger}, \citenamefont
  {Grimm},\ and\ \citenamefont {Schwalm}}]{Eik00}%
  \BibitemOpen
  \bibfield  {author} {\bibinfo {author} {\bibfnamefont {B.}~\bibnamefont
  {Eike}}, \bibinfo {author} {\bibfnamefont {V.}~\bibnamefont {Luger}},
  \bibinfo {author} {\bibfnamefont {I.}~\bibnamefont {Manek-Hönninger}},
  \bibinfo {author} {\bibfnamefont {R.}~\bibnamefont {Grimm}},\ and\ \bibinfo
  {author} {\bibfnamefont {D.}~\bibnamefont {Schwalm}},\ }\bibfield  {title}
  {\bibinfo {title} {Laser-trapped atoms as a precision target for the storage
  ring {TSR}},\ }\href
  {https://doi.org/https://doi.org/10.1016/S0168-9002(99)01113-4} {\bibfield
  {journal} {\bibinfo  {journal} {Nucl. Instr. Meth. Phys. Res. A}\ }\textbf
  {\bibinfo {volume} {441}},\ \bibinfo {pages} {81 } (\bibinfo {year}
  {2000})}\BibitemShut {NoStop}%
\bibitem [{\citenamefont {Fried}\ \emph {et~al.}(1998)\citenamefont {Fried},
  \citenamefont {Killian}, \citenamefont {Willmann}, \citenamefont {Landhuis},
  \citenamefont {Moss}, \citenamefont {Kleppner},\ and\ \citenamefont
  {Greytak}}]{Fri98}%
  \BibitemOpen
  \bibfield  {author} {\bibinfo {author} {\bibfnamefont {D.~G.}\ \bibnamefont
  {Fried}}, \bibinfo {author} {\bibfnamefont {T.~C.}\ \bibnamefont {Killian}},
  \bibinfo {author} {\bibfnamefont {L.}~\bibnamefont {Willmann}}, \bibinfo
  {author} {\bibfnamefont {D.}~\bibnamefont {Landhuis}}, \bibinfo {author}
  {\bibfnamefont {S.~C.}\ \bibnamefont {Moss}}, \bibinfo {author}
  {\bibfnamefont {D.}~\bibnamefont {Kleppner}},\ and\ \bibinfo {author}
  {\bibfnamefont {T.~J.}\ \bibnamefont {Greytak}},\ }\bibfield  {title}
  {\bibinfo {title} {{Bose}-{Einstein} condensation of atomic hydrogen},\
  }\href {https://doi.org/10.1103/PhysRevLett.81.3811} {\bibfield  {journal}
  {\bibinfo  {journal} {Phys. Rev. Lett.}\ }\textbf {\bibinfo {volume} {81}},\
  \bibinfo {pages} {3811} (\bibinfo {year} {1998})}\BibitemShut {NoStop}%
\bibitem [{\citenamefont {Doret}\ \emph {et~al.}(2009)\citenamefont {Doret},
  \citenamefont {Connolly}, \citenamefont {Ketterle},\ and\ \citenamefont
  {Doyle}}]{Dor09}%
  \BibitemOpen
  \bibfield  {author} {\bibinfo {author} {\bibfnamefont {S.~C.}\ \bibnamefont
  {Doret}}, \bibinfo {author} {\bibfnamefont {C.~B.}\ \bibnamefont {Connolly}},
  \bibinfo {author} {\bibfnamefont {W.}~\bibnamefont {Ketterle}},\ and\
  \bibinfo {author} {\bibfnamefont {J.~M.}\ \bibnamefont {Doyle}},\ }\bibfield
  {title} {\bibinfo {title} {Buffer-gas cooled {Bose}-{Einstein} condensate},\
  }\href {https://doi.org/10.1103/PhysRevLett.103.103005} {\bibfield  {journal}
  {\bibinfo  {journal} {Phys. Rev. Lett.}\ }\textbf {\bibinfo {volume} {103}},\
  \bibinfo {pages} {103005} (\bibinfo {year} {2009})}\BibitemShut {NoStop}%
\bibitem [{\citenamefont {Blume}(2012)}]{Blu12}%
  \BibitemOpen
  \bibfield  {author} {\bibinfo {author} {\bibfnamefont {D.}~\bibnamefont
  {Blume}},\ }\bibfield  {title} {\bibinfo {title} {Few-body physics with
  ultracold atomic and molecular systems in traps},\ }\href
  {https://doi.org/10.1088/0034-4885/75/4/046401} {\bibfield  {journal}
  {\bibinfo  {journal} {Rep. Prog. Phys.}\ }\textbf {\bibinfo {volume} {75}},\
  \bibinfo {pages} {046401} (\bibinfo {year} {2012})}\BibitemShut {NoStop}%
\bibitem [{\citenamefont {Bloch}\ \emph {et~al.}(2012)\citenamefont {Bloch},
  \citenamefont {Dalibard},\ and\ \citenamefont {Nascimb{\`e}ne}}]{Blo12}%
  \BibitemOpen
  \bibfield  {author} {\bibinfo {author} {\bibfnamefont {I.}~\bibnamefont
  {Bloch}}, \bibinfo {author} {\bibfnamefont {J.}~\bibnamefont {Dalibard}},\
  and\ \bibinfo {author} {\bibfnamefont {S.}~\bibnamefont {Nascimb{\`e}ne}},\
  }\bibfield  {title} {\bibinfo {title} {Quantum simulations with ultracold
  quantum gases},\ }\href {https://doi.org/10.1038/nphys2259} {\bibfield
  {journal} {\bibinfo  {journal} {Nature Physics}\ }\textbf {\bibinfo {volume}
  {8}},\ \bibinfo {pages} {267} (\bibinfo {year} {2012})}\BibitemShut {NoStop}%
\bibitem [{\citenamefont {Phillips}(1998)}]{Phi98}%
  \BibitemOpen
  \bibfield  {author} {\bibinfo {author} {\bibfnamefont {W.~D.}\ \bibnamefont
  {Phillips}},\ }\bibfield  {title} {\bibinfo {title} {Nobel lecture: Laser
  cooling and trapping of neutral atoms},\ }\href
  {https://doi.org/10.1103/RevModPhys.70.721} {\bibfield  {journal} {\bibinfo
  {journal} {Rev. Mod. Phys.}\ }\textbf {\bibinfo {volume} {70}},\ \bibinfo
  {pages} {721} (\bibinfo {year} {1998})}\BibitemShut {NoStop}%
\bibitem [{\citenamefont {Chu}(1998)}]{Chu98}%
  \BibitemOpen
  \bibfield  {author} {\bibinfo {author} {\bibfnamefont {S.}~\bibnamefont
  {Chu}},\ }\bibfield  {title} {\bibinfo {title} {Nobel lecture: The
  manipulation of neutral particles},\ }\href
  {https://doi.org/10.1103/RevModPhys.70.685} {\bibfield  {journal} {\bibinfo
  {journal} {Rev. Mod. Phys.}\ }\textbf {\bibinfo {volume} {70}},\ \bibinfo
  {pages} {685} (\bibinfo {year} {1998})}\BibitemShut {NoStop}%
\bibitem [{\citenamefont {Cohen-Tannoudji}(1998)}]{Coh98}%
  \BibitemOpen
  \bibfield  {author} {\bibinfo {author} {\bibfnamefont {C.~N.}\ \bibnamefont
  {Cohen-Tannoudji}},\ }\bibfield  {title} {\bibinfo {title} {Nobel lecture:
  Manipulating atoms with photons},\ }\href
  {https://doi.org/10.1103/RevModPhys.70.707} {\bibfield  {journal} {\bibinfo
  {journal} {Rev. Mod. Phys.}\ }\textbf {\bibinfo {volume} {70}},\ \bibinfo
  {pages} {707} (\bibinfo {year} {1998})}\BibitemShut {NoStop}%
\bibitem [{\citenamefont {Cornell}\ and\ \citenamefont {Wieman}(2002)}]{Cor02}%
  \BibitemOpen
  \bibfield  {author} {\bibinfo {author} {\bibfnamefont {E.~A.}\ \bibnamefont
  {Cornell}}\ and\ \bibinfo {author} {\bibfnamefont {C.~E.}\ \bibnamefont
  {Wieman}},\ }\bibfield  {title} {\bibinfo {title} {Nobel lecture:
  Bose-einstein condensation in a dilute gas, the first 70 years and some
  recent experiments},\ }\href {https://doi.org/10.1103/RevModPhys.74.875}
  {\bibfield  {journal} {\bibinfo  {journal} {Rev. Mod. Phys.}\ }\textbf
  {\bibinfo {volume} {74}},\ \bibinfo {pages} {875} (\bibinfo {year}
  {2002})}\BibitemShut {NoStop}%
\bibitem [{\citenamefont {Ketterle}(2002)}]{Ket02}%
  \BibitemOpen
  \bibfield  {author} {\bibinfo {author} {\bibfnamefont {W.}~\bibnamefont
  {Ketterle}},\ }\bibfield  {title} {\bibinfo {title} {Nobel lecture: When
  atoms behave as waves: Bose-einstein condensation and the atom laser},\
  }\href {https://doi.org/10.1103/RevModPhys.74.1131} {\bibfield  {journal}
  {\bibinfo  {journal} {Rev. Mod. Phys.}\ }\textbf {\bibinfo {volume} {74}},\
  \bibinfo {pages} {1131} (\bibinfo {year} {2002})}\BibitemShut {NoStop}%
\bibitem [{\citenamefont {Phillips}\ and\ \citenamefont
  {Metcalf}(1982)}]{Phi82}%
  \BibitemOpen
  \bibfield  {author} {\bibinfo {author} {\bibfnamefont {W.~D.}\ \bibnamefont
  {Phillips}}\ and\ \bibinfo {author} {\bibfnamefont {H.}~\bibnamefont
  {Metcalf}},\ }\bibfield  {title} {\bibinfo {title} {Laser deceleration of an
  atomic beam},\ }\href {https://doi.org/10.1103/PhysRevLett.48.596} {\bibfield
   {journal} {\bibinfo  {journal} {Phys. Rev. Lett.}\ }\textbf {\bibinfo
  {volume} {48}},\ \bibinfo {pages} {596} (\bibinfo {year} {1982})}\BibitemShut
  {NoStop}%
\bibitem [{\citenamefont {Hänsch}\ and\ \citenamefont
  {Schawlow}(1975)}]{Han75}%
  \BibitemOpen
  \bibfield  {author} {\bibinfo {author} {\bibfnamefont {T.}~\bibnamefont
  {Hänsch}}\ and\ \bibinfo {author} {\bibfnamefont {A.}~\bibnamefont
  {Schawlow}},\ }\bibfield  {title} {\bibinfo {title} {Cooling of gases by
  laser radiation},\ }\href
  {https://doi.org/https://doi.org/10.1016/0030-4018(75)90159-5} {\bibfield
  {journal} {\bibinfo  {journal} {Optics Communications}\ }\textbf {\bibinfo
  {volume} {13}},\ \bibinfo {pages} {68 } (\bibinfo {year} {1975})}\BibitemShut
  {NoStop}%
\bibitem [{\citenamefont {Wineland}\ and\ \citenamefont {Itano}(1979)}]{Win79}%
  \BibitemOpen
  \bibfield  {author} {\bibinfo {author} {\bibfnamefont {D.~J.}\ \bibnamefont
  {Wineland}}\ and\ \bibinfo {author} {\bibfnamefont {W.~M.}\ \bibnamefont
  {Itano}},\ }\bibfield  {title} {\bibinfo {title} {Laser cooling of atoms},\
  }\href {https://doi.org/10.1103/PhysRevA.20.1521} {\bibfield  {journal}
  {\bibinfo  {journal} {Phys. Rev. A}\ }\textbf {\bibinfo {volume} {20}},\
  \bibinfo {pages} {1521} (\bibinfo {year} {1979})}\BibitemShut {NoStop}%

\bibitem [{\citenamefont {Raab}\ \emph {et~al.}(1987)\citenamefont {Raab},
  \citenamefont {Prentiss}, \citenamefont {Cable}, \citenamefont {Chu},\ and\
  \citenamefont {Pritchard}}]{Raa87}%
  \BibitemOpen
  \bibfield  {author} {\bibinfo {author} {\bibfnamefont {E.~L.}\ \bibnamefont
  {Raab}}, \bibinfo {author} {\bibfnamefont {M.}~\bibnamefont {Prentiss}},
  \bibinfo {author} {\bibfnamefont {A.}~\bibnamefont {Cable}}, \bibinfo
  {author} {\bibfnamefont {S.}~\bibnamefont {Chu}},\ and\ \bibinfo {author}
  {\bibfnamefont {D.~E.}\ \bibnamefont {Pritchard}},\ }\bibfield  {title}
  {\bibinfo {title} {Trapping of neutral sodium atoms with radiation
  pressure},\ }\href {https://doi.org/10.1103/PhysRevLett.59.2631} {\bibfield
  {journal} {\bibinfo  {journal} {Phys. Rev. Lett.}\ }\textbf {\bibinfo
  {volume} {59}},\ \bibinfo {pages} {2631} (\bibinfo {year}
  {1987})}\BibitemShut {NoStop}%
\bibitem [{\citenamefont {Ketterle}\ and\ \citenamefont {van
  Druten}(1996)}]{KET96b}%
  \BibitemOpen
  \bibfield  {author} {\bibinfo {author} {\bibfnamefont {W.}~\bibnamefont
  {Ketterle}}\ and\ \bibinfo {author} {\bibfnamefont {N.~J.}\ \bibnamefont {van
  Druten}},\ }\bibfield  {title} {\bibinfo {title} {Evaporative cooling of
  trapped atoms},\ }in\ \href@noop {} {\emph {\bibinfo {booktitle} {Adv. At.
  Mol. Opt. Phys.}}},\ Vol.~\bibinfo {volume} {37},\ \bibinfo {editor} {edited
  by\ \bibinfo {editor} {\bibfnamefont {B.}~\bibnamefont {Bederson}}\ and\
  \bibinfo {editor} {\bibfnamefont {H.}~\bibnamefont {Walther}}}\ (\bibinfo
  {publisher} {Academic Press},\ \bibinfo {address} {San Diego},\ \bibinfo
  {year} {1996})\ pp.\ \bibinfo {pages} {181--236}\BibitemShut {NoStop}%
\bibitem [{\citenamefont {Pérez-Ríos}\ and\ \citenamefont
  {Sanz}(2013)}]{Per13}%
  \BibitemOpen
  \bibfield  {author} {\bibinfo {author} {\bibfnamefont {J.}~\bibnamefont
  {Pérez-Ríos}}\ and\ \bibinfo {author} {\bibfnamefont {A.~S.}\ \bibnamefont
  {Sanz}},\ }\bibfield  {title} {\bibinfo {title} {How does a magnetic trap
  work?},\ }\href {https://doi.org/10.1119/1.4819167} {\bibfield  {journal}
  {\bibinfo  {journal} {American Journal of Physics}\ }\textbf {\bibinfo
  {volume} {81}},\ \bibinfo {pages} {836} (\bibinfo {year} {2013})}\BibitemShut
  {NoStop}%
\bibitem [{\citenamefont {Grimm}\ \emph {et~al.}(2000)\citenamefont {Grimm},
  \citenamefont {Weidemüller},\ and\ \citenamefont {Ovchinnikov}}]{Gri00}%
  \BibitemOpen
  \bibfield  {author} {\bibinfo {author} {\bibfnamefont {R.}~\bibnamefont
  {Grimm}}, \bibinfo {author} {\bibfnamefont {M.}~\bibnamefont
  {Weidemüller}},\ and\ \bibinfo {author} {\bibfnamefont {Y.~B.}\ \bibnamefont
  {Ovchinnikov}},\ }\bibfield  {title} {\bibinfo {title} {Optical dipole traps
  for neutral atoms},\ }in\ \href
  {https://doi.org/10.1016/S1049-250X(08)60186-X} {\emph {\bibinfo {booktitle}
  {Adv. At. Mol. Opt. Phys.}}},\ Vol.~\bibinfo {volume} {42},\ \bibinfo
  {editor} {edited by\ \bibinfo {editor} {\bibfnamefont {B.}~\bibnamefont
  {Bederson}}\ and\ \bibinfo {editor} {\bibfnamefont {H.}~\bibnamefont
  {Walther}}}\ (\bibinfo  {publisher} {Academic Press},\ \bibinfo {year}
  {2000})\ p.~\bibinfo {pages} {95}\BibitemShut {NoStop}%
\bibitem [{\citenamefont {Dalfovo}\ \emph {et~al.}(1999)\citenamefont
  {Dalfovo}, \citenamefont {Giorgini}, \citenamefont {Pitaevskii},\ and\
  \citenamefont {Stringari}}]{Dal99}%
  \BibitemOpen
  \bibfield  {author} {\bibinfo {author} {\bibfnamefont {F.}~\bibnamefont
  {Dalfovo}}, \bibinfo {author} {\bibfnamefont {S.}~\bibnamefont {Giorgini}},
  \bibinfo {author} {\bibfnamefont {L.~P.}\ \bibnamefont {Pitaevskii}},\ and\
  \bibinfo {author} {\bibfnamefont {S.}~\bibnamefont {Stringari}},\ }\bibfield
  {title} {\bibinfo {title} {Theory of {Bose}-{Einstein} condensation in
  trapped gases},\ }\href {https://doi.org/10.1103/RevModPhys.71.463}
  {\bibfield  {journal} {\bibinfo  {journal} {Rev. Mod. Phys.}\ }\textbf
  {\bibinfo {volume} {71}},\ \bibinfo {pages} {463} (\bibinfo {year}
  {1999})}\BibitemShut {NoStop}%
\bibitem [{\citenamefont {Giorgini}\ \emph {et~al.}(2008)\citenamefont
  {Giorgini}, \citenamefont {Pitaevskii},\ and\ \citenamefont
  {Stringari}}]{Gio08}%
  \BibitemOpen
  \bibfield  {author} {\bibinfo {author} {\bibfnamefont {S.}~\bibnamefont
  {Giorgini}}, \bibinfo {author} {\bibfnamefont {L.~P.}\ \bibnamefont
  {Pitaevskii}},\ and\ \bibinfo {author} {\bibfnamefont {S.}~\bibnamefont
  {Stringari}},\ }\bibfield  {title} {\bibinfo {title} {Theory of ultracold
  atomic {Fermi} gases},\ }\href {https://doi.org/10.1103/RevModPhys.80.1215}
  {\bibfield  {journal} {\bibinfo  {journal} {Rev. Mod. Phys.}\ }\textbf
  {\bibinfo {volume} {80}},\ \bibinfo {pages} {1215} (\bibinfo {year}
  {2008})}\BibitemShut {NoStop}%
\bibitem [{\citenamefont {Anderson}\ \emph {et~al.}(1995)\citenamefont
  {Anderson}, \citenamefont {Ensher}, \citenamefont {Matthews}, \citenamefont
  {Wieman},\ and\ \citenamefont {Cornell}}]{And95}%
  \BibitemOpen
  \bibfield  {author} {\bibinfo {author} {\bibfnamefont {M.~H.}\ \bibnamefont
  {Anderson}}, \bibinfo {author} {\bibfnamefont {J.~R.}\ \bibnamefont
  {Ensher}}, \bibinfo {author} {\bibfnamefont {M.~R.}\ \bibnamefont
  {Matthews}}, \bibinfo {author} {\bibfnamefont {C.~E.}\ \bibnamefont
  {Wieman}},\ and\ \bibinfo {author} {\bibfnamefont {E.~A.}\ \bibnamefont
  {Cornell}},\ }\bibfield  {title} {\bibinfo {title} {Observation of
  {Bose}-{Einstein} condensation in a dilute atomic vapor},\ }\href
  {https://doi.org/10.1126/science.269.5221.198} {\bibfield  {journal}
  {\bibinfo  {journal} {Science}\ }\textbf {\bibinfo {volume} {269}},\ \bibinfo
  {pages} {198} (\bibinfo {year} {1995})}\BibitemShut {NoStop}%
\bibitem [{\citenamefont {Davis}\ \emph {et~al.}(1995)\citenamefont {Davis},
  \citenamefont {Mewes}, \citenamefont {Andrews}, \citenamefont {van Druten},
  \citenamefont {Durfee}, \citenamefont {Kurn},\ and\ \citenamefont
  {Ketterle}}]{Dav95}%
  \BibitemOpen
  \bibfield  {author} {\bibinfo {author} {\bibfnamefont {K.~B.}\ \bibnamefont
  {Davis}}, \bibinfo {author} {\bibfnamefont {M.~O.}\ \bibnamefont {Mewes}},
  \bibinfo {author} {\bibfnamefont {M.~R.}\ \bibnamefont {Andrews}}, \bibinfo
  {author} {\bibfnamefont {N.~J.}\ \bibnamefont {van Druten}}, \bibinfo
  {author} {\bibfnamefont {D.~S.}\ \bibnamefont {Durfee}}, \bibinfo {author}
  {\bibfnamefont {D.~M.}\ \bibnamefont {Kurn}},\ and\ \bibinfo {author}
  {\bibfnamefont {W.}~\bibnamefont {Ketterle}},\ }\bibfield  {title} {\bibinfo
  {title} {{Bose}-{Einstein} condensation in a gas of sodium atoms},\ }\href
  {https://doi.org/10.1103/PhysRevLett.75.3969} {\bibfield  {journal} {\bibinfo
   {journal} {Phys. Rev. Lett.}\ }\textbf {\bibinfo {volume} {75}},\ \bibinfo
  {pages} {3969} (\bibinfo {year} {1995})}\BibitemShut {NoStop}%
\bibitem [{\citenamefont {Myatt}\ \emph {et~al.}(1997)\citenamefont {Myatt},
  \citenamefont {Burt}, \citenamefont {Ghrist}, \citenamefont {Cornell},\ and\
  \citenamefont {Wieman}}]{Mya97}%
  \BibitemOpen
  \bibfield  {author} {\bibinfo {author} {\bibfnamefont {C.~J.}\ \bibnamefont
  {Myatt}}, \bibinfo {author} {\bibfnamefont {E.~A.}\ \bibnamefont {Burt}},
  \bibinfo {author} {\bibfnamefont {R.~W.}\ \bibnamefont {Ghrist}}, \bibinfo
  {author} {\bibfnamefont {E.~A.}\ \bibnamefont {Cornell}},\ and\ \bibinfo
  {author} {\bibfnamefont {C.~E.}\ \bibnamefont {Wieman}},\ }\bibfield  {title}
  {\bibinfo {title} {Production of two overlapping {Bose}-{Einstein}
  condensates by sympathetic cooling},\ }\href
  {https://doi.org/10.1103/PhysRevLett.78.586} {\bibfield  {journal} {\bibinfo
  {journal} {Phys. Rev. Lett.}\ }\textbf {\bibinfo {volume} {78}},\ \bibinfo
  {pages} {586} (\bibinfo {year} {1997})}\BibitemShut {NoStop}%
\bibitem [{\citenamefont {Weyers}\ \emph {et~al.}(1997)\citenamefont {Weyers},
  \citenamefont {Aucouturier}, \citenamefont {Valentin},\ and\ \citenamefont
  {Dimarcq}}]{Wey97}%
  \BibitemOpen
  \bibfield  {author} {\bibinfo {author} {\bibfnamefont {S.}~\bibnamefont
  {Weyers}}, \bibinfo {author} {\bibfnamefont {E.}~\bibnamefont {Aucouturier}},
  \bibinfo {author} {\bibfnamefont {C.}~\bibnamefont {Valentin}},\ and\
  \bibinfo {author} {\bibfnamefont {N.}~\bibnamefont {Dimarcq}},\ }\bibfield
  {title} {\bibinfo {title} {A continuous beam of cold cesium atoms extracted
  from a two-dimensional magneto-optical trap},\ }\href
  {https://doi.org/https://doi.org/10.1016/S0030-4018(97)00312-X} {\bibfield
  {journal} {\bibinfo  {journal} {Opt. Commun.}\ }\textbf {\bibinfo {volume}
  {143}},\ \bibinfo {pages} {30} (\bibinfo {year} {1997})}\BibitemShut
  {NoStop}%
\bibitem [{\citenamefont {Dieckmann}\ \emph {et~al.}(1998)\citenamefont
  {Dieckmann}, \citenamefont {Spreeuw}, \citenamefont {Weidem\"uller},\ and\
  \citenamefont {Walraven}}]{Die98}%
  \BibitemOpen
  \bibfield  {author} {\bibinfo {author} {\bibfnamefont {K.}~\bibnamefont
  {Dieckmann}}, \bibinfo {author} {\bibfnamefont {R.~J.~C.}\ \bibnamefont
  {Spreeuw}}, \bibinfo {author} {\bibfnamefont {M.}~\bibnamefont
  {Weidem\"uller}},\ and\ \bibinfo {author} {\bibfnamefont {J.~T.~M.}\
  \bibnamefont {Walraven}},\ }\bibfield  {title} {\bibinfo {title}
  {Two-dimensional magneto-optical trap as a source of slow atoms},\ }\href
  {https://doi.org/10.1103/PhysRevA.58.3891} {\bibfield  {journal} {\bibinfo
  {journal} {Phys. Rev. A}\ }\textbf {\bibinfo {volume} {58}},\ \bibinfo
  {pages} {3891} (\bibinfo {year} {1998})}\BibitemShut {NoStop}%
\bibitem [{\citenamefont {Greiner}\ \emph {et~al.}(2001)\citenamefont
  {Greiner}, \citenamefont {Bloch}, \citenamefont {H\"ansch},\ and\
  \citenamefont {Esslinger}}]{Gre01}%
  \BibitemOpen
  \bibfield  {author} {\bibinfo {author} {\bibfnamefont {M.}~\bibnamefont
  {Greiner}}, \bibinfo {author} {\bibfnamefont {I.}~\bibnamefont {Bloch}},
  \bibinfo {author} {\bibfnamefont {T.~W.}\ \bibnamefont {H\"ansch}},\ and\
  \bibinfo {author} {\bibfnamefont {T.}~\bibnamefont {Esslinger}},\ }\bibfield
  {title} {\bibinfo {title} {Magnetic transport of trapped cold atoms over a
  large distance},\ }\href {https://doi.org/10.1103/PhysRevA.63.031401}
  {\bibfield  {journal} {\bibinfo  {journal} {Phys. Rev. A}\ }\textbf {\bibinfo
  {volume} {63}},\ \bibinfo {pages} {031401} (\bibinfo {year}
  {2001})}\BibitemShut {NoStop}%
\bibitem [{\citenamefont {Lewandowski}\ \emph {et~al.}(2003)\citenamefont
  {Lewandowski}, \citenamefont {Harber}, \citenamefont {Whitaker},\ and\
  \citenamefont {Cornell}}]{Lew03}%
  \BibitemOpen
  \bibfield  {author} {\bibinfo {author} {\bibfnamefont {H.~J.}\ \bibnamefont
  {Lewandowski}}, \bibinfo {author} {\bibfnamefont {D.~M.}\ \bibnamefont
  {Harber}}, \bibinfo {author} {\bibfnamefont {D.~L.}\ \bibnamefont
  {Whitaker}},\ and\ \bibinfo {author} {\bibfnamefont {E.~A.}\ \bibnamefont
  {Cornell}},\ }\bibfield  {title} {\bibinfo {title} {Simplified system for
  creating a {Bose}-{Einstein} condensate},\ }\href
  {https://doi.org/10.1023/A:1024800600621} {\bibfield  {journal} {\bibinfo
  {journal} {Journal of Low Temperature Physics}\ }\textbf {\bibinfo {volume}
  {132}},\ \bibinfo {pages} {309} (\bibinfo {year} {2003})}\BibitemShut
  {NoStop}%
\bibitem [{\citenamefont {Fort\'agh}\ and\ \citenamefont
  {Zimmermann}(2007)}]{For07}%
  \BibitemOpen
  \bibfield  {author} {\bibinfo {author} {\bibfnamefont {J.}~\bibnamefont
  {Fort\'agh}}\ and\ \bibinfo {author} {\bibfnamefont {C.}~\bibnamefont
  {Zimmermann}},\ }\bibfield  {title} {\bibinfo {title} {Magnetic microtraps
  for ultracold atoms},\ }\href {https://doi.org/10.1103/RevModPhys.79.235}
  {\bibfield  {journal} {\bibinfo  {journal} {Rev. Mod. Phys.}\ }\textbf
  {\bibinfo {volume} {79}},\ \bibinfo {pages} {235} (\bibinfo {year}
  {2007})}\BibitemShut {NoStop}%
\bibitem [{\citenamefont {Lee}\ \emph {et~al.}(1996)\citenamefont {Lee},
  \citenamefont {Kim}, \citenamefont {Noh},\ and\ \citenamefont {Jhe}}]{Lee96}%
  \BibitemOpen
  \bibfield  {author} {\bibinfo {author} {\bibfnamefont {K.~I.}\ \bibnamefont
  {Lee}}, \bibinfo {author} {\bibfnamefont {J.~A.}\ \bibnamefont {Kim}},
  \bibinfo {author} {\bibfnamefont {H.~R.}\ \bibnamefont {Noh}},\ and\ \bibinfo
  {author} {\bibfnamefont {W.}~\bibnamefont {Jhe}},\ }\bibfield  {title}
  {\bibinfo {title} {Single-beam atom trap in a pyramidal and conical hollow
  mirror},\ }\href {https://doi.org/10.1364/OL.21.001177} {\bibfield  {journal}
  {\bibinfo  {journal} {Opt. Lett.}\ }\textbf {\bibinfo {volume} {21}},\
  \bibinfo {pages} {1177} (\bibinfo {year} {1996})}\BibitemShut {NoStop}%
\bibitem [{\citenamefont {Vangeleyn}\ \emph {et~al.}(2010)\citenamefont
  {Vangeleyn}, \citenamefont {Griffin}, \citenamefont {Riis},\ and\
  \citenamefont {Arnold}}]{Van10}%
  \BibitemOpen
  \bibfield  {author} {\bibinfo {author} {\bibfnamefont {M.}~\bibnamefont
  {Vangeleyn}}, \bibinfo {author} {\bibfnamefont {P.~F.}\ \bibnamefont
  {Griffin}}, \bibinfo {author} {\bibfnamefont {E.}~\bibnamefont {Riis}},\ and\
  \bibinfo {author} {\bibfnamefont {A.~S.}\ \bibnamefont {Arnold}},\ }\bibfield
   {title} {\bibinfo {title} {Laser cooling with a single laser beam and a
  planar diffractor},\ }\href {https://doi.org/10.1364/OL.35.003453} {\bibfield
   {journal} {\bibinfo  {journal} {Opt. Lett.}\ }\textbf {\bibinfo {volume}
  {35}},\ \bibinfo {pages} {3453} (\bibinfo {year} {2010})}\BibitemShut
  {NoStop}%
\bibitem [{\citenamefont {Becker}\ \emph {et~al.}(2018)\citenamefont {Becker},
  \citenamefont {Lachmann}, \citenamefont {Seidel}, \citenamefont {Ahlers},
  \citenamefont {Dinkelaker}, \citenamefont {Grosse}, \citenamefont {Hellmig},
  \citenamefont {M{\"u}ntinga}, \citenamefont {Schkolnik}, \citenamefont
  {Wendrich}, \citenamefont {Wenzlawski}, \citenamefont {Weps}, \citenamefont
  {Corgier}, \citenamefont {Franz}, \citenamefont {Gaaloul}, \citenamefont
  {Herr}, \citenamefont {L{\"u}dtke}, \citenamefont {Popp}, \citenamefont
  {Amri}, \citenamefont {Duncker}, \citenamefont {Erbe}, \citenamefont
  {Kohfeldt}, \citenamefont {Kubelka-Lange}, \citenamefont {Braxmaier},
  \citenamefont {Charron}, \citenamefont {Ertmer}, \citenamefont {Krutzik},
  \citenamefont {L{\"a}mmerzahl}, \citenamefont {Peters}, \citenamefont
  {Schleich}, \citenamefont {Sengstock}, \citenamefont {Walser}, \citenamefont
  {Wicht}, \citenamefont {Windpassinger},\ and\ \citenamefont {Rasel}}]{Bec18}%
  \BibitemOpen
  \bibfield  {author} {\bibinfo {author} {\bibfnamefont {D.}~\bibnamefont
  {Becker}}, \bibinfo {author} {\bibfnamefont {M.~D.}\ \bibnamefont
  {Lachmann}}, \bibinfo {author} {\bibfnamefont {S.~T.}\ \bibnamefont
  {Seidel}}, \bibinfo {author} {\bibfnamefont {H.}~\bibnamefont {Ahlers}},
  \bibinfo {author} {\bibfnamefont {A.~N.}\ \bibnamefont {Dinkelaker}},
  \bibinfo {author} {\bibfnamefont {J.}~\bibnamefont {Grosse}}, \bibinfo
  {author} {\bibfnamefont {O.}~\bibnamefont {Hellmig}}, \bibinfo {author}
  {\bibfnamefont {H.}~\bibnamefont {M{\"u}ntinga}}, \bibinfo {author}
  {\bibfnamefont {V.}~\bibnamefont {Schkolnik}}, \bibinfo {author}
  {\bibfnamefont {T.}~\bibnamefont {Wendrich}}, \bibinfo {author}
  {\bibfnamefont {A.}~\bibnamefont {Wenzlawski}}, \bibinfo {author}
  {\bibfnamefont {B.}~\bibnamefont {Weps}}, \bibinfo {author} {\bibfnamefont
  {R.}~\bibnamefont {Corgier}}, \bibinfo {author} {\bibfnamefont
  {T.}~\bibnamefont {Franz}}, \bibinfo {author} {\bibfnamefont
  {N.}~\bibnamefont {Gaaloul}}, \bibinfo {author} {\bibfnamefont
  {W.}~\bibnamefont {Herr}}, \bibinfo {author} {\bibfnamefont {D.}~\bibnamefont
  {L{\"u}dtke}}, \bibinfo {author} {\bibfnamefont {M.}~\bibnamefont {Popp}},
  \bibinfo {author} {\bibfnamefont {S.}~\bibnamefont {Amri}}, \bibinfo {author}
  {\bibfnamefont {H.}~\bibnamefont {Duncker}}, \bibinfo {author} {\bibfnamefont
  {M.}~\bibnamefont {Erbe}}, \bibinfo {author} {\bibfnamefont {A.}~\bibnamefont
  {Kohfeldt}}, \bibinfo {author} {\bibfnamefont {A.}~\bibnamefont
  {Kubelka-Lange}}, \bibinfo {author} {\bibfnamefont {C.}~\bibnamefont
  {Braxmaier}}, \bibinfo {author} {\bibfnamefont {E.}~\bibnamefont {Charron}},
  \bibinfo {author} {\bibfnamefont {W.}~\bibnamefont {Ertmer}}, \bibinfo
  {author} {\bibfnamefont {M.}~\bibnamefont {Krutzik}}, \bibinfo {author}
  {\bibfnamefont {C.}~\bibnamefont {L{\"a}mmerzahl}}, \bibinfo {author}
  {\bibfnamefont {A.}~\bibnamefont {Peters}}, \bibinfo {author} {\bibfnamefont
  {W.~P.}\ \bibnamefont {Schleich}}, \bibinfo {author} {\bibfnamefont
  {K.}~\bibnamefont {Sengstock}}, \bibinfo {author} {\bibfnamefont
  {R.}~\bibnamefont {Walser}}, \bibinfo {author} {\bibfnamefont
  {A.}~\bibnamefont {Wicht}}, \bibinfo {author} {\bibfnamefont
  {P.}~\bibnamefont {Windpassinger}},\ and\ \bibinfo {author} {\bibfnamefont
  {E.~M.}\ \bibnamefont {Rasel}},\ }\bibfield  {title} {\bibinfo {title}
  {Space-borne {Bose}--{Einstein} condensation for precision interferometry},\
  }\href {https://doi.org/10.1038/s41586-018-0605-1} {\bibfield  {journal}
  {\bibinfo  {journal} {Nature}\ }\textbf {\bibinfo {volume} {562}},\ \bibinfo
  {pages} {391} (\bibinfo {year} {2018})}\BibitemShut {NoStop}%
\bibitem [{\citenamefont {Aveline}\ \emph {et~al.}(2020)\citenamefont
  {Aveline}, \citenamefont {Williams}, \citenamefont {Elliott}, \citenamefont
  {Dutenhoffer}, \citenamefont {Kellogg}, \citenamefont {Kohel}, \citenamefont
  {Lay}, \citenamefont {Oudrhiri}, \citenamefont {Shotwell}, \citenamefont
  {Yu},\ and\ \citenamefont {Thompson}}]{Ave20}%
  \BibitemOpen
  \bibfield  {author} {\bibinfo {author} {\bibfnamefont {D.~C.}\ \bibnamefont
  {Aveline}}, \bibinfo {author} {\bibfnamefont {J.~R.}\ \bibnamefont
  {Williams}}, \bibinfo {author} {\bibfnamefont {E.~R.}\ \bibnamefont
  {Elliott}}, \bibinfo {author} {\bibfnamefont {C.}~\bibnamefont
  {Dutenhoffer}}, \bibinfo {author} {\bibfnamefont {J.~R.}\ \bibnamefont
  {Kellogg}}, \bibinfo {author} {\bibfnamefont {J.~M.}\ \bibnamefont {Kohel}},
  \bibinfo {author} {\bibfnamefont {N.~E.}\ \bibnamefont {Lay}}, \bibinfo
  {author} {\bibfnamefont {K.}~\bibnamefont {Oudrhiri}}, \bibinfo {author}
  {\bibfnamefont {R.~F.}\ \bibnamefont {Shotwell}}, \bibinfo {author}
  {\bibfnamefont {N.}~\bibnamefont {Yu}},\ and\ \bibinfo {author}
  {\bibfnamefont {R.~J.}\ \bibnamefont {Thompson}},\ }\bibfield  {title}
  {\bibinfo {title} {Observation of {Bose}--{Einstein} condensates in an
  earth-orbiting research lab},\ }\href
  {https://doi.org/10.1038/s41586-020-2346-1} {\bibfield  {journal} {\bibinfo
  {journal} {Nature}\ }\textbf {\bibinfo {volume} {582}},\ \bibinfo {pages}
  {193} (\bibinfo {year} {2020})}\BibitemShut {NoStop}%
\bibitem [{\citenamefont {Ketterle}\ \emph {et~al.}(1993)\citenamefont
  {Ketterle}, \citenamefont {Davis}, \citenamefont {Joffe}, \citenamefont
  {Martin},\ and\ \citenamefont {Pritchard}}]{Ket93}%
  \BibitemOpen
  \bibfield  {author} {\bibinfo {author} {\bibfnamefont {W.}~\bibnamefont
  {Ketterle}}, \bibinfo {author} {\bibfnamefont {K.~B.}\ \bibnamefont {Davis}},
  \bibinfo {author} {\bibfnamefont {M.~A.}\ \bibnamefont {Joffe}}, \bibinfo
  {author} {\bibfnamefont {A.}~\bibnamefont {Martin}},\ and\ \bibinfo {author}
  {\bibfnamefont {D.~E.}\ \bibnamefont {Pritchard}},\ }\bibfield  {title}
  {\bibinfo {title} {High densities of cold atoms in a dark spontaneous-force
  optical trap},\ }\href {https://doi.org/10.1103/PhysRevLett.70.2253}
  {\bibfield  {journal} {\bibinfo  {journal} {Phys. Rev. Lett.}\ }\textbf
  {\bibinfo {volume} {70}},\ \bibinfo {pages} {2253} (\bibinfo {year}
  {1993})}\BibitemShut {NoStop}%
\bibitem [{\citenamefont {Anderson}\ \emph {et~al.}(1994)\citenamefont
  {Anderson}, \citenamefont {Petrich}, \citenamefont {Ensher},\ and\
  \citenamefont {Cornell}}]{And94}%
  \BibitemOpen
  \bibfield  {author} {\bibinfo {author} {\bibfnamefont {M.~H.}\ \bibnamefont
  {Anderson}}, \bibinfo {author} {\bibfnamefont {W.}~\bibnamefont {Petrich}},
  \bibinfo {author} {\bibfnamefont {J.~R.}\ \bibnamefont {Ensher}},\ and\
  \bibinfo {author} {\bibfnamefont {E.~A.}\ \bibnamefont {Cornell}},\
  }\bibfield  {title} {\bibinfo {title} {Reduction of light-assisted
  collisional loss rate from a low-pressure vapor-cell trap},\ }\href
  {https://doi.org/10.1103/PhysRevA.50.R3597} {\bibfield  {journal} {\bibinfo
  {journal} {Phys. Rev. A}\ }\textbf {\bibinfo {volume} {50}},\ \bibinfo
  {pages} {R3597} (\bibinfo {year} {1994})}\BibitemShut {NoStop}%
\bibitem [{\citenamefont {DeMarco}\ and\ \citenamefont {Jin}(1999)}]{DeM99}%
  \BibitemOpen
  \bibfield  {author} {\bibinfo {author} {\bibfnamefont {B.}~\bibnamefont
  {DeMarco}}\ and\ \bibinfo {author} {\bibfnamefont {D.~S.}\ \bibnamefont
  {Jin}},\ }\bibfield  {title} {\bibinfo {title} {Onset of {Fermi} degeneracy
  in a trapped atomic gas},\ }\href
  {https://doi.org/10.1126/science.285.5434.1703} {\bibfield  {journal}
  {\bibinfo  {journal} {Science}\ }\textbf {\bibinfo {volume} {285}},\ \bibinfo
  {pages} {1703} (\bibinfo {year} {1999})}\BibitemShut {NoStop}%
\bibitem [{\citenamefont {Truscott}\ \emph {et~al.}(2001)\citenamefont
  {Truscott}, \citenamefont {Strecker}, \citenamefont {McAlexander},
  \citenamefont {Partridge},\ and\ \citenamefont {Hulet}}]{Tru01}%
  \BibitemOpen
  \bibfield  {author} {\bibinfo {author} {\bibfnamefont {A.~G.}\ \bibnamefont
  {Truscott}}, \bibinfo {author} {\bibfnamefont {K.~E.}\ \bibnamefont
  {Strecker}}, \bibinfo {author} {\bibfnamefont {W.~I.}\ \bibnamefont
  {McAlexander}}, \bibinfo {author} {\bibfnamefont {G.~B.}\ \bibnamefont
  {Partridge}},\ and\ \bibinfo {author} {\bibfnamefont {R.~G.}\ \bibnamefont
  {Hulet}},\ }\bibfield  {title} {\bibinfo {title} {Observation of {Fermi}
  pressure in a gas of trapped atoms},\ }\href
  {https://doi.org/10.1126/science.1059318} {\bibfield  {journal} {\bibinfo
  {journal} {Science}\ }\textbf {\bibinfo {volume} {291}},\ \bibinfo {pages}
  {2570} (\bibinfo {year} {2001})}\BibitemShut {NoStop}%
\bibitem [{\citenamefont {Bradley}\ \emph {et~al.}(1995)\citenamefont
  {Bradley}, \citenamefont {Sackett}, \citenamefont {Tollett},\ and\
  \citenamefont {Hulet}}]{Bra95}%
  \BibitemOpen
  \bibfield  {author} {\bibinfo {author} {\bibfnamefont {C.~C.}\ \bibnamefont
  {Bradley}}, \bibinfo {author} {\bibfnamefont {C.~A.}\ \bibnamefont
  {Sackett}}, \bibinfo {author} {\bibfnamefont {J.~J.}\ \bibnamefont
  {Tollett}},\ and\ \bibinfo {author} {\bibfnamefont {R.~G.}\ \bibnamefont
  {Hulet}},\ }\bibfield  {title} {\bibinfo {title} {Evidence of
  {Bose}-{Einstein} condensation in an atomic gas with attractive
  interactions},\ }\href {https://doi.org/10.1103/PhysRevLett.75.1687}
  {\bibfield  {journal} {\bibinfo  {journal} {Phys. Rev. Lett.}\ }\textbf
  {\bibinfo {volume} {75}},\ \bibinfo {pages} {1687} (\bibinfo {year}
  {1995})}\BibitemShut {NoStop}%
\bibitem [{\citenamefont {Bradley}\ \emph {et~al.}(1997)\citenamefont
  {Bradley}, \citenamefont {Sackett}, \citenamefont {Tollett},\ and\
  \citenamefont {Hulet}}]{BRA97b}%
  \BibitemOpen
  \bibfield  {author} {\bibinfo {author} {\bibfnamefont {C.~C.}\ \bibnamefont
  {Bradley}}, \bibinfo {author} {\bibfnamefont {C.~A.}\ \bibnamefont
  {Sackett}}, \bibinfo {author} {\bibfnamefont {J.~J.}\ \bibnamefont
  {Tollett}},\ and\ \bibinfo {author} {\bibfnamefont {R.~G.}\ \bibnamefont
  {Hulet}},\ }\bibfield  {title} {\bibinfo {title} {Evidence of
  {Bose}-{Einstein} condensation in an atomic gas with attractive interactions
  [{Phys. Rev. Lett.} 75, 1687 (1995)]},\ }\href
  {https://doi.org/10.1103/PhysRevLett.79.1170} {\bibfield  {journal} {\bibinfo
   {journal} {Phys. Rev. Lett.}\ }\textbf {\bibinfo {volume} {79}},\ \bibinfo
  {pages} {1170} (\bibinfo {year} {1997})}\BibitemShut {NoStop}%
\bibitem [{\citenamefont {Duarte}\ \emph {et~al.}(2011)\citenamefont {Duarte},
  \citenamefont {Hart}, \citenamefont {Hitchcock}, \citenamefont {Corcovilos},
  \citenamefont {Yang}, \citenamefont {Reed},\ and\ \citenamefont
  {Hulet}}]{Dua11}%
  \BibitemOpen
  \bibfield  {author} {\bibinfo {author} {\bibfnamefont {P.~M.}\ \bibnamefont
  {Duarte}}, \bibinfo {author} {\bibfnamefont {R.~A.}\ \bibnamefont {Hart}},
  \bibinfo {author} {\bibfnamefont {J.~M.}\ \bibnamefont {Hitchcock}}, \bibinfo
  {author} {\bibfnamefont {T.~A.}\ \bibnamefont {Corcovilos}}, \bibinfo
  {author} {\bibfnamefont {T.-L.}\ \bibnamefont {Yang}}, \bibinfo {author}
  {\bibfnamefont {A.}~\bibnamefont {Reed}},\ and\ \bibinfo {author}
  {\bibfnamefont {R.~G.}\ \bibnamefont {Hulet}},\ }\bibfield  {title} {\bibinfo
  {title} {All-optical production of a lithium quantum gas using narrow-line
  laser cooling},\ }\href {https://doi.org/10.1103/PhysRevA.84.061406}
  {\bibfield  {journal} {\bibinfo  {journal} {Phys. Rev. A}\ }\textbf {\bibinfo
  {volume} {84}},\ \bibinfo {pages} {061406} (\bibinfo {year}
  {2011})}\BibitemShut {NoStop}%
\bibitem [{\citenamefont {McKay}\ \emph {et~al.}(2011)\citenamefont {McKay},
  \citenamefont {Jervis}, \citenamefont {Fine}, \citenamefont {Simpson-Porco},
  \citenamefont {Edge},\ and\ \citenamefont {Thywissen}}]{McK11}%
  \BibitemOpen
  \bibfield  {author} {\bibinfo {author} {\bibfnamefont {D.~C.}\ \bibnamefont
  {McKay}}, \bibinfo {author} {\bibfnamefont {D.}~\bibnamefont {Jervis}},
  \bibinfo {author} {\bibfnamefont {D.~J.}\ \bibnamefont {Fine}}, \bibinfo
  {author} {\bibfnamefont {J.~W.}\ \bibnamefont {Simpson-Porco}}, \bibinfo
  {author} {\bibfnamefont {G.~J.~A.}\ \bibnamefont {Edge}},\ and\ \bibinfo
  {author} {\bibfnamefont {J.~H.}\ \bibnamefont {Thywissen}},\ }\bibfield
  {title} {\bibinfo {title} {Low-temperature high-density magneto-optical
  trapping of potassium using the open $4s\ensuremath{\rightarrow}5p$
  transition at 405 nm},\ }\href {https://doi.org/10.1103/PhysRevA.84.063420}
  {\bibfield  {journal} {\bibinfo  {journal} {Phys. Rev. A}\ }\textbf {\bibinfo
  {volume} {84}},\ \bibinfo {pages} {063420} (\bibinfo {year}
  {2011})}\BibitemShut {NoStop}%
\bibitem [{\citenamefont {Salomon}\ \emph {et~al.}(2013)\citenamefont
  {Salomon}, \citenamefont {Fouch{\'{e}}}, \citenamefont {Wang}, \citenamefont
  {Aspect}, \citenamefont {Bouyer},\ and\ \citenamefont {Bourdel}}]{Sal13}%
  \BibitemOpen
  \bibfield  {author} {\bibinfo {author} {\bibfnamefont {G.}~\bibnamefont
  {Salomon}}, \bibinfo {author} {\bibfnamefont {L.}~\bibnamefont
  {Fouch{\'{e}}}}, \bibinfo {author} {\bibfnamefont {P.}~\bibnamefont {Wang}},
  \bibinfo {author} {\bibfnamefont {A.}~\bibnamefont {Aspect}}, \bibinfo
  {author} {\bibfnamefont {P.}~\bibnamefont {Bouyer}},\ and\ \bibinfo {author}
  {\bibfnamefont {T.}~\bibnamefont {Bourdel}},\ }\bibfield  {title} {\bibinfo
  {title} {Gray-molasses cooling of $^{39}${K} to a high phase-space density},\
  }\href {https://doi.org/10.1209/0295-5075/104/63002} {\bibfield  {journal}
  {\bibinfo  {journal} {Europhys. Lett.}\ }\textbf {\bibinfo {volume} {104}},\
  \bibinfo {pages} {63002} (\bibinfo {year} {2013})}\BibitemShut {NoStop}%
\bibitem [{\citenamefont {Aspect}\ \emph {et~al.}(1988)\citenamefont {Aspect},
  \citenamefont {Arimondo}, \citenamefont {Kaiser}, \citenamefont
  {Vansteenkiste},\ and\ \citenamefont {Cohen-Tannoudji}}]{Asp88}%
  \BibitemOpen
  \bibfield  {author} {\bibinfo {author} {\bibfnamefont {A.}~\bibnamefont
  {Aspect}}, \bibinfo {author} {\bibfnamefont {E.}~\bibnamefont {Arimondo}},
  \bibinfo {author} {\bibfnamefont {R.}~\bibnamefont {Kaiser}}, \bibinfo
  {author} {\bibfnamefont {N.}~\bibnamefont {Vansteenkiste}},\ and\ \bibinfo
  {author} {\bibfnamefont {C.}~\bibnamefont {Cohen-Tannoudji}},\ }\bibfield
  {title} {\bibinfo {title} {Laser cooling below the one-photon recoil energy
  by velocity-selective coherent population trapping},\ }\href
  {https://doi.org/10.1103/PhysRevLett.61.826} {\bibfield  {journal} {\bibinfo
  {journal} {Phys. Rev. Lett.}\ }\textbf {\bibinfo {volume} {61}},\ \bibinfo
  {pages} {826} (\bibinfo {year} {1988})}\BibitemShut {NoStop}%
\bibitem [{\citenamefont {Weber}\ \emph {et~al.}(2003)\citenamefont {Weber},
  \citenamefont {Herbig}, \citenamefont {Mark}, \citenamefont {N{\"a}gerl},\
  and\ \citenamefont {Grimm}}]{Web03}%
  \BibitemOpen
  \bibfield  {author} {\bibinfo {author} {\bibfnamefont {T.}~\bibnamefont
  {Weber}}, \bibinfo {author} {\bibfnamefont {J.}~\bibnamefont {Herbig}},
  \bibinfo {author} {\bibfnamefont {M.}~\bibnamefont {Mark}}, \bibinfo {author}
  {\bibfnamefont {H.-C.}\ \bibnamefont {N{\"a}gerl}},\ and\ \bibinfo {author}
  {\bibfnamefont {R.}~\bibnamefont {Grimm}},\ }\bibfield  {title} {\bibinfo
  {title} {{Bose}-{Einstein} condensation of cesium},\ }\href
  {https://doi.org/10.1126/science.1079699} {\bibfield  {journal} {\bibinfo
  {journal} {Science}\ }\textbf {\bibinfo {volume} {299}},\ \bibinfo {pages}
  {232} (\bibinfo {year} {2003})}\BibitemShut {NoStop}%
\bibitem [{\citenamefont {Hamann}\ \emph {et~al.}(1998)\citenamefont {Hamann},
  \citenamefont {Haycock}, \citenamefont {Klose}, \citenamefont {Pax},
  \citenamefont {Deutsch},\ and\ \citenamefont {Jessen}}]{Ham98}%
  \BibitemOpen
  \bibfield  {author} {\bibinfo {author} {\bibfnamefont {S.~E.}\ \bibnamefont
  {Hamann}}, \bibinfo {author} {\bibfnamefont {D.~L.}\ \bibnamefont {Haycock}},
  \bibinfo {author} {\bibfnamefont {G.}~\bibnamefont {Klose}}, \bibinfo
  {author} {\bibfnamefont {P.~H.}\ \bibnamefont {Pax}}, \bibinfo {author}
  {\bibfnamefont {I.~H.}\ \bibnamefont {Deutsch}},\ and\ \bibinfo {author}
  {\bibfnamefont {P.~S.}\ \bibnamefont {Jessen}},\ }\bibfield  {title}
  {\bibinfo {title} {Resolved-sideband {Raman} cooling to the ground state of
  an optical lattice},\ }\href {https://doi.org/10.1103/PhysRevLett.80.4149}
  {\bibfield  {journal} {\bibinfo  {journal} {Phys. Rev. Lett.}\ }\textbf
  {\bibinfo {volume} {80}},\ \bibinfo {pages} {4149} (\bibinfo {year}
  {1998})}\BibitemShut {NoStop}%
\bibitem [{\citenamefont {Vuleti\ifmmode~\acute{c}\else \'{c}\fi{}}\ \emph
  {et~al.}(1998)\citenamefont {Vuleti\ifmmode~\acute{c}\else \'{c}\fi{}},
  \citenamefont {Chin}, \citenamefont {Kerman},\ and\ \citenamefont
  {Chu}}]{Vul98}%
  \BibitemOpen
  \bibfield  {author} {\bibinfo {author} {\bibfnamefont {V.}~\bibnamefont
  {Vuleti\ifmmode~\acute{c}\else \'{c}\fi{}}}, \bibinfo {author} {\bibfnamefont
  {C.}~\bibnamefont {Chin}}, \bibinfo {author} {\bibfnamefont {A.~J.}\
  \bibnamefont {Kerman}},\ and\ \bibinfo {author} {\bibfnamefont
  {S.}~\bibnamefont {Chu}},\ }\bibfield  {title} {\bibinfo {title} {Degenerate
  {Raman} sideband cooling of trapped cesium atoms at very high atomic
  densities},\ }\href {https://doi.org/10.1103/PhysRevLett.81.5768} {\bibfield
  {journal} {\bibinfo  {journal} {Phys. Rev. Lett.}\ }\textbf {\bibinfo
  {volume} {81}},\ \bibinfo {pages} {5768} (\bibinfo {year}
  {1998})}\BibitemShut {NoStop}%
\bibitem [{\citenamefont {Han}\ \emph {et~al.}(2000)\citenamefont {Han},
  \citenamefont {Wolf}, \citenamefont {Oliver}, \citenamefont {McCormick},
  \citenamefont {DePue},\ and\ \citenamefont {Weiss}}]{Han00}%
  \BibitemOpen
  \bibfield  {author} {\bibinfo {author} {\bibfnamefont {D.-J.}\ \bibnamefont
  {Han}}, \bibinfo {author} {\bibfnamefont {S.}~\bibnamefont {Wolf}}, \bibinfo
  {author} {\bibfnamefont {S.}~\bibnamefont {Oliver}}, \bibinfo {author}
  {\bibfnamefont {C.}~\bibnamefont {McCormick}}, \bibinfo {author}
  {\bibfnamefont {M.~T.}\ \bibnamefont {DePue}},\ and\ \bibinfo {author}
  {\bibfnamefont {D.~S.}\ \bibnamefont {Weiss}},\ }\bibfield  {title} {\bibinfo
  {title} {{3D} {Raman} sideband cooling of cesium atoms at high density},\
  }\href {https://doi.org/10.1103/PhysRevLett.85.724} {\bibfield  {journal}
  {\bibinfo  {journal} {Phys. Rev. Lett.}\ }\textbf {\bibinfo {volume} {85}},\
  \bibinfo {pages} {724} (\bibinfo {year} {2000})}\BibitemShut {NoStop}%
\bibitem [{\citenamefont {Kerman}\ \emph {et~al.}(2000)\citenamefont {Kerman},
  \citenamefont {Vuleti\ifmmode~\acute{c}\else \'{c}\fi{}}, \citenamefont
  {Chin},\ and\ \citenamefont {Chu}}]{Ker00}%
  \BibitemOpen
  \bibfield  {author} {\bibinfo {author} {\bibfnamefont {A.~J.}\ \bibnamefont
  {Kerman}}, \bibinfo {author} {\bibfnamefont {V.}~\bibnamefont
  {Vuleti\ifmmode~\acute{c}\else \'{c}\fi{}}}, \bibinfo {author} {\bibfnamefont
  {C.}~\bibnamefont {Chin}},\ and\ \bibinfo {author} {\bibfnamefont
  {S.}~\bibnamefont {Chu}},\ }\bibfield  {title} {\bibinfo {title} {Beyond
  optical molasses: {3D} {Raman} sideband cooling of atomic cesium to high
  phase-space density},\ }\href {https://doi.org/10.1103/PhysRevLett.84.439}
  {\bibfield  {journal} {\bibinfo  {journal} {Phys. Rev. Lett.}\ }\textbf
  {\bibinfo {volume} {84}},\ \bibinfo {pages} {439} (\bibinfo {year}
  {2000})}\BibitemShut {NoStop}%
\bibitem [{\citenamefont {Treutlein}\ \emph {et~al.}(2001)\citenamefont
  {Treutlein}, \citenamefont {Chung},\ and\ \citenamefont {Chu}}]{Tre01}%
  \BibitemOpen
  \bibfield  {author} {\bibinfo {author} {\bibfnamefont {P.}~\bibnamefont
  {Treutlein}}, \bibinfo {author} {\bibfnamefont {K.~Y.}\ \bibnamefont
  {Chung}},\ and\ \bibinfo {author} {\bibfnamefont {S.}~\bibnamefont {Chu}},\
  }\bibfield  {title} {\bibinfo {title} {High-brightness atom source for atomic
  fountains},\ }\href {https://doi.org/10.1103/PhysRevA.63.051401} {\bibfield
  {journal} {\bibinfo  {journal} {Phys. Rev. A}\ }\textbf {\bibinfo {volume}
  {63}},\ \bibinfo {pages} {051401} (\bibinfo {year} {2001})}\BibitemShut
  {NoStop}%
\bibitem [{\citenamefont {Dicke}(1953)}]{Dic53}%
  \BibitemOpen
  \bibfield  {author} {\bibinfo {author} {\bibfnamefont {R.~H.}\ \bibnamefont
  {Dicke}},\ }\bibfield  {title} {\bibinfo {title} {The effect of collisions
  upon the {Doppler} width of spectral lines},\ }\href
  {https://doi.org/10.1103/PhysRev.89.472} {\bibfield  {journal} {\bibinfo
  {journal} {Phys. Rev.}\ }\textbf {\bibinfo {volume} {89}},\ \bibinfo {pages}
  {472} (\bibinfo {year} {1953})}\BibitemShut {NoStop}%
\bibitem [{\citenamefont {Setija}\ \emph {et~al.}(1993)\citenamefont {Setija},
  \citenamefont {Werij}, \citenamefont {Luiten}, \citenamefont {Reynolds},
  \citenamefont {Hijmans},\ and\ \citenamefont {Walraven}}]{Set93}%
  \BibitemOpen
  \bibfield  {author} {\bibinfo {author} {\bibfnamefont {I.~D.}\ \bibnamefont
  {Setija}}, \bibinfo {author} {\bibfnamefont {H.~G.~C.}\ \bibnamefont
  {Werij}}, \bibinfo {author} {\bibfnamefont {O.~J.}\ \bibnamefont {Luiten}},
  \bibinfo {author} {\bibfnamefont {M.~W.}\ \bibnamefont {Reynolds}}, \bibinfo
  {author} {\bibfnamefont {T.~W.}\ \bibnamefont {Hijmans}},\ and\ \bibinfo
  {author} {\bibfnamefont {J.~T.~M.}\ \bibnamefont {Walraven}},\ }\bibfield
  {title} {\bibinfo {title} {Optical cooling of atomic hydrogen in a magnetic
  trap},\ }\href {https://doi.org/10.1103/PhysRevLett.70.2257} {\bibfield
  {journal} {\bibinfo  {journal} {Phys. Rev. Lett.}\ }\textbf {\bibinfo
  {volume} {70}},\ \bibinfo {pages} {2257} (\bibinfo {year}
  {1993})}\BibitemShut {NoStop}%
\bibitem [{\citenamefont {Baker}\ \emph {et~al.}(2021)\citenamefont {Baker},
  \citenamefont {Bertsche}, \citenamefont {Capra}, \citenamefont {Carruth},
  \citenamefont {Cesar}, \citenamefont {Charlton}, \citenamefont {Christensen},
  \citenamefont {Collister}, \citenamefont {Mathad}, \citenamefont {Eriksson},
  \citenamefont {Evans}, \citenamefont {Evetts}, \citenamefont {Fajans},
  \citenamefont {Friesen}, \citenamefont {Fujiwara}, \citenamefont {Gill},
  \citenamefont {Grandemange}, \citenamefont {Granum}, \citenamefont {Hangst},
  \citenamefont {Hardy}, \citenamefont {Hayden}, \citenamefont {Hodgkinson},
  \citenamefont {Hunter}, \citenamefont {Isaac}, \citenamefont {Johnson},
  \citenamefont {Jones}, \citenamefont {Jones}, \citenamefont {Jonsell},
  \citenamefont {Khramov}, \citenamefont {Knapp}, \citenamefont {Kurchaninov},
  \citenamefont {Madsen}, \citenamefont {Maxwell}, \citenamefont {McKenna},
  \citenamefont {Menary}, \citenamefont {Michan}, \citenamefont {Momose},
  \citenamefont {Mullan}, \citenamefont {Munich}, \citenamefont {Olchanski},
  \citenamefont {Olin}, \citenamefont {Peszka}, \citenamefont {Powell},
  \citenamefont {Pusa}, \citenamefont {Rasmussen}, \citenamefont {Robicheaux},
  \citenamefont {Sacramento}, \citenamefont {Sameed}, \citenamefont {Sarid},
  \citenamefont {Silveira}, \citenamefont {Starko}, \citenamefont {So},
  \citenamefont {Stutter}, \citenamefont {Tharp}, \citenamefont {Thibeault},
  \citenamefont {Thompson}, \citenamefont {van~der Werf},\ and\ \citenamefont
  {Wurtele}}]{Bak21}%
  \BibitemOpen
  \bibfield  {author} {\bibinfo {author} {\bibfnamefont {C.~J.}\ \bibnamefont
  {Baker}}, \bibinfo {author} {\bibfnamefont {W.}~\bibnamefont {Bertsche}},
  \bibinfo {author} {\bibfnamefont {A.}~\bibnamefont {Capra}}, \bibinfo
  {author} {\bibfnamefont {C.}~\bibnamefont {Carruth}}, \bibinfo {author}
  {\bibfnamefont {C.~L.}\ \bibnamefont {Cesar}}, \bibinfo {author}
  {\bibfnamefont {M.}~\bibnamefont {Charlton}}, \bibinfo {author}
  {\bibfnamefont {A.}~\bibnamefont {Christensen}}, \bibinfo {author}
  {\bibfnamefont {R.}~\bibnamefont {Collister}}, \bibinfo {author}
  {\bibfnamefont {A.~C.}\ \bibnamefont {Mathad}}, \bibinfo {author}
  {\bibfnamefont {S.}~\bibnamefont {Eriksson}}, \bibinfo {author}
  {\bibfnamefont {A.}~\bibnamefont {Evans}}, \bibinfo {author} {\bibfnamefont
  {N.}~\bibnamefont {Evetts}}, \bibinfo {author} {\bibfnamefont
  {J.}~\bibnamefont {Fajans}}, \bibinfo {author} {\bibfnamefont
  {T.}~\bibnamefont {Friesen}}, \bibinfo {author} {\bibfnamefont {M.~C.}\
  \bibnamefont {Fujiwara}}, \bibinfo {author} {\bibfnamefont {D.~R.}\
  \bibnamefont {Gill}}, \bibinfo {author} {\bibfnamefont {P.}~\bibnamefont
  {Grandemange}}, \bibinfo {author} {\bibfnamefont {P.}~\bibnamefont {Granum}},
  \bibinfo {author} {\bibfnamefont {J.~S.}\ \bibnamefont {Hangst}}, \bibinfo
  {author} {\bibfnamefont {W.~N.}\ \bibnamefont {Hardy}}, \bibinfo {author}
  {\bibfnamefont {M.~E.}\ \bibnamefont {Hayden}}, \bibinfo {author}
  {\bibfnamefont {D.}~\bibnamefont {Hodgkinson}}, \bibinfo {author}
  {\bibfnamefont {E.}~\bibnamefont {Hunter}}, \bibinfo {author} {\bibfnamefont
  {C.~A.}\ \bibnamefont {Isaac}}, \bibinfo {author} {\bibfnamefont {M.~A.}\
  \bibnamefont {Johnson}}, \bibinfo {author} {\bibfnamefont {J.~M.}\
  \bibnamefont {Jones}}, \bibinfo {author} {\bibfnamefont {S.~A.}\ \bibnamefont
  {Jones}}, \bibinfo {author} {\bibfnamefont {S.}~\bibnamefont {Jonsell}},
  \bibinfo {author} {\bibfnamefont {A.}~\bibnamefont {Khramov}}, \bibinfo
  {author} {\bibfnamefont {P.}~\bibnamefont {Knapp}}, \bibinfo {author}
  {\bibfnamefont {L.}~\bibnamefont {Kurchaninov}}, \bibinfo {author}
  {\bibfnamefont {N.}~\bibnamefont {Madsen}}, \bibinfo {author} {\bibfnamefont
  {D.}~\bibnamefont {Maxwell}}, \bibinfo {author} {\bibfnamefont {J.~T.~K.}\
  \bibnamefont {McKenna}}, \bibinfo {author} {\bibfnamefont {S.}~\bibnamefont
  {Menary}}, \bibinfo {author} {\bibfnamefont {J.~M.}\ \bibnamefont {Michan}},
  \bibinfo {author} {\bibfnamefont {T.}~\bibnamefont {Momose}}, \bibinfo
  {author} {\bibfnamefont {P.~S.}\ \bibnamefont {Mullan}}, \bibinfo {author}
  {\bibfnamefont {J.~J.}\ \bibnamefont {Munich}}, \bibinfo {author}
  {\bibfnamefont {K.}~\bibnamefont {Olchanski}}, \bibinfo {author}
  {\bibfnamefont {A.}~\bibnamefont {Olin}}, \bibinfo {author} {\bibfnamefont
  {J.}~\bibnamefont {Peszka}}, \bibinfo {author} {\bibfnamefont
  {A.}~\bibnamefont {Powell}}, \bibinfo {author} {\bibfnamefont
  {P.}~\bibnamefont {Pusa}}, \bibinfo {author} {\bibfnamefont {C.~{\O}.}\
  \bibnamefont {Rasmussen}}, \bibinfo {author} {\bibfnamefont {F.}~\bibnamefont
  {Robicheaux}}, \bibinfo {author} {\bibfnamefont {R.~L.}\ \bibnamefont
  {Sacramento}}, \bibinfo {author} {\bibfnamefont {M.}~\bibnamefont {Sameed}},
  \bibinfo {author} {\bibfnamefont {E.}~\bibnamefont {Sarid}}, \bibinfo
  {author} {\bibfnamefont {D.~M.}\ \bibnamefont {Silveira}}, \bibinfo {author}
  {\bibfnamefont {D.~M.}\ \bibnamefont {Starko}}, \bibinfo {author}
  {\bibfnamefont {C.}~\bibnamefont {So}}, \bibinfo {author} {\bibfnamefont
  {G.}~\bibnamefont {Stutter}}, \bibinfo {author} {\bibfnamefont {T.~D.}\
  \bibnamefont {Tharp}}, \bibinfo {author} {\bibfnamefont {A.}~\bibnamefont
  {Thibeault}}, \bibinfo {author} {\bibfnamefont {R.~I.}\ \bibnamefont
  {Thompson}}, \bibinfo {author} {\bibfnamefont {D.~P.}\ \bibnamefont {van~der
  Werf}},\ and\ \bibinfo {author} {\bibfnamefont {J.}~\bibnamefont {Wurtele}},\
  }\bibfield  {title} {\bibinfo {title} {Laser cooling of antihydrogen atoms},\
  }\href {https://doi.org/10.1038/s41586-021-03289-6} {\bibfield  {journal}
  {\bibinfo  {journal} {Nature}\ }\textbf {\bibinfo {volume} {592}},\ \bibinfo
  {pages} {35} (\bibinfo {year} {2021})}\BibitemShut {NoStop}%
\bibitem [{\citenamefont {Vassen}\ \emph {et~al.}(2012)\citenamefont {Vassen},
  \citenamefont {Cohen-Tannoudji}, \citenamefont {Leduc}, \citenamefont
  {Boiron}, \citenamefont {Westbrook}, \citenamefont {Truscott}, \citenamefont
  {Baldwin}, \citenamefont {Birkl}, \citenamefont {Cancio},\ and\ \citenamefont
  {Trippenbach}}]{Vas12}%
  \BibitemOpen
  \bibfield  {author} {\bibinfo {author} {\bibfnamefont {W.}~\bibnamefont
  {Vassen}}, \bibinfo {author} {\bibfnamefont {C.}~\bibnamefont
  {Cohen-Tannoudji}}, \bibinfo {author} {\bibfnamefont {M.}~\bibnamefont
  {Leduc}}, \bibinfo {author} {\bibfnamefont {D.}~\bibnamefont {Boiron}},
  \bibinfo {author} {\bibfnamefont {C.~I.}\ \bibnamefont {Westbrook}}, \bibinfo
  {author} {\bibfnamefont {A.}~\bibnamefont {Truscott}}, \bibinfo {author}
  {\bibfnamefont {K.}~\bibnamefont {Baldwin}}, \bibinfo {author} {\bibfnamefont
  {G.}~\bibnamefont {Birkl}}, \bibinfo {author} {\bibfnamefont
  {P.}~\bibnamefont {Cancio}},\ and\ \bibinfo {author} {\bibfnamefont
  {M.}~\bibnamefont {Trippenbach}},\ }\bibfield  {title} {\bibinfo {title}
  {Cold and trapped metastable noble gases},\ }\href
  {https://doi.org/10.1103/RevModPhys.84.175} {\bibfield  {journal} {\bibinfo
  {journal} {Rev. Mod. Phys.}\ }\textbf {\bibinfo {volume} {84}},\ \bibinfo
  {pages} {175} (\bibinfo {year} {2012})}\BibitemShut {NoStop}%
\bibitem [{\citenamefont {Takasu}\ \emph {et~al.}(2003)\citenamefont {Takasu},
  \citenamefont {Maki}, \citenamefont {Komori}, \citenamefont {Takano},
  \citenamefont {Honda}, \citenamefont {Kumakura}, \citenamefont {Yabuzaki},\
  and\ \citenamefont {Takahashi}}]{Tak03b}%
  \BibitemOpen
  \bibfield  {author} {\bibinfo {author} {\bibfnamefont {Y.}~\bibnamefont
  {Takasu}}, \bibinfo {author} {\bibfnamefont {K.}~\bibnamefont {Maki}},
  \bibinfo {author} {\bibfnamefont {K.}~\bibnamefont {Komori}}, \bibinfo
  {author} {\bibfnamefont {T.}~\bibnamefont {Takano}}, \bibinfo {author}
  {\bibfnamefont {K.}~\bibnamefont {Honda}}, \bibinfo {author} {\bibfnamefont
  {M.}~\bibnamefont {Kumakura}}, \bibinfo {author} {\bibfnamefont
  {T.}~\bibnamefont {Yabuzaki}},\ and\ \bibinfo {author} {\bibfnamefont
  {Y.}~\bibnamefont {Takahashi}},\ }\bibfield  {title} {\bibinfo {title}
  {Spin-singlet {Bose}-{Einstein} condensation of two-electron atoms},\ }\href
  {https://doi.org/10.1103/PhysRevLett.91.040404} {\bibfield  {journal}
  {\bibinfo  {journal} {Phys. Rev. Lett.}\ }\textbf {\bibinfo {volume} {91}},\
  \bibinfo {pages} {040404} (\bibinfo {year} {2003})}\BibitemShut {NoStop}%
\bibitem [{\citenamefont {Kraft}\ \emph {et~al.}(2009)\citenamefont {Kraft},
  \citenamefont {Vogt}, \citenamefont {Appel}, \citenamefont {Riehle},\ and\
  \citenamefont {Sterr}}]{Kra09}%
  \BibitemOpen
  \bibfield  {author} {\bibinfo {author} {\bibfnamefont {S.}~\bibnamefont
  {Kraft}}, \bibinfo {author} {\bibfnamefont {F.}~\bibnamefont {Vogt}},
  \bibinfo {author} {\bibfnamefont {O.}~\bibnamefont {Appel}}, \bibinfo
  {author} {\bibfnamefont {F.}~\bibnamefont {Riehle}},\ and\ \bibinfo {author}
  {\bibfnamefont {U.}~\bibnamefont {Sterr}},\ }\bibfield  {title} {\bibinfo
  {title} {{Bose}-{Einstein} condensation of alkaline earth atoms:
  $^{40}\mathrm{Ca}$},\ }\href {https://doi.org/10.1103/PhysRevLett.103.130401}
  {\bibfield  {journal} {\bibinfo  {journal} {Phys. Rev. Lett.}\ }\textbf
  {\bibinfo {volume} {103}},\ \bibinfo {pages} {130401} (\bibinfo {year}
  {2009})}\BibitemShut {NoStop}%
\bibitem [{\citenamefont {Stellmer}\ \emph {et~al.}(2009)\citenamefont
  {Stellmer}, \citenamefont {Tey}, \citenamefont {Huang}, \citenamefont
  {Grimm},\ and\ \citenamefont {Schreck}}]{Ste09}%
  \BibitemOpen
  \bibfield  {author} {\bibinfo {author} {\bibfnamefont {S.}~\bibnamefont
  {Stellmer}}, \bibinfo {author} {\bibfnamefont {M.~K.}\ \bibnamefont {Tey}},
  \bibinfo {author} {\bibfnamefont {B.}~\bibnamefont {Huang}}, \bibinfo
  {author} {\bibfnamefont {R.}~\bibnamefont {Grimm}},\ and\ \bibinfo {author}
  {\bibfnamefont {F.}~\bibnamefont {Schreck}},\ }\bibfield  {title} {\bibinfo
  {title} {{Bose}-{Einstein} condensation of strontium},\ }\href
  {https://doi.org/10.1103/PhysRevLett.103.200401} {\bibfield  {journal}
  {\bibinfo  {journal} {Phys. Rev. Lett.}\ }\textbf {\bibinfo {volume} {103}},\
  \bibinfo {pages} {200401} (\bibinfo {year} {2009})}\BibitemShut {NoStop}%
\bibitem [{\citenamefont {de~Escobar}\ \emph {et~al.}(2009)\citenamefont
  {de~Escobar}, \citenamefont {Mickelson}, \citenamefont {Yan}, \citenamefont
  {DeSalvo}, \citenamefont {Nagel},\ and\ \citenamefont {Killian}}]{Mar09}%
  \BibitemOpen
  \bibfield  {author} {\bibinfo {author} {\bibfnamefont {Y.~N.~M.}\
  \bibnamefont {de~Escobar}}, \bibinfo {author} {\bibfnamefont {P.~G.}\
  \bibnamefont {Mickelson}}, \bibinfo {author} {\bibfnamefont {M.}~\bibnamefont
  {Yan}}, \bibinfo {author} {\bibfnamefont {B.~J.}\ \bibnamefont {DeSalvo}},
  \bibinfo {author} {\bibfnamefont {S.~B.}\ \bibnamefont {Nagel}},\ and\
  \bibinfo {author} {\bibfnamefont {T.~C.}\ \bibnamefont {Killian}},\
  }\bibfield  {title} {\bibinfo {title} {{Bose}-{Einstein} condensation of
  $^{84}\mathrm{Sr}$},\ }\href {https://doi.org/10.1103/PhysRevLett.103.200402}
  {\bibfield  {journal} {\bibinfo  {journal} {Phys. Rev. Lett.}\ }\textbf
  {\bibinfo {volume} {103}},\ \bibinfo {pages} {200402} (\bibinfo {year}
  {2009})}\BibitemShut {NoStop}%
\bibitem [{\citenamefont {Yu}\ and\ \citenamefont {Tinto}(2011)}]{Yu11}%
  \BibitemOpen
  \bibfield  {author} {\bibinfo {author} {\bibfnamefont {N.}~\bibnamefont
  {Yu}}\ and\ \bibinfo {author} {\bibfnamefont {M.}~\bibnamefont {Tinto}},\
  }\bibfield  {title} {\bibinfo {title} {Gravitational wave detection with
  single-laser atom interferometers},\ }\href
  {https://doi.org/10.1007/s10714-010-1055-8} {\bibfield  {journal} {\bibinfo
  {journal} {General Relativity and Gravitation}\ }\textbf {\bibinfo {volume}
  {43}},\ \bibinfo {pages} {1943} (\bibinfo {year} {2011})}\BibitemShut
  {NoStop}%
\bibitem [{\citenamefont {Gorshkov}\ \emph {et~al.}(2010)\citenamefont
  {Gorshkov}, \citenamefont {Hermele}, \citenamefont {Gurarie}, \citenamefont
  {Xu}, \citenamefont {Julienne}, \citenamefont {Ye}, \citenamefont {Zoller},
  \citenamefont {Demler}, \citenamefont {Lukin},\ and\ \citenamefont
  {Rey}}]{Gor10}%
  \BibitemOpen
  \bibfield  {author} {\bibinfo {author} {\bibfnamefont {A.~V.}\ \bibnamefont
  {Gorshkov}}, \bibinfo {author} {\bibfnamefont {M.}~\bibnamefont {Hermele}},
  \bibinfo {author} {\bibfnamefont {V.}~\bibnamefont {Gurarie}}, \bibinfo
  {author} {\bibfnamefont {C.}~\bibnamefont {Xu}}, \bibinfo {author}
  {\bibfnamefont {P.~S.}\ \bibnamefont {Julienne}}, \bibinfo {author}
  {\bibfnamefont {J.}~\bibnamefont {Ye}}, \bibinfo {author} {\bibfnamefont
  {P.}~\bibnamefont {Zoller}}, \bibinfo {author} {\bibfnamefont
  {E.}~\bibnamefont {Demler}}, \bibinfo {author} {\bibfnamefont {M.~D.}\
  \bibnamefont {Lukin}},\ and\ \bibinfo {author} {\bibfnamefont {A.~M.}\
  \bibnamefont {Rey}},\ }\bibfield  {title} {\bibinfo {title} {Two-orbital
  {SU(N)} magnetism with ultracold alkaline-earth atoms},\ }\href
  {https://doi.org/10.1038/nphys1535} {\bibfield  {journal} {\bibinfo
  {journal} {Nature Physics}\ }\textbf {\bibinfo {volume} {6}},\ \bibinfo
  {pages} {289} (\bibinfo {year} {2010})}\BibitemShut {NoStop}%
\bibitem [{\citenamefont {Cooper}\ \emph {et~al.}(2019)\citenamefont {Cooper},
  \citenamefont {Dalibard},\ and\ \citenamefont {Spielman}}]{Coo19}%
  \BibitemOpen
  \bibfield  {author} {\bibinfo {author} {\bibfnamefont {N.~R.}\ \bibnamefont
  {Cooper}}, \bibinfo {author} {\bibfnamefont {J.}~\bibnamefont {Dalibard}},\
  and\ \bibinfo {author} {\bibfnamefont {I.~B.}\ \bibnamefont {Spielman}},\
  }\bibfield  {title} {\bibinfo {title} {Topological bands for ultracold
  atoms},\ }\href {https://doi.org/10.1103/RevModPhys.91.015005} {\bibfield
  {journal} {\bibinfo  {journal} {Rev. Mod. Phys.}\ }\textbf {\bibinfo {volume}
  {91}},\ \bibinfo {pages} {015005} (\bibinfo {year} {2019})}\BibitemShut
  {NoStop}%
\bibitem [{\citenamefont {Kuwamoto}\ \emph {et~al.}(1999)\citenamefont
  {Kuwamoto}, \citenamefont {Honda}, \citenamefont {Takahashi},\ and\
  \citenamefont {Yabuzaki}}]{Kuw99}%
  \BibitemOpen
  \bibfield  {author} {\bibinfo {author} {\bibfnamefont {T.}~\bibnamefont
  {Kuwamoto}}, \bibinfo {author} {\bibfnamefont {K.}~\bibnamefont {Honda}},
  \bibinfo {author} {\bibfnamefont {Y.}~\bibnamefont {Takahashi}},\ and\
  \bibinfo {author} {\bibfnamefont {T.}~\bibnamefont {Yabuzaki}},\ }\bibfield
  {title} {\bibinfo {title} {Magneto-optical trapping of {Yb} atoms using an
  intercombination transition},\ }\href
  {https://doi.org/10.1103/PhysRevA.60.R745} {\bibfield  {journal} {\bibinfo
  {journal} {Phys. Rev. A}\ }\textbf {\bibinfo {volume} {60}},\ \bibinfo
  {pages} {R745} (\bibinfo {year} {1999})}\BibitemShut {NoStop}%
\bibitem [{\citenamefont {Binnewies}\ \emph {et~al.}(2001)\citenamefont
  {Binnewies}, \citenamefont {Wilpers}, \citenamefont {Sterr}, \citenamefont
  {Riehle}, \citenamefont {Helmcke}, \citenamefont {Mehlst\"aubler},
  \citenamefont {Rasel},\ and\ \citenamefont {Ertmer}}]{Bin01}%
  \BibitemOpen
  \bibfield  {author} {\bibinfo {author} {\bibfnamefont {T.}~\bibnamefont
  {Binnewies}}, \bibinfo {author} {\bibfnamefont {G.}~\bibnamefont {Wilpers}},
  \bibinfo {author} {\bibfnamefont {U.}~\bibnamefont {Sterr}}, \bibinfo
  {author} {\bibfnamefont {F.}~\bibnamefont {Riehle}}, \bibinfo {author}
  {\bibfnamefont {J.}~\bibnamefont {Helmcke}}, \bibinfo {author} {\bibfnamefont
  {T.~E.}\ \bibnamefont {Mehlst\"aubler}}, \bibinfo {author} {\bibfnamefont
  {E.~M.}\ \bibnamefont {Rasel}},\ and\ \bibinfo {author} {\bibfnamefont
  {W.}~\bibnamefont {Ertmer}},\ }\bibfield  {title} {\bibinfo {title} {Doppler
  cooling and trapping on forbidden transitions},\ }\href
  {https://doi.org/10.1103/PhysRevLett.87.123002} {\bibfield  {journal}
  {\bibinfo  {journal} {Phys. Rev. Lett.}\ }\textbf {\bibinfo {volume} {87}},\
  \bibinfo {pages} {123002} (\bibinfo {year} {2001})}\BibitemShut {NoStop}%
\bibitem [{\citenamefont {Katori}\ \emph {et~al.}(1999)\citenamefont {Katori},
  \citenamefont {Ido}, \citenamefont {Isoya},\ and\ \citenamefont
  {Kuwata-Gonokami}}]{Kat99}%
  \BibitemOpen
  \bibfield  {author} {\bibinfo {author} {\bibfnamefont {H.}~\bibnamefont
  {Katori}}, \bibinfo {author} {\bibfnamefont {T.}~\bibnamefont {Ido}},
  \bibinfo {author} {\bibfnamefont {Y.}~\bibnamefont {Isoya}},\ and\ \bibinfo
  {author} {\bibfnamefont {M.}~\bibnamefont {Kuwata-Gonokami}},\ }\bibfield
  {title} {\bibinfo {title} {Magneto-optical trapping and cooling of strontium
  atoms down to the photon recoil temperature},\ }\href
  {https://doi.org/10.1103/PhysRevLett.82.1116} {\bibfield  {journal} {\bibinfo
   {journal} {Phys. Rev. Lett.}\ }\textbf {\bibinfo {volume} {82}},\ \bibinfo
  {pages} {1116} (\bibinfo {year} {1999})}\BibitemShut {NoStop}%
\bibitem [{\citenamefont {Ido}\ \emph {et~al.}(2000)\citenamefont {Ido},
  \citenamefont {Isoya},\ and\ \citenamefont {Katori}}]{Ido00}%
  \BibitemOpen
  \bibfield  {author} {\bibinfo {author} {\bibfnamefont {T.}~\bibnamefont
  {Ido}}, \bibinfo {author} {\bibfnamefont {Y.}~\bibnamefont {Isoya}},\ and\
  \bibinfo {author} {\bibfnamefont {H.}~\bibnamefont {Katori}},\ }\bibfield
  {title} {\bibinfo {title} {Optical-dipole trapping of {Sr} atoms at a high
  phase-space density},\ }\href {https://doi.org/10.1103/PhysRevA.61.061403}
  {\bibfield  {journal} {\bibinfo  {journal} {Phys. Rev. A}\ }\textbf {\bibinfo
  {volume} {61}},\ \bibinfo {pages} {061403} (\bibinfo {year}
  {2000})}\BibitemShut {NoStop}%
\bibitem [{\citenamefont {Cooper}\ \emph {et~al.}(2018)\citenamefont {Cooper},
  \citenamefont {Covey}, \citenamefont {Madjarov}, \citenamefont {Porsev},
  \citenamefont {Safronova},\ and\ \citenamefont {Endres}}]{Coo18}%
  \BibitemOpen
  \bibfield  {author} {\bibinfo {author} {\bibfnamefont {A.}~\bibnamefont
  {Cooper}}, \bibinfo {author} {\bibfnamefont {J.~P.}\ \bibnamefont {Covey}},
  \bibinfo {author} {\bibfnamefont {I.~S.}\ \bibnamefont {Madjarov}}, \bibinfo
  {author} {\bibfnamefont {S.~G.}\ \bibnamefont {Porsev}}, \bibinfo {author}
  {\bibfnamefont {M.~S.}\ \bibnamefont {Safronova}},\ and\ \bibinfo {author}
  {\bibfnamefont {M.}~\bibnamefont {Endres}},\ }\bibfield  {title} {\bibinfo
  {title} {Alkaline-earth atoms in optical tweezers},\ }\href
  {https://doi.org/10.1103/PhysRevX.8.041055} {\bibfield  {journal} {\bibinfo
  {journal} {Phys. Rev. X}\ }\textbf {\bibinfo {volume} {8}},\ \bibinfo {pages}
  {041055} (\bibinfo {year} {2018})}\BibitemShut {NoStop}%
\bibitem [{\citenamefont {Chen}\ \emph {et~al.}(2019)\citenamefont {Chen},
  \citenamefont {Bennetts}, \citenamefont {Gonz\'alez~Escudero}, \citenamefont
  {Schreck},\ and\ \citenamefont {Pasquiou}}]{Che19}%
  \BibitemOpen
  \bibfield  {author} {\bibinfo {author} {\bibfnamefont {C.-C.}\ \bibnamefont
  {Chen}}, \bibinfo {author} {\bibfnamefont {S.}~\bibnamefont {Bennetts}},
  \bibinfo {author} {\bibfnamefont {R.}~\bibnamefont {Gonz\'alez~Escudero}},
  \bibinfo {author} {\bibfnamefont {F.}~\bibnamefont {Schreck}},\ and\ \bibinfo
  {author} {\bibfnamefont {B.}~\bibnamefont {Pasquiou}},\ }\bibfield  {title}
  {\bibinfo {title} {{Sisyphus} optical lattice decelerator},\ }\href
  {https://doi.org/10.1103/PhysRevA.100.023401} {\bibfield  {journal} {\bibinfo
   {journal} {Phys. Rev. A}\ }\textbf {\bibinfo {volume} {100}},\ \bibinfo
  {pages} {023401} (\bibinfo {year} {2019})}\BibitemShut {NoStop}%
\bibitem [{\citenamefont {Covey}\ \emph {et~al.}(2019)\citenamefont {Covey},
  \citenamefont {Madjarov}, \citenamefont {Cooper},\ and\ \citenamefont
  {Endres}}]{Cov19}%
  \BibitemOpen
  \bibfield  {author} {\bibinfo {author} {\bibfnamefont {J.~P.}\ \bibnamefont
  {Covey}}, \bibinfo {author} {\bibfnamefont {I.~S.}\ \bibnamefont {Madjarov}},
  \bibinfo {author} {\bibfnamefont {A.}~\bibnamefont {Cooper}},\ and\ \bibinfo
  {author} {\bibfnamefont {M.}~\bibnamefont {Endres}},\ }\bibfield  {title}
  {\bibinfo {title} {2000-times repeated imaging of strontium atoms in
  clock-magic tweezer arrays},\ }\href
  {https://doi.org/10.1103/PhysRevLett.122.173201} {\bibfield  {journal}
  {\bibinfo  {journal} {Phys. Rev. Lett.}\ }\textbf {\bibinfo {volume} {122}},\
  \bibinfo {pages} {173201} (\bibinfo {year} {2019})}\BibitemShut {NoStop}%
\bibitem [{\citenamefont {Fukuhara}\ \emph {et~al.}(2007)\citenamefont
  {Fukuhara}, \citenamefont {Takasu}, \citenamefont {Kumakura},\ and\
  \citenamefont {Takahashi}}]{Fuk07}%
  \BibitemOpen
  \bibfield  {author} {\bibinfo {author} {\bibfnamefont {T.}~\bibnamefont
  {Fukuhara}}, \bibinfo {author} {\bibfnamefont {Y.}~\bibnamefont {Takasu}},
  \bibinfo {author} {\bibfnamefont {M.}~\bibnamefont {Kumakura}},\ and\
  \bibinfo {author} {\bibfnamefont {Y.}~\bibnamefont {Takahashi}},\ }\bibfield
  {title} {\bibinfo {title} {Degenerate {Fermi} gases of ytterbium},\ }\href
  {https://doi.org/10.1103/PhysRevLett.98.030401} {\bibfield  {journal}
  {\bibinfo  {journal} {Phys. Rev. Lett.}\ }\textbf {\bibinfo {volume} {98}},\
  \bibinfo {pages} {030401} (\bibinfo {year} {2007})}\BibitemShut {NoStop}%
\bibitem [{\citenamefont {DeSalvo}\ \emph {et~al.}(2010)\citenamefont
  {DeSalvo}, \citenamefont {Yan}, \citenamefont {Mickelson}, \citenamefont
  {Martinez~de Escobar},\ and\ \citenamefont {Killian}}]{Des10}%
  \BibitemOpen
  \bibfield  {author} {\bibinfo {author} {\bibfnamefont {B.~J.}\ \bibnamefont
  {DeSalvo}}, \bibinfo {author} {\bibfnamefont {M.}~\bibnamefont {Yan}},
  \bibinfo {author} {\bibfnamefont {P.~G.}\ \bibnamefont {Mickelson}}, \bibinfo
  {author} {\bibfnamefont {Y.~N.}\ \bibnamefont {Martinez~de Escobar}},\ and\
  \bibinfo {author} {\bibfnamefont {T.~C.}\ \bibnamefont {Killian}},\
  }\bibfield  {title} {\bibinfo {title} {Degenerate {Fermi} gas of
  $^{87}\mathrm{Sr}$},\ }\href {https://doi.org/10.1103/PhysRevLett.105.030402}
  {\bibfield  {journal} {\bibinfo  {journal} {Phys. Rev. Lett.}\ }\textbf
  {\bibinfo {volume} {105}},\ \bibinfo {pages} {030402} (\bibinfo {year}
  {2010})}\BibitemShut {NoStop}%
\bibitem [{\citenamefont {Mukaiyama}\ \emph {et~al.}(2003)\citenamefont
  {Mukaiyama}, \citenamefont {Katori}, \citenamefont {Ido}, \citenamefont
  {Li},\ and\ \citenamefont {Kuwata-Gonokami}}]{Muk03}%
  \BibitemOpen
  \bibfield  {author} {\bibinfo {author} {\bibfnamefont {T.}~\bibnamefont
  {Mukaiyama}}, \bibinfo {author} {\bibfnamefont {H.}~\bibnamefont {Katori}},
  \bibinfo {author} {\bibfnamefont {T.}~\bibnamefont {Ido}}, \bibinfo {author}
  {\bibfnamefont {Y.}~\bibnamefont {Li}},\ and\ \bibinfo {author}
  {\bibfnamefont {M.}~\bibnamefont {Kuwata-Gonokami}},\ }\bibfield  {title}
  {\bibinfo {title} {Recoil-limited laser cooling of $^{87}${Sr} atoms near the
  {Fermi} temperature},\ }\href {https://doi.org/10.1103/PhysRevLett.90.113002}
  {\bibfield  {journal} {\bibinfo  {journal} {Phys. Rev. Lett.}\ }\textbf
  {\bibinfo {volume} {90}},\ \bibinfo {pages} {113002} (\bibinfo {year}
  {2003})}\BibitemShut {NoStop}%
\bibitem [{\citenamefont {Norcia}\ \emph {et~al.}(2018)\citenamefont {Norcia},
  \citenamefont {Cline}, \citenamefont {Bartolotta}, \citenamefont {Holland},\
  and\ \citenamefont {Thompson}}]{Nor18SWAP}%
  \BibitemOpen
  \bibfield  {author} {\bibinfo {author} {\bibfnamefont {M.~A.}\ \bibnamefont
  {Norcia}}, \bibinfo {author} {\bibfnamefont {J.~R.~K.}\ \bibnamefont
  {Cline}}, \bibinfo {author} {\bibfnamefont {J.~P.}\ \bibnamefont
  {Bartolotta}}, \bibinfo {author} {\bibfnamefont {M.~J.}\ \bibnamefont
  {Holland}},\ and\ \bibinfo {author} {\bibfnamefont {J.~K.}\ \bibnamefont
  {Thompson}},\ }\bibfield  {title} {\bibinfo {title} {Narrow-line laser
  cooling by adiabatic transfer},\ }\href
  {https://doi.org/10.1088/1367-2630/aaa950} {\bibfield  {journal} {\bibinfo
  {journal} {New J. Phys.}\ }\textbf {\bibinfo {volume} {20}},\ \bibinfo
  {pages} {023021} (\bibinfo {year} {2018})}\BibitemShut {NoStop}%
\bibitem [{\citenamefont {Muniz}\ \emph {et~al.}(2018)\citenamefont {Muniz},
  \citenamefont {Norcia}, \citenamefont {Cline},\ and\ \citenamefont
  {Thompson}}]{Mun18}%
  \BibitemOpen
  \bibfield  {author} {\bibinfo {author} {\bibfnamefont {J.~A.}\ \bibnamefont
  {Muniz}}, \bibinfo {author} {\bibfnamefont {M.~A.}\ \bibnamefont {Norcia}},
  \bibinfo {author} {\bibfnamefont {J.~R.~K.}\ \bibnamefont {Cline}},\ and\
  \bibinfo {author} {\bibfnamefont {J.~K.}\ \bibnamefont {Thompson}},\
  }\bibfield  {title} {\bibinfo {title} {A robust narrow-line magneto-optical
  trap using adiabatic transfer},\ }\href {https://arxiv.org/abs/1806.00838}
  {\bibfield  {journal} {\bibinfo  {journal} {arXiv:1806.00838}\ } (\bibinfo
  {year} {2018})}\BibitemShut {NoStop}%
\bibitem [{\citenamefont {Bartolotta}\ \emph {et~al.}(2018)\citenamefont
  {Bartolotta}, \citenamefont {Norcia}, \citenamefont {Cline}, \citenamefont
  {Thompson},\ and\ \citenamefont {Holland}}]{Bar18}%
  \BibitemOpen
  \bibfield  {author} {\bibinfo {author} {\bibfnamefont {J.~P.}\ \bibnamefont
  {Bartolotta}}, \bibinfo {author} {\bibfnamefont {M.~A.}\ \bibnamefont
  {Norcia}}, \bibinfo {author} {\bibfnamefont {J.~R.~K.}\ \bibnamefont
  {Cline}}, \bibinfo {author} {\bibfnamefont {J.~K.}\ \bibnamefont
  {Thompson}},\ and\ \bibinfo {author} {\bibfnamefont {M.~J.}\ \bibnamefont
  {Holland}},\ }\bibfield  {title} {\bibinfo {title} {Laser cooling by
  sawtooth-wave adiabatic passage},\ }\href
  {https://doi.org/10.1103/PhysRevA.98.023404} {\bibfield  {journal} {\bibinfo
  {journal} {Phys. Rev. A}\ }\textbf {\bibinfo {volume} {98}},\ \bibinfo
  {pages} {023404} (\bibinfo {year} {2018})}\BibitemShut {NoStop}%
\bibitem [{\citenamefont {Stellmer}\ \emph {et~al.}(2014)\citenamefont
  {Stellmer}, \citenamefont {Schreck},\ and\ \citenamefont {Killian}}]{Ste14}%
  \BibitemOpen
  \bibfield  {author} {\bibinfo {author} {\bibfnamefont {S.}~\bibnamefont
  {Stellmer}}, \bibinfo {author} {\bibfnamefont {F.}~\bibnamefont {Schreck}},\
  and\ \bibinfo {author} {\bibfnamefont {T.}~\bibnamefont {Killian}},\
  }\bibfield  {title} {\bibinfo {title} {Degenerate quantum gases of
  strontium},\ }in\ \href {https://doi.org/10.1142/9100} {\emph {\bibinfo
  {booktitle} {Annual Review of Cold Atoms and Molecules}}},\ \bibinfo {editor}
  {edited by\ \bibinfo {editor} {\bibfnamefont {K.}~\bibnamefont {Madison}},
  \bibinfo {editor} {\bibfnamefont {K.}~\bibnamefont {Bongs}}, \bibinfo
  {editor} {\bibfnamefont {L.~D.}\ \bibnamefont {Carr}}, \bibinfo {editor}
  {\bibfnamefont {A.~M.}\ \bibnamefont {Rey}},\ and\ \bibinfo {editor}
  {\bibfnamefont {H.}~\bibnamefont {Zhai}}}\ (\bibinfo  {publisher} {World
  Scientific},\ \bibinfo {year} {2014})\ Chap.~\bibinfo {chapter} {1},
  p.~\bibinfo {pages} {1}\BibitemShut {NoStop}%
\bibitem [{\citenamefont {Gr\"unert}\ and\ \citenamefont
  {Hemmerich}(2002)}]{Gru02}%
  \BibitemOpen
  \bibfield  {author} {\bibinfo {author} {\bibfnamefont {J.}~\bibnamefont
  {Gr\"unert}}\ and\ \bibinfo {author} {\bibfnamefont {A.}~\bibnamefont
  {Hemmerich}},\ }\bibfield  {title} {\bibinfo {title} {Sub-{Doppler}
  magneto-optical trap for calcium},\ }\href
  {https://doi.org/10.1103/PhysRevA.65.041401} {\bibfield  {journal} {\bibinfo
  {journal} {Phys. Rev. A}\ }\textbf {\bibinfo {volume} {65}},\ \bibinfo
  {pages} {041401} (\bibinfo {year} {2002})}\BibitemShut {NoStop}%
\bibitem [{\citenamefont {Hobson}\ \emph {et~al.}(2020)\citenamefont {Hobson},
  \citenamefont {Bowden}, \citenamefont {Vianello}, \citenamefont {Hill},\ and\
  \citenamefont {Gill}}]{Hob20}%
  \BibitemOpen
  \bibfield  {author} {\bibinfo {author} {\bibfnamefont {R.}~\bibnamefont
  {Hobson}}, \bibinfo {author} {\bibfnamefont {W.}~\bibnamefont {Bowden}},
  \bibinfo {author} {\bibfnamefont {A.}~\bibnamefont {Vianello}}, \bibinfo
  {author} {\bibfnamefont {I.~R.}\ \bibnamefont {Hill}},\ and\ \bibinfo
  {author} {\bibfnamefont {P.}~\bibnamefont {Gill}},\ }\bibfield  {title}
  {\bibinfo {title} {Midinfrared magneto-optical trap of metastable strontium
  for an optical lattice clock},\ }\href
  {https://doi.org/10.1103/PhysRevA.101.013420} {\bibfield  {journal} {\bibinfo
   {journal} {Phys. Rev. A}\ }\textbf {\bibinfo {volume} {101}},\ \bibinfo
  {pages} {013420} (\bibinfo {year} {2020})}\BibitemShut {NoStop}%
\bibitem [{\citenamefont {Riedmann}\ \emph {et~al.}(2012)\citenamefont
  {Riedmann}, \citenamefont {Kelkar}, \citenamefont {W\"ubbena}, \citenamefont
  {Pape}, \citenamefont {Kulosa}, \citenamefont {Zipfel}, \citenamefont {Fim},
  \citenamefont {R\"uhmann}, \citenamefont {Friebe}, \citenamefont {Ertmer},\
  and\ \citenamefont {Rasel}}]{Rie12}%
  \BibitemOpen
  \bibfield  {author} {\bibinfo {author} {\bibfnamefont {M.}~\bibnamefont
  {Riedmann}}, \bibinfo {author} {\bibfnamefont {H.}~\bibnamefont {Kelkar}},
  \bibinfo {author} {\bibfnamefont {T.}~\bibnamefont {W\"ubbena}}, \bibinfo
  {author} {\bibfnamefont {A.}~\bibnamefont {Pape}}, \bibinfo {author}
  {\bibfnamefont {A.}~\bibnamefont {Kulosa}}, \bibinfo {author} {\bibfnamefont
  {K.}~\bibnamefont {Zipfel}}, \bibinfo {author} {\bibfnamefont
  {D.}~\bibnamefont {Fim}}, \bibinfo {author} {\bibfnamefont {S.}~\bibnamefont
  {R\"uhmann}}, \bibinfo {author} {\bibfnamefont {J.}~\bibnamefont {Friebe}},
  \bibinfo {author} {\bibfnamefont {W.}~\bibnamefont {Ertmer}},\ and\ \bibinfo
  {author} {\bibfnamefont {E.}~\bibnamefont {Rasel}},\ }\bibfield  {title}
  {\bibinfo {title} {Beating the density limit by continuously loading a dipole
  trap from millikelvin-hot magnesium atoms},\ }\href
  {https://doi.org/10.1103/PhysRevA.86.043416} {\bibfield  {journal} {\bibinfo
  {journal} {Phys. Rev. A}\ }\textbf {\bibinfo {volume} {86}},\ \bibinfo
  {pages} {043416} (\bibinfo {year} {2012})}\BibitemShut {NoStop}%
\bibitem [{\citenamefont {Griesmaier}\ \emph {et~al.}(2005)\citenamefont
  {Griesmaier}, \citenamefont {Werner}, \citenamefont {Hensler}, \citenamefont
  {Stuhler},\ and\ \citenamefont {Pfau}}]{Gri05}%
  \BibitemOpen
  \bibfield  {author} {\bibinfo {author} {\bibfnamefont {A.}~\bibnamefont
  {Griesmaier}}, \bibinfo {author} {\bibfnamefont {J.}~\bibnamefont {Werner}},
  \bibinfo {author} {\bibfnamefont {S.}~\bibnamefont {Hensler}}, \bibinfo
  {author} {\bibfnamefont {J.}~\bibnamefont {Stuhler}},\ and\ \bibinfo {author}
  {\bibfnamefont {T.}~\bibnamefont {Pfau}},\ }\bibfield  {title} {\bibinfo
  {title} {{Bose}-{Einstein} condensation of chromium},\ }\href
  {https://doi.org/10.1103/PhysRevLett.94.160401} {\bibfield  {journal}
  {\bibinfo  {journal} {Phys. Rev. Lett.}\ }\textbf {\bibinfo {volume} {94}},\
  \bibinfo {pages} {160401} (\bibinfo {year} {2005})}\BibitemShut {NoStop}%
\bibitem [{\citenamefont {Lu}\ \emph {et~al.}(2011)\citenamefont {Lu},
  \citenamefont {Burdick}, \citenamefont {Youn},\ and\ \citenamefont
  {Lev}}]{Lu11}%
  \BibitemOpen
  \bibfield  {author} {\bibinfo {author} {\bibfnamefont {M.}~\bibnamefont
  {Lu}}, \bibinfo {author} {\bibfnamefont {N.~Q.}\ \bibnamefont {Burdick}},
  \bibinfo {author} {\bibfnamefont {S.~H.}\ \bibnamefont {Youn}},\ and\
  \bibinfo {author} {\bibfnamefont {B.~L.}\ \bibnamefont {Lev}},\ }\bibfield
  {title} {\bibinfo {title} {Strongly dipolar {Bose}-{Einstein} condensate of
  dysprosium},\ }\href {https://doi.org/10.1103/PhysRevLett.107.190401}
  {\bibfield  {journal} {\bibinfo  {journal} {Phys. Rev. Lett.}\ }\textbf
  {\bibinfo {volume} {107}},\ \bibinfo {pages} {190401} (\bibinfo {year}
  {2011})}\BibitemShut {NoStop}%
\bibitem [{\citenamefont {Aikawa}\ \emph {et~al.}(2012)\citenamefont {Aikawa},
  \citenamefont {Frisch}, \citenamefont {Mark}, \citenamefont {Baier},
  \citenamefont {Rietzler}, \citenamefont {Grimm},\ and\ \citenamefont
  {Ferlaino}}]{Aik12}%
  \BibitemOpen
  \bibfield  {author} {\bibinfo {author} {\bibfnamefont {K.}~\bibnamefont
  {Aikawa}}, \bibinfo {author} {\bibfnamefont {A.}~\bibnamefont {Frisch}},
  \bibinfo {author} {\bibfnamefont {M.}~\bibnamefont {Mark}}, \bibinfo {author}
  {\bibfnamefont {S.}~\bibnamefont {Baier}}, \bibinfo {author} {\bibfnamefont
  {A.}~\bibnamefont {Rietzler}}, \bibinfo {author} {\bibfnamefont
  {R.}~\bibnamefont {Grimm}},\ and\ \bibinfo {author} {\bibfnamefont
  {F.}~\bibnamefont {Ferlaino}},\ }\bibfield  {title} {\bibinfo {title}
  {{Bose}-{Einstein} condensation of erbium},\ }\href
  {https://doi.org/10.1103/PhysRevLett.108.210401} {\bibfield  {journal}
  {\bibinfo  {journal} {Phys. Rev. Lett.}\ }\textbf {\bibinfo {volume} {108}},\
  \bibinfo {pages} {210401} (\bibinfo {year} {2012})}\BibitemShut {NoStop}%
\bibitem [{\citenamefont {Davletov}\ \emph {et~al.}(2020)\citenamefont
  {Davletov}, \citenamefont {Tsyganok}, \citenamefont {Khlebnikov},
  \citenamefont {Pershin}, \citenamefont {Shaykin},\ and\ \citenamefont
  {Akimov}}]{Dav20}%
  \BibitemOpen
  \bibfield  {author} {\bibinfo {author} {\bibfnamefont {E.~T.}\ \bibnamefont
  {Davletov}}, \bibinfo {author} {\bibfnamefont {V.~V.}\ \bibnamefont
  {Tsyganok}}, \bibinfo {author} {\bibfnamefont {V.~A.}\ \bibnamefont
  {Khlebnikov}}, \bibinfo {author} {\bibfnamefont {D.~A.}\ \bibnamefont
  {Pershin}}, \bibinfo {author} {\bibfnamefont {D.~V.}\ \bibnamefont
  {Shaykin}},\ and\ \bibinfo {author} {\bibfnamefont {A.~V.}\ \bibnamefont
  {Akimov}},\ }\bibfield  {title} {\bibinfo {title} {Machine learning for
  achieving {Bose}-{Einstein} condensation of thulium atoms},\ }\href
  {https://doi.org/10.1103/PhysRevA.102.011302} {\bibfield  {journal} {\bibinfo
   {journal} {Phys. Rev. A}\ }\textbf {\bibinfo {volume} {102}},\ \bibinfo
  {pages} {011302} (\bibinfo {year} {2020})}\BibitemShut {NoStop}%
\bibitem [{\citenamefont {Lahaye}\ \emph {et~al.}(2009)\citenamefont {Lahaye},
  \citenamefont {Menotti}, \citenamefont {Santos}, \citenamefont {Lewenstein},\
  and\ \citenamefont {Pfau}}]{Lah09}%
  \BibitemOpen
  \bibfield  {author} {\bibinfo {author} {\bibfnamefont {T.}~\bibnamefont
  {Lahaye}}, \bibinfo {author} {\bibfnamefont {C.}~\bibnamefont {Menotti}},
  \bibinfo {author} {\bibfnamefont {L.}~\bibnamefont {Santos}}, \bibinfo
  {author} {\bibfnamefont {M.}~\bibnamefont {Lewenstein}},\ and\ \bibinfo
  {author} {\bibfnamefont {T.}~\bibnamefont {Pfau}},\ }\bibfield  {title}
  {\bibinfo {title} {The physics of dipolar bosonic quantum gases},\ }\href
  {https://doi.org/10.1088/0034-4885/72/12/126401} {\bibfield  {journal}
  {\bibinfo  {journal} {Reports on Progress in Physics}\ }\textbf {\bibinfo
  {volume} {72}},\ \bibinfo {pages} {126401} (\bibinfo {year}
  {2009})}\BibitemShut {NoStop}%
\bibitem [{\citenamefont {Baranov}\ \emph {et~al.}(2012)\citenamefont
  {Baranov}, \citenamefont {Dalmonte}, \citenamefont {Pupillo},\ and\
  \citenamefont {Zoller}}]{Bar12}%
  \BibitemOpen
  \bibfield  {author} {\bibinfo {author} {\bibfnamefont {M.~A.}\ \bibnamefont
  {Baranov}}, \bibinfo {author} {\bibfnamefont {M.}~\bibnamefont {Dalmonte}},
  \bibinfo {author} {\bibfnamefont {G.}~\bibnamefont {Pupillo}},\ and\ \bibinfo
  {author} {\bibfnamefont {P.}~\bibnamefont {Zoller}},\ }\bibfield  {title}
  {\bibinfo {title} {Condensed matter theory of dipolar quantum gases},\ }\href
  {https://doi.org/10.1021/cr2003568} {\bibfield  {journal} {\bibinfo
  {journal} {Chemical Reviews}\ }\textbf {\bibinfo {volume} {112}},\ \bibinfo
  {pages} {5012} (\bibinfo {year} {2012})}\BibitemShut {NoStop}%
\bibitem [{\citenamefont {Burdick}\ \emph {et~al.}(2016)\citenamefont
  {Burdick}, \citenamefont {Tang},\ and\ \citenamefont {Lev}}]{Bur16}%
  \BibitemOpen
  \bibfield  {author} {\bibinfo {author} {\bibfnamefont {N.~Q.}\ \bibnamefont
  {Burdick}}, \bibinfo {author} {\bibfnamefont {Y.}~\bibnamefont {Tang}},\ and\
  \bibinfo {author} {\bibfnamefont {B.~L.}\ \bibnamefont {Lev}},\ }\bibfield
  {title} {\bibinfo {title} {Long-lived spin-orbit-coupled degenerate dipolar
  fermi gas},\ }\href {https://doi.org/10.1103/PhysRevX.6.031022} {\bibfield
  {journal} {\bibinfo  {journal} {Phys. Rev. X}\ }\textbf {\bibinfo {volume}
  {6}},\ \bibinfo {pages} {031022} (\bibinfo {year} {2016})}\BibitemShut
  {NoStop}%
\bibitem [{\citenamefont {Chalopin}\ \emph {et~al.}(2020)\citenamefont
  {Chalopin}, \citenamefont {Satoor}, \citenamefont {Evrard}, \citenamefont
  {Makhalov}, \citenamefont {Dalibard}, \citenamefont {Lopes},\ and\
  \citenamefont {Nascimbene}}]{Cha20}%
  \BibitemOpen
  \bibfield  {author} {\bibinfo {author} {\bibfnamefont {T.}~\bibnamefont
  {Chalopin}}, \bibinfo {author} {\bibfnamefont {T.}~\bibnamefont {Satoor}},
  \bibinfo {author} {\bibfnamefont {A.}~\bibnamefont {Evrard}}, \bibinfo
  {author} {\bibfnamefont {V.}~\bibnamefont {Makhalov}}, \bibinfo {author}
  {\bibfnamefont {J.}~\bibnamefont {Dalibard}}, \bibinfo {author}
  {\bibfnamefont {R.}~\bibnamefont {Lopes}},\ and\ \bibinfo {author}
  {\bibfnamefont {S.}~\bibnamefont {Nascimbene}},\ }\bibfield  {title}
  {\bibinfo {title} {Probing chiral edge dynamics and bulk topology of a
  synthetic {Hall} system},\ }\href {https://doi.org/10.1038/s41567-020-0942-5}
  {\bibfield  {journal} {\bibinfo  {journal} {Nature Physics}\ }\textbf
  {\bibinfo {volume} {16}},\ \bibinfo {pages} {1017} (\bibinfo {year}
  {2020})}\BibitemShut {NoStop}%
\bibitem [{\citenamefont {Schmidt}\ \emph
  {et~al.}(2003{\natexlab{a}})\citenamefont {Schmidt}, \citenamefont {Hensler},
  \citenamefont {Werner}, \citenamefont {Binhammer}, \citenamefont
  {G\"orlitz},\ and\ \citenamefont {Pfau}}]{Sch03a}%
  \BibitemOpen
  \bibfield  {author} {\bibinfo {author} {\bibfnamefont {P.~O.}\ \bibnamefont
  {Schmidt}}, \bibinfo {author} {\bibfnamefont {S.}~\bibnamefont {Hensler}},
  \bibinfo {author} {\bibfnamefont {J.}~\bibnamefont {Werner}}, \bibinfo
  {author} {\bibfnamefont {T.}~\bibnamefont {Binhammer}}, \bibinfo {author}
  {\bibfnamefont {A.}~\bibnamefont {G\"orlitz}},\ and\ \bibinfo {author}
  {\bibfnamefont {T.}~\bibnamefont {Pfau}},\ }\bibfield  {title} {\bibinfo
  {title} {Continuous loading of cold atoms into a {Ioffe}-{Pritchard} magnetic
  trap},\ }\href {https://doi.org/10.1088/1464-4266/5/2/376} {\bibfield
  {journal} {\bibinfo  {journal} {Journal of Optics B: Quantum and
  Semiclassical Optics}\ }\textbf {\bibinfo {volume} {5}},\ \bibinfo {pages}
  {S170} (\bibinfo {year} {2003}{\natexlab{a}})}\BibitemShut {NoStop}%
\bibitem [{\citenamefont {Schmidt}\ \emph
  {et~al.}(2003{\natexlab{b}})\citenamefont {Schmidt}, \citenamefont {Hensler},
  \citenamefont {Werner}, \citenamefont {Binhammer}, \citenamefont
  {G\"{o}rlitz},\ and\ \citenamefont {Pfau}}]{Sch03b}%
  \BibitemOpen
  \bibfield  {author} {\bibinfo {author} {\bibfnamefont {P.~O.}\ \bibnamefont
  {Schmidt}}, \bibinfo {author} {\bibfnamefont {S.}~\bibnamefont {Hensler}},
  \bibinfo {author} {\bibfnamefont {J.}~\bibnamefont {Werner}}, \bibinfo
  {author} {\bibfnamefont {T.}~\bibnamefont {Binhammer}}, \bibinfo {author}
  {\bibfnamefont {A.}~\bibnamefont {G\"{o}rlitz}},\ and\ \bibinfo {author}
  {\bibfnamefont {T.}~\bibnamefont {Pfau}},\ }\bibfield  {title} {\bibinfo
  {title} {Doppler cooling of an optically dense cloud of magnetically trapped
  atoms},\ }\href {https://doi.org/10.1364/JOSAB.20.000960} {\bibfield
  {journal} {\bibinfo  {journal} {J. Opt. Soc. Am. B}\ }\textbf {\bibinfo
  {volume} {20}},\ \bibinfo {pages} {960} (\bibinfo {year}
  {2003}{\natexlab{b}})}\BibitemShut {NoStop}%
\bibitem [{\citenamefont {McClelland}\ and\ \citenamefont
  {Hanssen}(2006)}]{McC06}%
  \BibitemOpen
  \bibfield  {author} {\bibinfo {author} {\bibfnamefont {J.~J.}\ \bibnamefont
  {McClelland}}\ and\ \bibinfo {author} {\bibfnamefont {J.~L.}\ \bibnamefont
  {Hanssen}},\ }\bibfield  {title} {\bibinfo {title} {Laser cooling without
  repumping: A magneto-optical trap for erbium atoms},\ }\href
  {https://doi.org/10.1103/PhysRevLett.96.143005} {\bibfield  {journal}
  {\bibinfo  {journal} {Phys. Rev. Lett.}\ }\textbf {\bibinfo {volume} {96}},\
  \bibinfo {pages} {143005} (\bibinfo {year} {2006})}\BibitemShut {NoStop}%
\bibitem [{\citenamefont {Berglund}\ \emph {et~al.}(2008)\citenamefont
  {Berglund}, \citenamefont {Hanssen},\ and\ \citenamefont
  {McClelland}}]{Ber08}%
  \BibitemOpen
  \bibfield  {author} {\bibinfo {author} {\bibfnamefont {A.~J.}\ \bibnamefont
  {Berglund}}, \bibinfo {author} {\bibfnamefont {J.~L.}\ \bibnamefont
  {Hanssen}},\ and\ \bibinfo {author} {\bibfnamefont {J.~J.}\ \bibnamefont
  {McClelland}},\ }\bibfield  {title} {\bibinfo {title} {Narrow-line
  magneto-optical cooling and trapping of strongly magnetic atoms},\ }\href
  {https://doi.org/10.1103/PhysRevLett.100.113002} {\bibfield  {journal}
  {\bibinfo  {journal} {Phys. Rev. Lett.}\ }\textbf {\bibinfo {volume} {100}},\
  \bibinfo {pages} {113002} (\bibinfo {year} {2008})}\BibitemShut {NoStop}%
\bibitem [{\citenamefont {Lu}\ \emph {et~al.}(2012)\citenamefont {Lu},
  \citenamefont {Burdick},\ and\ \citenamefont {Lev}}]{Lu12}%
  \BibitemOpen
  \bibfield  {author} {\bibinfo {author} {\bibfnamefont {M.}~\bibnamefont
  {Lu}}, \bibinfo {author} {\bibfnamefont {N.~Q.}\ \bibnamefont {Burdick}},\
  and\ \bibinfo {author} {\bibfnamefont {B.~L.}\ \bibnamefont {Lev}},\
  }\bibfield  {title} {\bibinfo {title} {Quantum degenerate dipolar fermi
  gas},\ }\href {https://doi.org/10.1103/PhysRevLett.108.215301} {\bibfield
  {journal} {\bibinfo  {journal} {Phys. Rev. Lett.}\ }\textbf {\bibinfo
  {volume} {108}},\ \bibinfo {pages} {215301} (\bibinfo {year}
  {2012})}\BibitemShut {NoStop}%
\bibitem [{\citenamefont {Ilzh\"ofer}\ \emph {et~al.}(2018)\citenamefont
  {Ilzh\"ofer}, \citenamefont {Durastante}, \citenamefont {Patscheider},
  \citenamefont {Trautmann}, \citenamefont {Mark},\ and\ \citenamefont
  {Ferlaino}}]{Ilz18}%
  \BibitemOpen
  \bibfield  {author} {\bibinfo {author} {\bibfnamefont {P.}~\bibnamefont
  {Ilzh\"ofer}}, \bibinfo {author} {\bibfnamefont {G.}~\bibnamefont
  {Durastante}}, \bibinfo {author} {\bibfnamefont {A.}~\bibnamefont
  {Patscheider}}, \bibinfo {author} {\bibfnamefont {A.}~\bibnamefont
  {Trautmann}}, \bibinfo {author} {\bibfnamefont {M.~J.}\ \bibnamefont
  {Mark}},\ and\ \bibinfo {author} {\bibfnamefont {F.}~\bibnamefont
  {Ferlaino}},\ }\bibfield  {title} {\bibinfo {title} {Two-species five-beam
  magneto-optical trap for erbium and dysprosium},\ }\href
  {https://doi.org/10.1103/PhysRevA.97.023633} {\bibfield  {journal} {\bibinfo
  {journal} {Phys. Rev. A}\ }\textbf {\bibinfo {volume} {97}},\ \bibinfo
  {pages} {023633} (\bibinfo {year} {2018})}\BibitemShut {NoStop}%
\bibitem [{\citenamefont {Fattori}\ \emph {et~al.}(2006)\citenamefont
  {Fattori}, \citenamefont {Koch}, \citenamefont {Goetz}, \citenamefont
  {Griesmaier}, \citenamefont {Hensler}, \citenamefont {Stuhler},\ and\
  \citenamefont {Pfau}}]{Fat06}%
  \BibitemOpen
  \bibfield  {author} {\bibinfo {author} {\bibfnamefont {M.}~\bibnamefont
  {Fattori}}, \bibinfo {author} {\bibfnamefont {T.}~\bibnamefont {Koch}},
  \bibinfo {author} {\bibfnamefont {S.}~\bibnamefont {Goetz}}, \bibinfo
  {author} {\bibfnamefont {A.}~\bibnamefont {Griesmaier}}, \bibinfo {author}
  {\bibfnamefont {S.}~\bibnamefont {Hensler}}, \bibinfo {author} {\bibfnamefont
  {J.}~\bibnamefont {Stuhler}},\ and\ \bibinfo {author} {\bibfnamefont
  {T.}~\bibnamefont {Pfau}},\ }\bibfield  {title} {\bibinfo {title}
  {Demagnetization cooling of a gas},\ }\href
  {https://doi.org/10.1038/nphys443} {\bibfield  {journal} {\bibinfo  {journal}
  {Nature Physics}\ }\textbf {\bibinfo {volume} {2}},\ \bibinfo {pages} {765}
  (\bibinfo {year} {2006})}\BibitemShut {NoStop}%
\bibitem [{\citenamefont {Sengstock}\ \emph {et~al.}(1994)\citenamefont
  {Sengstock}, \citenamefont {Sterr}, \citenamefont {Müller}, \citenamefont
  {Rieger}, \citenamefont {Bettermann},\ and\ \citenamefont {Ertmer}}]{Sen94}%
  \BibitemOpen
  \bibfield  {author} {\bibinfo {author} {\bibfnamefont {K.}~\bibnamefont
  {Sengstock}}, \bibinfo {author} {\bibfnamefont {U.}~\bibnamefont {Sterr}},
  \bibinfo {author} {\bibfnamefont {J.}~\bibnamefont {Müller}}, \bibinfo
  {author} {\bibfnamefont {V.}~\bibnamefont {Rieger}}, \bibinfo {author}
  {\bibfnamefont {D.}~\bibnamefont {Bettermann}},\ and\ \bibinfo {author}
  {\bibfnamefont {W.}~\bibnamefont {Ertmer}},\ }\bibfield  {title} {\bibinfo
  {title} {Optical {Ramsey} spectroscopy on laser-trapped and thermal {Mg}
  atoms},\ }\href {https://doi.org/10.1007/BF01081160} {\bibfield  {journal}
  {\bibinfo  {journal} {Appl. Phys. B}\ }\textbf {\bibinfo {volume} {59}},\
  \bibinfo {pages} {99} (\bibinfo {year} {1994})}\BibitemShut {NoStop}%
\bibitem [{\citenamefont {De}\ \emph {et~al.}(2009)\citenamefont {De},
  \citenamefont {Dammalapati}, \citenamefont {Jungmann},\ and\ \citenamefont
  {Willmann}}]{De09}%
  \BibitemOpen
  \bibfield  {author} {\bibinfo {author} {\bibfnamefont {S.}~\bibnamefont
  {De}}, \bibinfo {author} {\bibfnamefont {U.}~\bibnamefont {Dammalapati}},
  \bibinfo {author} {\bibfnamefont {K.}~\bibnamefont {Jungmann}},\ and\
  \bibinfo {author} {\bibfnamefont {L.}~\bibnamefont {Willmann}},\ }\bibfield
  {title} {\bibinfo {title} {Magneto-optical trapping of barium},\ }\href
  {https://doi.org/10.1103/PhysRevA.79.041402} {\bibfield  {journal} {\bibinfo
  {journal} {Phys. Rev. A}\ }\textbf {\bibinfo {volume} {79}},\ \bibinfo
  {pages} {041402} (\bibinfo {year} {2009})}\BibitemShut {NoStop}%
\bibitem [{\citenamefont {Guest}\ \emph {et~al.}(2007)\citenamefont {Guest},
  \citenamefont {Scielzo}, \citenamefont {Ahmad}, \citenamefont {Bailey},
  \citenamefont {Greene}, \citenamefont {Holt}, \citenamefont {Lu},
  \citenamefont {O'Connor},\ and\ \citenamefont {Potterveld}}]{Gue07}%
  \BibitemOpen
  \bibfield  {author} {\bibinfo {author} {\bibfnamefont {J.~R.}\ \bibnamefont
  {Guest}}, \bibinfo {author} {\bibfnamefont {N.~D.}\ \bibnamefont {Scielzo}},
  \bibinfo {author} {\bibfnamefont {I.}~\bibnamefont {Ahmad}}, \bibinfo
  {author} {\bibfnamefont {K.}~\bibnamefont {Bailey}}, \bibinfo {author}
  {\bibfnamefont {J.~P.}\ \bibnamefont {Greene}}, \bibinfo {author}
  {\bibfnamefont {R.~J.}\ \bibnamefont {Holt}}, \bibinfo {author}
  {\bibfnamefont {Z.-T.}\ \bibnamefont {Lu}}, \bibinfo {author} {\bibfnamefont
  {T.~P.}\ \bibnamefont {O'Connor}},\ and\ \bibinfo {author} {\bibfnamefont
  {D.~H.}\ \bibnamefont {Potterveld}},\ }\bibfield  {title} {\bibinfo {title}
  {Laser trapping of $^{225}\mathrm{Ra}$ and $^{226}\mathrm{Ra}$ with repumping
  by room-temperature blackbody radiation},\ }\href
  {https://doi.org/10.1103/PhysRevLett.98.093001} {\bibfield  {journal}
  {\bibinfo  {journal} {Phys. Rev. Lett.}\ }\textbf {\bibinfo {volume} {98}},\
  \bibinfo {pages} {093001} (\bibinfo {year} {2007})}\BibitemShut {NoStop}%
\bibitem [{\citenamefont {Simsarian}\ \emph {et~al.}(1996)\citenamefont
  {Simsarian}, \citenamefont {Ghosh}, \citenamefont {Gwinner}, \citenamefont
  {Orozco}, \citenamefont {Sprouse},\ and\ \citenamefont {Voytas}}]{Sim96}%
  \BibitemOpen
  \bibfield  {author} {\bibinfo {author} {\bibfnamefont {J.~E.}\ \bibnamefont
  {Simsarian}}, \bibinfo {author} {\bibfnamefont {A.}~\bibnamefont {Ghosh}},
  \bibinfo {author} {\bibfnamefont {G.}~\bibnamefont {Gwinner}}, \bibinfo
  {author} {\bibfnamefont {L.~A.}\ \bibnamefont {Orozco}}, \bibinfo {author}
  {\bibfnamefont {G.~D.}\ \bibnamefont {Sprouse}},\ and\ \bibinfo {author}
  {\bibfnamefont {P.~A.}\ \bibnamefont {Voytas}},\ }\bibfield  {title}
  {\bibinfo {title} {Magneto-optic trapping of ${}^{210}${Fr}},\ }\href
  {https://doi.org/10.1103/PhysRevLett.76.3522} {\bibfield  {journal} {\bibinfo
   {journal} {Phys. Rev. Lett.}\ }\textbf {\bibinfo {volume} {76}},\ \bibinfo
  {pages} {3522} (\bibinfo {year} {1996})}\BibitemShut {NoStop}%
\bibitem [{\citenamefont {Brickman}\ \emph {et~al.}(2007)\citenamefont
  {Brickman}, \citenamefont {Chang}, \citenamefont {Acton}, \citenamefont
  {Chew}, \citenamefont {Matsukevich}, \citenamefont {Haljan}, \citenamefont
  {Bagnato},\ and\ \citenamefont {Monroe}}]{Bri07}%
  \BibitemOpen
  \bibfield  {author} {\bibinfo {author} {\bibfnamefont {K.-A.}\ \bibnamefont
  {Brickman}}, \bibinfo {author} {\bibfnamefont {M.-S.}\ \bibnamefont {Chang}},
  \bibinfo {author} {\bibfnamefont {M.}~\bibnamefont {Acton}}, \bibinfo
  {author} {\bibfnamefont {A.}~\bibnamefont {Chew}}, \bibinfo {author}
  {\bibfnamefont {D.}~\bibnamefont {Matsukevich}}, \bibinfo {author}
  {\bibfnamefont {P.~C.}\ \bibnamefont {Haljan}}, \bibinfo {author}
  {\bibfnamefont {V.~S.}\ \bibnamefont {Bagnato}},\ and\ \bibinfo {author}
  {\bibfnamefont {C.}~\bibnamefont {Monroe}},\ }\bibfield  {title} {\bibinfo
  {title} {Magneto-optical trapping of cadmium},\ }\href
  {https://doi.org/10.1103/PhysRevA.76.043411} {\bibfield  {journal} {\bibinfo
  {journal} {Phys. Rev. A}\ }\textbf {\bibinfo {volume} {76}},\ \bibinfo
  {pages} {043411} (\bibinfo {year} {2007})}\BibitemShut {NoStop}%
\bibitem [{\citenamefont {Hachisu}\ \emph {et~al.}(2008)\citenamefont
  {Hachisu}, \citenamefont {Miyagishi}, \citenamefont {Porsev}, \citenamefont
  {Derevianko}, \citenamefont {Ovsiannikov}, \citenamefont {Pal'chikov},
  \citenamefont {Takamoto},\ and\ \citenamefont {Katori}}]{Hac08}%
  \BibitemOpen
  \bibfield  {author} {\bibinfo {author} {\bibfnamefont {H.}~\bibnamefont
  {Hachisu}}, \bibinfo {author} {\bibfnamefont {K.}~\bibnamefont {Miyagishi}},
  \bibinfo {author} {\bibfnamefont {S.~G.}\ \bibnamefont {Porsev}}, \bibinfo
  {author} {\bibfnamefont {A.}~\bibnamefont {Derevianko}}, \bibinfo {author}
  {\bibfnamefont {V.~D.}\ \bibnamefont {Ovsiannikov}}, \bibinfo {author}
  {\bibfnamefont {V.~G.}\ \bibnamefont {Pal'chikov}}, \bibinfo {author}
  {\bibfnamefont {M.}~\bibnamefont {Takamoto}},\ and\ \bibinfo {author}
  {\bibfnamefont {H.}~\bibnamefont {Katori}},\ }\bibfield  {title} {\bibinfo
  {title} {Trapping of neutral mercury atoms and prospects for optical lattice
  clocks},\ }\href {https://doi.org/10.1103/PhysRevLett.100.053001} {\bibfield
  {journal} {\bibinfo  {journal} {Phys. Rev. Lett.}\ }\textbf {\bibinfo
  {volume} {100}},\ \bibinfo {pages} {053001} (\bibinfo {year}
  {2008})}\BibitemShut {NoStop}%
\bibitem [{\citenamefont {Miao}\ \emph {et~al.}(2014)\citenamefont {Miao},
  \citenamefont {Hostetter}, \citenamefont {Stratis},\ and\ \citenamefont
  {Saffman}}]{Mia14}%
  \BibitemOpen
  \bibfield  {author} {\bibinfo {author} {\bibfnamefont {J.}~\bibnamefont
  {Miao}}, \bibinfo {author} {\bibfnamefont {J.}~\bibnamefont {Hostetter}},
  \bibinfo {author} {\bibfnamefont {G.}~\bibnamefont {Stratis}},\ and\ \bibinfo
  {author} {\bibfnamefont {M.}~\bibnamefont {Saffman}},\ }\bibfield  {title}
  {\bibinfo {title} {Magneto-optical trapping of holmium atoms},\ }\href
  {https://doi.org/10.1103/PhysRevA.89.041401} {\bibfield  {journal} {\bibinfo
  {journal} {Phys. Rev. A}\ }\textbf {\bibinfo {volume} {89}},\ \bibinfo
  {pages} {041401} (\bibinfo {year} {2014})}\BibitemShut {NoStop}%
\bibitem [{\citenamefont {Uhlenberg}\ \emph {et~al.}(2000)\citenamefont
  {Uhlenberg}, \citenamefont {Dirscherl},\ and\ \citenamefont
  {Walther}}]{Uhl00}%
  \BibitemOpen
  \bibfield  {author} {\bibinfo {author} {\bibfnamefont {G.}~\bibnamefont
  {Uhlenberg}}, \bibinfo {author} {\bibfnamefont {J.}~\bibnamefont
  {Dirscherl}},\ and\ \bibinfo {author} {\bibfnamefont {H.}~\bibnamefont
  {Walther}},\ }\bibfield  {title} {\bibinfo {title} {Magneto-optical trapping
  of silver atoms},\ }\href {https://doi.org/10.1103/PhysRevA.62.063404}
  {\bibfield  {journal} {\bibinfo  {journal} {Phys. Rev. A}\ }\textbf {\bibinfo
  {volume} {62}},\ \bibinfo {pages} {063404} (\bibinfo {year}
  {2000})}\BibitemShut {NoStop}%
\bibitem [{\citenamefont {Eustice}\ \emph {et~al.}(2020)\citenamefont
  {Eustice}, \citenamefont {Cassella},\ and\ \citenamefont
  {Stamper-Kurn}}]{Eus20}%
  \BibitemOpen
  \bibfield  {author} {\bibinfo {author} {\bibfnamefont {S.}~\bibnamefont
  {Eustice}}, \bibinfo {author} {\bibfnamefont {K.}~\bibnamefont {Cassella}},\
  and\ \bibinfo {author} {\bibfnamefont {D.}~\bibnamefont {Stamper-Kurn}},\
  }\bibfield  {title} {\bibinfo {title} {Laser cooling of transition-metal
  atoms},\ }\href {https://doi.org/10.1103/PhysRevA.102.053327} {\bibfield
  {journal} {\bibinfo  {journal} {Phys. Rev. A}\ }\textbf {\bibinfo {volume}
  {102}},\ \bibinfo {pages} {053327} (\bibinfo {year} {2020})}\BibitemShut
  {NoStop}%
\bibitem [{\citenamefont {Stellmer}\ \emph {et~al.}(2013)\citenamefont
  {Stellmer}, \citenamefont {Pasquiou}, \citenamefont {Grimm},\ and\
  \citenamefont {Schreck}}]{Ste13}%
  \BibitemOpen
  \bibfield  {author} {\bibinfo {author} {\bibfnamefont {S.}~\bibnamefont
  {Stellmer}}, \bibinfo {author} {\bibfnamefont {B.}~\bibnamefont {Pasquiou}},
  \bibinfo {author} {\bibfnamefont {R.}~\bibnamefont {Grimm}},\ and\ \bibinfo
  {author} {\bibfnamefont {F.}~\bibnamefont {Schreck}},\ }\bibfield  {title}
  {\bibinfo {title} {Laser cooling to quantum degeneracy},\ }\href
  {https://doi.org/10.1103/PhysRevLett.110.263003} {\bibfield  {journal}
  {\bibinfo  {journal} {Phys. Rev. Lett.}\ }\textbf {\bibinfo {volume} {110}},\
  \bibinfo {pages} {263003} (\bibinfo {year} {2013})}\BibitemShut {NoStop}%
\bibitem [{\citenamefont {Hu}\ \emph {et~al.}(2017)\citenamefont {Hu},
  \citenamefont {Urvoy}, \citenamefont {Vendeiro}, \citenamefont {Cr{\'e}pel},
  \citenamefont {Chen},\ and\ \citenamefont {Vuleti{\'c}}}]{Hu17}%
  \BibitemOpen
  \bibfield  {author} {\bibinfo {author} {\bibfnamefont {J.}~\bibnamefont
  {Hu}}, \bibinfo {author} {\bibfnamefont {A.}~\bibnamefont {Urvoy}}, \bibinfo
  {author} {\bibfnamefont {Z.}~\bibnamefont {Vendeiro}}, \bibinfo {author}
  {\bibfnamefont {V.}~\bibnamefont {Cr{\'e}pel}}, \bibinfo {author}
  {\bibfnamefont {W.}~\bibnamefont {Chen}},\ and\ \bibinfo {author}
  {\bibfnamefont {V.}~\bibnamefont {Vuleti{\'c}}},\ }\bibfield  {title}
  {\bibinfo {title} {Creation of a {Bose}-condensed gas of $^{87}${Rb} by laser
  cooling},\ }\href {https://doi.org/10.1126/science.aan5614} {\bibfield
  {journal} {\bibinfo  {journal} {Science}\ }\textbf {\bibinfo {volume}
  {358}},\ \bibinfo {pages} {1078} (\bibinfo {year} {2017})}\BibitemShut
  {NoStop}%
\bibitem [{\citenamefont {Urvoy}\ \emph {et~al.}(2019)\citenamefont {Urvoy},
  \citenamefont {Vendeiro}, \citenamefont {Ramette}, \citenamefont
  {Adiyatullin},\ and\ \citenamefont {Vuleti\ifmmode~\acute{c}\else
  \'{c}\fi{}}}]{Urv19}%
  \BibitemOpen
  \bibfield  {author} {\bibinfo {author} {\bibfnamefont {A.}~\bibnamefont
  {Urvoy}}, \bibinfo {author} {\bibfnamefont {Z.}~\bibnamefont {Vendeiro}},
  \bibinfo {author} {\bibfnamefont {J.}~\bibnamefont {Ramette}}, \bibinfo
  {author} {\bibfnamefont {A.}~\bibnamefont {Adiyatullin}},\ and\ \bibinfo
  {author} {\bibfnamefont {V.}~\bibnamefont {Vuleti\ifmmode~\acute{c}\else
  \'{c}\fi{}}},\ }\bibfield  {title} {\bibinfo {title} {Direct laser cooling to
  {Bose}-{Einstein} condensation in a dipole trap},\ }\href
  {https://doi.org/10.1103/PhysRevLett.122.203202} {\bibfield  {journal}
  {\bibinfo  {journal} {Phys. Rev. Lett.}\ }\textbf {\bibinfo {volume} {122}},\
  \bibinfo {pages} {203202} (\bibinfo {year} {2019})}\BibitemShut {NoStop}%
\bibitem [{\citenamefont {Pinkse}\ \emph {et~al.}(1997)\citenamefont {Pinkse},
  \citenamefont {Mosk}, \citenamefont {Weidem\"uller}, \citenamefont
  {Reynolds}, \citenamefont {Hijmans},\ and\ \citenamefont {Walraven}}]{Pin97}%
  \BibitemOpen
  \bibfield  {author} {\bibinfo {author} {\bibfnamefont {P.~W.~H.}\
  \bibnamefont {Pinkse}}, \bibinfo {author} {\bibfnamefont {A.}~\bibnamefont
  {Mosk}}, \bibinfo {author} {\bibfnamefont {M.}~\bibnamefont {Weidem\"uller}},
  \bibinfo {author} {\bibfnamefont {M.~W.}\ \bibnamefont {Reynolds}}, \bibinfo
  {author} {\bibfnamefont {T.~W.}\ \bibnamefont {Hijmans}},\ and\ \bibinfo
  {author} {\bibfnamefont {J.~T.~M.}\ \bibnamefont {Walraven}},\ }\bibfield
  {title} {\bibinfo {title} {Adiabatically changing the phase-space density of
  a trapped {Bose} gas},\ }\href {https://doi.org/10.1103/PhysRevLett.78.990}
  {\bibfield  {journal} {\bibinfo  {journal} {Phys. Rev. Lett.}\ }\textbf
  {\bibinfo {volume} {78}},\ \bibinfo {pages} {990} (\bibinfo {year}
  {1997})}\BibitemShut {NoStop}%
\bibitem [{\citenamefont {Stamper-Kurn}\ \emph {et~al.}(1998)\citenamefont
  {Stamper-Kurn}, \citenamefont {Miesner}, \citenamefont {Chikkatur},
  \citenamefont {Inouye}, \citenamefont {Stenger},\ and\ \citenamefont
  {Ketterle}}]{Sta98}%
  \BibitemOpen
  \bibfield  {author} {\bibinfo {author} {\bibfnamefont {D.~M.}\ \bibnamefont
  {Stamper-Kurn}}, \bibinfo {author} {\bibfnamefont {H.-J.}\ \bibnamefont
  {Miesner}}, \bibinfo {author} {\bibfnamefont {A.~P.}\ \bibnamefont
  {Chikkatur}}, \bibinfo {author} {\bibfnamefont {S.}~\bibnamefont {Inouye}},
  \bibinfo {author} {\bibfnamefont {J.}~\bibnamefont {Stenger}},\ and\ \bibinfo
  {author} {\bibfnamefont {W.}~\bibnamefont {Ketterle}},\ }\bibfield  {title}
  {\bibinfo {title} {Reversible formation of a {Bose}-{Einstein} condensate},\
  }\href {https://doi.org/10.1103/PhysRevLett.81.2194} {\bibfield  {journal}
  {\bibinfo  {journal} {Phys. Rev. Lett.}\ }\textbf {\bibinfo {volume} {81}},\
  \bibinfo {pages} {2194} (\bibinfo {year} {1998})}\BibitemShut {NoStop}%
\bibitem [{\citenamefont {Sonderhouse}\ \emph {et~al.}(2020)\citenamefont
  {Sonderhouse}, \citenamefont {Sanner}, \citenamefont {Hutson}, \citenamefont
  {Goban}, \citenamefont {Bilitewski}, \citenamefont {Yan}, \citenamefont
  {Milner}, \citenamefont {Rey},\ and\ \citenamefont {Ye}}]{Son20}%
  \BibitemOpen
  \bibfield  {author} {\bibinfo {author} {\bibfnamefont {L.}~\bibnamefont
  {Sonderhouse}}, \bibinfo {author} {\bibfnamefont {C.}~\bibnamefont {Sanner}},
  \bibinfo {author} {\bibfnamefont {R.~B.}\ \bibnamefont {Hutson}}, \bibinfo
  {author} {\bibfnamefont {A.}~\bibnamefont {Goban}}, \bibinfo {author}
  {\bibfnamefont {T.}~\bibnamefont {Bilitewski}}, \bibinfo {author}
  {\bibfnamefont {L.}~\bibnamefont {Yan}}, \bibinfo {author} {\bibfnamefont
  {W.~R.}\ \bibnamefont {Milner}}, \bibinfo {author} {\bibfnamefont {A.~M.}\
  \bibnamefont {Rey}},\ and\ \bibinfo {author} {\bibfnamefont {J.}~\bibnamefont
  {Ye}},\ }\bibfield  {title} {\bibinfo {title} {Thermodynamics of a deeply
  degenerate {SU(N)}-symmetric {Fermi} gas},\ }\bibfield  {journal} {\bibinfo
  {journal} {Nature Physics}\ }\href
  {https://doi.org/10.1038/s41567-020-0986-6} {10.1038/s41567-020-0986-6}
  (\bibinfo {year} {2020})\BibitemShut {NoStop}%
\bibitem [{\citenamefont {Robins}\ \emph {et~al.}(2013)\citenamefont {Robins},
  \citenamefont {Altin}, \citenamefont {Debs},\ and\ \citenamefont
  {Close}}]{Rob13}%
  \BibitemOpen
  \bibfield  {author} {\bibinfo {author} {\bibfnamefont {N.}~\bibnamefont
  {Robins}}, \bibinfo {author} {\bibfnamefont {P.}~\bibnamefont {Altin}},
  \bibinfo {author} {\bibfnamefont {J.}~\bibnamefont {Debs}},\ and\ \bibinfo
  {author} {\bibfnamefont {J.}~\bibnamefont {Close}},\ }\bibfield  {title}
  {\bibinfo {title} {Atom lasers: Production, properties and prospects for
  precision inertial measurement},\ }\href
  {https://doi.org/https://doi.org/10.1016/j.physrep.2013.03.006} {\bibfield
  {journal} {\bibinfo  {journal} {Physics Reports}\ }\textbf {\bibinfo {volume}
  {529}},\ \bibinfo {pages} {265} (\bibinfo {year} {2013})}\BibitemShut
  {NoStop}%
\bibitem [{\citenamefont {Chikkatur}\ \emph {et~al.}(2002)\citenamefont
  {Chikkatur}, \citenamefont {Shin}, \citenamefont {Leanhardt}, \citenamefont
  {Kielpinski}, \citenamefont {Tsikata}, \citenamefont {Gustavson},
  \citenamefont {Pritchard},\ and\ \citenamefont {Ketterle}}]{Chi02}%
  \BibitemOpen
  \bibfield  {author} {\bibinfo {author} {\bibfnamefont {A.~P.}\ \bibnamefont
  {Chikkatur}}, \bibinfo {author} {\bibfnamefont {Y.}~\bibnamefont {Shin}},
  \bibinfo {author} {\bibfnamefont {A.~E.}\ \bibnamefont {Leanhardt}}, \bibinfo
  {author} {\bibfnamefont {D.}~\bibnamefont {Kielpinski}}, \bibinfo {author}
  {\bibfnamefont {E.}~\bibnamefont {Tsikata}}, \bibinfo {author} {\bibfnamefont
  {T.~L.}\ \bibnamefont {Gustavson}}, \bibinfo {author} {\bibfnamefont {D.~E.}\
  \bibnamefont {Pritchard}},\ and\ \bibinfo {author} {\bibfnamefont
  {W.}~\bibnamefont {Ketterle}},\ }\bibfield  {title} {\bibinfo {title} {A
  continuous source of {Bose}-{Einstein} condensed atoms},\ }\href
  {https://doi.org/10.1126/science.296.5576.2193} {\bibfield  {journal}
  {\bibinfo  {journal} {Science}\ }\textbf {\bibinfo {volume} {296}},\ \bibinfo
  {pages} {2193} (\bibinfo {year} {2002})}\BibitemShut {NoStop}%
\bibitem [{\citenamefont {Miesner}\ \emph {et~al.}(1998)\citenamefont
  {Miesner}, \citenamefont {Stamper-Kurn}, \citenamefont {Andrews},
  \citenamefont {Durfee}, \citenamefont {Inouye},\ and\ \citenamefont
  {Ketterle}}]{Mie98}%
  \BibitemOpen
  \bibfield  {author} {\bibinfo {author} {\bibfnamefont {H.-J.}\ \bibnamefont
  {Miesner}}, \bibinfo {author} {\bibfnamefont {D.~M.}\ \bibnamefont
  {Stamper-Kurn}}, \bibinfo {author} {\bibfnamefont {M.~R.}\ \bibnamefont
  {Andrews}}, \bibinfo {author} {\bibfnamefont {D.~S.}\ \bibnamefont {Durfee}},
  \bibinfo {author} {\bibfnamefont {S.}~\bibnamefont {Inouye}},\ and\ \bibinfo
  {author} {\bibfnamefont {W.}~\bibnamefont {Ketterle}},\ }\bibfield  {title}
  {\bibinfo {title} {Bosonic stimulation in the formation of a
  {Bose}-{Einstein} condensate},\ }\href
  {https://doi.org/10.1126/science.279.5353.1005} {\bibfield  {journal}
  {\bibinfo  {journal} {Science}\ }\textbf {\bibinfo {volume} {279}},\ \bibinfo
  {pages} {1005} (\bibinfo {year} {1998})}\BibitemShut {NoStop}%
\bibitem [{\citenamefont {Robins}\ \emph {et~al.}(2008)\citenamefont {Robins},
  \citenamefont {Figl}, \citenamefont {Jeppesen}, \citenamefont {Dennis},\ and\
  \citenamefont {Close}}]{Rob08}%
  \BibitemOpen
  \bibfield  {author} {\bibinfo {author} {\bibfnamefont {N.~P.}\ \bibnamefont
  {Robins}}, \bibinfo {author} {\bibfnamefont {C.}~\bibnamefont {Figl}},
  \bibinfo {author} {\bibfnamefont {M.}~\bibnamefont {Jeppesen}}, \bibinfo
  {author} {\bibfnamefont {G.~R.}\ \bibnamefont {Dennis}},\ and\ \bibinfo
  {author} {\bibfnamefont {J.~D.}\ \bibnamefont {Close}},\ }\bibfield  {title}
  {\bibinfo {title} {A pumped atom laser},\ }\href
  {https://doi.org/10.1038/nphys1027} {\bibfield  {journal} {\bibinfo
  {journal} {Nat. Phys.}\ }\textbf {\bibinfo {volume} {4}},\ \bibinfo {pages}
  {731} (\bibinfo {year} {2008})}\BibitemShut {NoStop}%
\bibitem [{\citenamefont {Chen}\ \emph {et~al.}(2022)\citenamefont {Chen},
  \citenamefont {Escudero}, \citenamefont {Minář}, \citenamefont {Pasquiou},
  \citenamefont {Bennetts},\ and\ \citenamefont {Schreck}}]{Che20}%
  \BibitemOpen
  \bibfield  {author} {\bibinfo {author} {\bibfnamefont {C.-C.}\ \bibnamefont
  {Chen}}, \bibinfo {author} {\bibfnamefont {R.~G.}\ \bibnamefont {Escudero}},
  \bibinfo {author} {\bibfnamefont {J.}~\bibnamefont {Minář}}, \bibinfo
  {author} {\bibfnamefont {B.}~\bibnamefont {Pasquiou}}, \bibinfo {author}
  {\bibfnamefont {S.}~\bibnamefont {Bennetts}},\ and\ \bibinfo {author}
  {\bibfnamefont {F.}~\bibnamefont {Schreck}},\ }\bibfield  {title} {\bibinfo
  {title} {Continuous {Bose}-{Einstein} condensation},\ }\href
  {https://doi.org/10.1038/s41586-022-04731-z} {\bibfield  {journal} {\bibinfo
  {journal} {Nature}\ }\textbf {\bibinfo {volume} {606}},\ \bibinfo {pages}
  {683} (\bibinfo {year} {2022})}\BibitemShut {NoStop}%
\bibitem [{\citenamefont {Kaufman}\ and\ \citenamefont
  {Ni}(2021)}]{KaufmanNi2021}%
  \BibitemOpen
  \bibfield  {author} {\bibinfo {author} {\bibfnamefont {A.~M.}\ \bibnamefont
  {Kaufman}}\ and\ \bibinfo {author} {\bibfnamefont {K.-K.}\ \bibnamefont
  {Ni}},\ }\bibfield  {title} {\bibinfo {title} {Quantum science with optical
  tweezer arrays of ultracold atoms and molecules},\ }\href
  {https://doi.org/10.1038/s41567-021-01357-2} {\bibfield  {journal} {\bibinfo
  {journal} {Nature Physics}\ }\textbf {\bibinfo {volume} {17}},\ \bibinfo
  {pages} {1324} (\bibinfo {year} {2021})}\BibitemShut {NoStop}%
\bibitem [{\citenamefont {Ott}(2016)}]{Ott16}%
  \BibitemOpen
  \bibfield  {author} {\bibinfo {author} {\bibfnamefont {H.}~\bibnamefont
  {Ott}},\ }\bibfield  {title} {\bibinfo {title} {Single atom detection in
  ultracold quantum gases: a review of current progress},\ }\href
  {https://doi.org/10.1088/0034-4885/79/5/054401} {\bibfield  {journal}
  {\bibinfo  {journal} {Rep. Prog. Phys.}\ }\textbf {\bibinfo {volume} {79}},\
  \bibinfo {pages} {054401} (\bibinfo {year} {2016})}\BibitemShut {NoStop}%
\bibitem [{\citenamefont {Barredo}\ \emph {et~al.}(2016)\citenamefont
  {Barredo}, \citenamefont {de~L{\'e}s{\'e}leuc}, \citenamefont {Lienhard},
  \citenamefont {Lahaye},\ and\ \citenamefont {Browaeys}}]{Bar16}%
  \BibitemOpen
  \bibfield  {author} {\bibinfo {author} {\bibfnamefont {D.}~\bibnamefont
  {Barredo}}, \bibinfo {author} {\bibfnamefont {S.}~\bibnamefont
  {de~L{\'e}s{\'e}leuc}}, \bibinfo {author} {\bibfnamefont {V.}~\bibnamefont
  {Lienhard}}, \bibinfo {author} {\bibfnamefont {T.}~\bibnamefont {Lahaye}},\
  and\ \bibinfo {author} {\bibfnamefont {A.}~\bibnamefont {Browaeys}},\
  }\bibfield  {title} {\bibinfo {title} {An atom-by-atom assembler of
  defect-free arbitrary two-dimensional atomic arrays},\ }\href
  {https://doi.org/10.1126/science.aah3778} {\bibfield  {journal} {\bibinfo
  {journal} {Science}\ }\textbf {\bibinfo {volume} {354}},\ \bibinfo {pages}
  {1021} (\bibinfo {year} {2016})}\BibitemShut {NoStop}%
\bibitem [{\citenamefont {Olshanii}\ and\ \citenamefont {Weiss}(2002)}]{Ols02}%
  \BibitemOpen
  \bibfield  {author} {\bibinfo {author} {\bibfnamefont {M.}~\bibnamefont
  {Olshanii}}\ and\ \bibinfo {author} {\bibfnamefont {D.}~\bibnamefont
  {Weiss}},\ }\bibfield  {title} {\bibinfo {title} {Producing {Bose}-{Einstein}
  condensates using optical lattices},\ }\href
  {https://doi.org/10.1103/PhysRevLett.89.090404} {\bibfield  {journal}
  {\bibinfo  {journal} {Phys. Rev. Lett.}\ }\textbf {\bibinfo {volume} {89}},\
  \bibinfo {pages} {090404} (\bibinfo {year} {2002})}\BibitemShut {NoStop}%
\bibitem [{\citenamefont {Tarbutt}(2018)}]{Tar18}%
  \BibitemOpen
  \bibfield  {author} {\bibinfo {author} {\bibfnamefont {M.~R.}\ \bibnamefont
  {Tarbutt}},\ }\bibfield  {title} {\bibinfo {title} {Laser cooling of
  molecules},\ }\href {https://doi.org/10.1080/00107514.2018.1576338}
  {\bibfield  {journal} {\bibinfo  {journal} {Contemporary Physics}\ }\textbf
  {\bibinfo {volume} {59}},\ \bibinfo {pages} {356} (\bibinfo {year}
  {2018})}\BibitemShut {NoStop}%
\bibitem [{\citenamefont {Fitch}\ and\ \citenamefont {Tarbutt}(2021)}]{Fit21}%
  \BibitemOpen
  \bibfield  {author} {\bibinfo {author} {\bibfnamefont {N.}~\bibnamefont
  {Fitch}}\ and\ \bibinfo {author} {\bibfnamefont {M.}~\bibnamefont
  {Tarbutt}},\ }\bibfield  {title} {\bibinfo {title} {Chapter three -
  laser-cooled molecules},\ }in\ \href
  {https://doi.org/https://doi.org/10.1016/bs.aamop.2021.04.003} {\emph
  {\bibinfo {booktitle} {Adv. At. Mol. Phys.}}},\ Vol.~\bibinfo {volume} {70},\
  \bibinfo {editor} {edited by\ \bibinfo {editor} {\bibfnamefont {L.~F.}\
  \bibnamefont {Dimauro}}, \bibinfo {editor} {\bibfnamefont {H.}~\bibnamefont
  {Perrin}},\ and\ \bibinfo {editor} {\bibfnamefont {S.~F.}\ \bibnamefont
  {Yelin}}}\ (\bibinfo  {publisher} {Academic Press},\ \bibinfo {year} {2021})\
  pp.\ \bibinfo {pages} {157--262}\BibitemShut {NoStop}%
\bibitem [{\citenamefont {Barry}\ \emph {et~al.}(2012)\citenamefont {Barry},
  \citenamefont {Shuman}, \citenamefont {Norrgard},\ and\ \citenamefont
  {DeMille}}]{Bar12b}%
  \BibitemOpen
  \bibfield  {author} {\bibinfo {author} {\bibfnamefont {J.~F.}\ \bibnamefont
  {Barry}}, \bibinfo {author} {\bibfnamefont {E.~S.}\ \bibnamefont {Shuman}},
  \bibinfo {author} {\bibfnamefont {E.~B.}\ \bibnamefont {Norrgard}},\ and\
  \bibinfo {author} {\bibfnamefont {D.}~\bibnamefont {DeMille}},\ }\bibfield
  {title} {\bibinfo {title} {Laser radiation pressure slowing of a molecular
  beam},\ }\href {https://doi.org/10.1103/PhysRevLett.108.103002} {\bibfield
  {journal} {\bibinfo  {journal} {Phys. Rev. Lett.}\ }\textbf {\bibinfo
  {volume} {108}},\ \bibinfo {pages} {103002} (\bibinfo {year}
  {2012})}\BibitemShut {NoStop}%
\bibitem [{\citenamefont {Hummon}\ \emph {et~al.}(2013)\citenamefont {Hummon},
  \citenamefont {Yeo}, \citenamefont {Stuhl}, \citenamefont {Collopy},
  \citenamefont {Xia},\ and\ \citenamefont {Ye}}]{Hum13}%
  \BibitemOpen
  \bibfield  {author} {\bibinfo {author} {\bibfnamefont {M.~T.}\ \bibnamefont
  {Hummon}}, \bibinfo {author} {\bibfnamefont {M.}~\bibnamefont {Yeo}},
  \bibinfo {author} {\bibfnamefont {B.~K.}\ \bibnamefont {Stuhl}}, \bibinfo
  {author} {\bibfnamefont {A.~L.}\ \bibnamefont {Collopy}}, \bibinfo {author}
  {\bibfnamefont {Y.}~\bibnamefont {Xia}},\ and\ \bibinfo {author}
  {\bibfnamefont {J.}~\bibnamefont {Ye}},\ }\bibfield  {title} {\bibinfo
  {title} {{2D} magneto-optical trapping of diatomic molecules},\ }\href
  {https://doi.org/10.1103/PhysRevLett.110.143001} {\bibfield  {journal}
  {\bibinfo  {journal} {Phys. Rev. Lett.}\ }\textbf {\bibinfo {volume} {110}},\
  \bibinfo {pages} {143001} (\bibinfo {year} {2013})}\BibitemShut {NoStop}%
\bibitem [{\citenamefont {Barry}\ \emph {et~al.}(2014)\citenamefont {Barry},
  \citenamefont {McCarron}, \citenamefont {Norrgard}, \citenamefont
  {Steinecker},\ and\ \citenamefont {DeMille}}]{Bar14}%
  \BibitemOpen
  \bibfield  {author} {\bibinfo {author} {\bibfnamefont {J.~F.}\ \bibnamefont
  {Barry}}, \bibinfo {author} {\bibfnamefont {D.~J.}\ \bibnamefont {McCarron}},
  \bibinfo {author} {\bibfnamefont {E.~B.}\ \bibnamefont {Norrgard}}, \bibinfo
  {author} {\bibfnamefont {M.~H.}\ \bibnamefont {Steinecker}},\ and\ \bibinfo
  {author} {\bibfnamefont {D.}~\bibnamefont {DeMille}},\ }\bibfield  {title}
  {\bibinfo {title} {Magneto-optical trapping of a diatomic molecule},\ }\href
  {https://doi.org/10.1038/nature13634} {\bibfield  {journal} {\bibinfo
  {journal} {Nature}\ }\textbf {\bibinfo {volume} {512}},\ \bibinfo {pages}
  {286} (\bibinfo {year} {2014})}\BibitemShut {NoStop}%
\bibitem [{\citenamefont {Norrgard}\ \emph {et~al.}(2016)\citenamefont
  {Norrgard}, \citenamefont {McCarron}, \citenamefont {Steinecker},
  \citenamefont {Tarbutt},\ and\ \citenamefont {DeMille}}]{Nor16}%
  \BibitemOpen
  \bibfield  {author} {\bibinfo {author} {\bibfnamefont {E.~B.}\ \bibnamefont
  {Norrgard}}, \bibinfo {author} {\bibfnamefont {D.~J.}\ \bibnamefont
  {McCarron}}, \bibinfo {author} {\bibfnamefont {M.~H.}\ \bibnamefont
  {Steinecker}}, \bibinfo {author} {\bibfnamefont {M.~R.}\ \bibnamefont
  {Tarbutt}},\ and\ \bibinfo {author} {\bibfnamefont {D.}~\bibnamefont
  {DeMille}},\ }\bibfield  {title} {\bibinfo {title} {Submillikelvin dipolar
  molecules in a radio-frequency magneto-optical trap},\ }\href
  {https://doi.org/10.1103/PhysRevLett.116.063004} {\bibfield  {journal}
  {\bibinfo  {journal} {Phys. Rev. Lett.}\ }\textbf {\bibinfo {volume} {116}},\
  \bibinfo {pages} {063004} (\bibinfo {year} {2016})}\BibitemShut {NoStop}%
\bibitem [{\citenamefont {Williams}\ \emph {et~al.}(2018)\citenamefont
  {Williams}, \citenamefont {Caldwell}, \citenamefont {Fitch}, \citenamefont
  {Truppe}, \citenamefont {Rodewald}, \citenamefont {Hinds}, \citenamefont
  {Sauer},\ and\ \citenamefont {Tarbutt}}]{Wil18}%
  \BibitemOpen
  \bibfield  {author} {\bibinfo {author} {\bibfnamefont {H.~J.}\ \bibnamefont
  {Williams}}, \bibinfo {author} {\bibfnamefont {L.}~\bibnamefont {Caldwell}},
  \bibinfo {author} {\bibfnamefont {N.~J.}\ \bibnamefont {Fitch}}, \bibinfo
  {author} {\bibfnamefont {S.}~\bibnamefont {Truppe}}, \bibinfo {author}
  {\bibfnamefont {J.}~\bibnamefont {Rodewald}}, \bibinfo {author}
  {\bibfnamefont {E.~A.}\ \bibnamefont {Hinds}}, \bibinfo {author}
  {\bibfnamefont {B.~E.}\ \bibnamefont {Sauer}},\ and\ \bibinfo {author}
  {\bibfnamefont {M.~R.}\ \bibnamefont {Tarbutt}},\ }\bibfield  {title}
  {\bibinfo {title} {Magnetic trapping and coherent control of laser-cooled
  molecules},\ }\href {https://doi.org/10.1103/PhysRevLett.120.163201}
  {\bibfield  {journal} {\bibinfo  {journal} {Phys. Rev. Lett.}\ }\textbf
  {\bibinfo {volume} {120}},\ \bibinfo {pages} {163201} (\bibinfo {year}
  {2018})}\BibitemShut {NoStop}%
\bibitem [{\citenamefont {McCarron}\ \emph {et~al.}(2018)\citenamefont
  {McCarron}, \citenamefont {Steinecker}, \citenamefont {Zhu},\ and\
  \citenamefont {DeMille}}]{McC18}%
  \BibitemOpen
  \bibfield  {author} {\bibinfo {author} {\bibfnamefont {D.~J.}\ \bibnamefont
  {McCarron}}, \bibinfo {author} {\bibfnamefont {M.~H.}\ \bibnamefont
  {Steinecker}}, \bibinfo {author} {\bibfnamefont {Y.}~\bibnamefont {Zhu}},\
  and\ \bibinfo {author} {\bibfnamefont {D.}~\bibnamefont {DeMille}},\
  }\bibfield  {title} {\bibinfo {title} {Magnetic trapping of an ultracold gas
  of polar molecules},\ }\href {https://doi.org/10.1103/PhysRevLett.121.013202}
  {\bibfield  {journal} {\bibinfo  {journal} {Phys. Rev. Lett.}\ }\textbf
  {\bibinfo {volume} {121}},\ \bibinfo {pages} {013202} (\bibinfo {year}
  {2018})}\BibitemShut {NoStop}%
\bibitem [{\citenamefont {Anderegg}\ \emph {et~al.}(2019)\citenamefont
  {Anderegg}, \citenamefont {Cheuk}, \citenamefont {Bao}, \citenamefont
  {Burchesky}, \citenamefont {Ketterle}, \citenamefont {Ni},\ and\
  \citenamefont {Doyle}}]{And19}%
  \BibitemOpen
  \bibfield  {author} {\bibinfo {author} {\bibfnamefont {L.}~\bibnamefont
  {Anderegg}}, \bibinfo {author} {\bibfnamefont {L.~W.}\ \bibnamefont {Cheuk}},
  \bibinfo {author} {\bibfnamefont {Y.}~\bibnamefont {Bao}}, \bibinfo {author}
  {\bibfnamefont {S.}~\bibnamefont {Burchesky}}, \bibinfo {author}
  {\bibfnamefont {W.}~\bibnamefont {Ketterle}}, \bibinfo {author}
  {\bibfnamefont {K.-K.}\ \bibnamefont {Ni}},\ and\ \bibinfo {author}
  {\bibfnamefont {J.~M.}\ \bibnamefont {Doyle}},\ }\bibfield  {title} {\bibinfo
  {title} {An optical tweezer array of ultracold molecules},\ }\href
  {https://doi.org/10.1126/science.aax1265} {\bibfield  {journal} {\bibinfo
  {journal} {Science}\ }\textbf {\bibinfo {volume} {365}},\ \bibinfo {pages}
  {1156} (\bibinfo {year} {2019})}\BibitemShut {NoStop}%
\bibitem [{\citenamefont {Ding}\ \emph {et~al.}(2020)\citenamefont {Ding},
  \citenamefont {Wu}, \citenamefont {Finneran}, \citenamefont {Burau},\ and\
  \citenamefont {Ye}}]{Shi20}%
  \BibitemOpen
  \bibfield  {author} {\bibinfo {author} {\bibfnamefont {S.}~\bibnamefont
  {Ding}}, \bibinfo {author} {\bibfnamefont {Y.}~\bibnamefont {Wu}}, \bibinfo
  {author} {\bibfnamefont {I.~A.}\ \bibnamefont {Finneran}}, \bibinfo {author}
  {\bibfnamefont {J.~J.}\ \bibnamefont {Burau}},\ and\ \bibinfo {author}
  {\bibfnamefont {J.}~\bibnamefont {Ye}},\ }\bibfield  {title} {\bibinfo
  {title} {Sub-{Doppler} cooling and compressed trapping of {YO} molecules at
  $\ensuremath{\mu}\mathrm{K}$ temperatures},\ }\href
  {https://doi.org/10.1103/PhysRevX.10.021049} {\bibfield  {journal} {\bibinfo
  {journal} {Phys. Rev. X}\ }\textbf {\bibinfo {volume} {10}},\ \bibinfo
  {pages} {021049} (\bibinfo {year} {2020})}\BibitemShut {NoStop}%
\bibitem [{\citenamefont {Qu{\'e}m{\'e}ner}\ and\ \citenamefont
  {Julienne}(2012)}]{Que12}%
  \BibitemOpen
  \bibfield  {author} {\bibinfo {author} {\bibfnamefont {G.}~\bibnamefont
  {Qu{\'e}m{\'e}ner}}\ and\ \bibinfo {author} {\bibfnamefont {P.~S.}\
  \bibnamefont {Julienne}},\ }\bibfield  {title} {\bibinfo {title} {Ultracold
  molecules under control!},\ }\href {https://doi.org/10.1021/cr300092g}
  {\bibfield  {journal} {\bibinfo  {journal} {Chemical Reviews}\ }\textbf
  {\bibinfo {volume} {112}},\ \bibinfo {pages} {4949} (\bibinfo {year}
  {2012})}\BibitemShut {NoStop}%
\bibitem [{\citenamefont {Lemeshko}\ \emph {et~al.}(2013)\citenamefont
  {Lemeshko}, \citenamefont {Krems}, \citenamefont {Doyle},\ and\ \citenamefont
  {Kais}}]{Lem13}%
  \BibitemOpen
  \bibfield  {author} {\bibinfo {author} {\bibfnamefont {M.}~\bibnamefont
  {Lemeshko}}, \bibinfo {author} {\bibfnamefont {R.~V.}\ \bibnamefont {Krems}},
  \bibinfo {author} {\bibfnamefont {J.~M.}\ \bibnamefont {Doyle}},\ and\
  \bibinfo {author} {\bibfnamefont {S.}~\bibnamefont {Kais}},\ }\bibfield
  {title} {\bibinfo {title} {Manipulation of molecules with electromagnetic
  fields},\ }\href {https://doi.org/10.1080/00268976.2013.813595} {\bibfield
  {journal} {\bibinfo  {journal} {Molecular Physics}\ }\textbf {\bibinfo
  {volume} {111}},\ \bibinfo {pages} {1648} (\bibinfo {year}
  {2013})}\BibitemShut {NoStop}%
\bibitem [{\citenamefont {Valtolina}\ \emph {et~al.}(2020)\citenamefont
  {Valtolina}, \citenamefont {Matsuda}, \citenamefont {Tobias}, \citenamefont
  {Li}, \citenamefont {Marco},\ and\ \citenamefont {Ye}}]{DeM19b}%
  \BibitemOpen
  \bibfield  {author} {\bibinfo {author} {\bibfnamefont {G.}~\bibnamefont
  {Valtolina}}, \bibinfo {author} {\bibfnamefont {K.}~\bibnamefont {Matsuda}},
  \bibinfo {author} {\bibfnamefont {W.~G.}\ \bibnamefont {Tobias}}, \bibinfo
  {author} {\bibfnamefont {J.-R.}\ \bibnamefont {Li}}, \bibinfo {author}
  {\bibfnamefont {L.~D.}\ \bibnamefont {Marco}},\ and\ \bibinfo {author}
  {\bibfnamefont {J.}~\bibnamefont {Ye}},\ }\bibfield  {title} {\bibinfo
  {title} {Dipolar evaporation of reactive molecules to below the {Fermi}
  temperature},\ }\href {https://www.nature.com/articles/s41586-020-2980-7}
  {\bibfield  {journal} {\bibinfo  {journal} {Nature}\ }\textbf {\bibinfo
  {volume} {588}},\ \bibinfo {pages} {239} (\bibinfo {year}
  {2020})}\BibitemShut {NoStop}%
\bibitem [{\citenamefont {Son}\ \emph {et~al.}(2020)\citenamefont {Son},
  \citenamefont {Park}, \citenamefont {Ketterle},\ and\ \citenamefont
  {Jamison}}]{Son20cco}%
  \BibitemOpen
  \bibfield  {author} {\bibinfo {author} {\bibfnamefont {H.}~\bibnamefont
  {Son}}, \bibinfo {author} {\bibfnamefont {J.~J.}\ \bibnamefont {Park}},
  \bibinfo {author} {\bibfnamefont {W.}~\bibnamefont {Ketterle}},\ and\
  \bibinfo {author} {\bibfnamefont {A.~O.}\ \bibnamefont {Jamison}},\
  }\bibfield  {title} {\bibinfo {title} {Collisional cooling of ultracold
  molecules},\ }\href {https://doi.org/10.1038/s41586-020-2141-z} {\bibfield
  {journal} {\bibinfo  {journal} {Nature}\ }\textbf {\bibinfo {volume} {580}},\
  \bibinfo {pages} {197} (\bibinfo {year} {2020})}\BibitemShut {NoStop}%
\bibitem [{\citenamefont {Matsuda}\ \emph {et~al.}(2020)\citenamefont
  {Matsuda}, \citenamefont {De~Marco}, \citenamefont {Li}, \citenamefont
  {Tobias}, \citenamefont {Valtolina}, \citenamefont {Qu{\'e}m{\'e}ner},\ and\
  \citenamefont {Ye}}]{Mat20}%
  \BibitemOpen
  \bibfield  {author} {\bibinfo {author} {\bibfnamefont {K.}~\bibnamefont
  {Matsuda}}, \bibinfo {author} {\bibfnamefont {L.}~\bibnamefont {De~Marco}},
  \bibinfo {author} {\bibfnamefont {J.-R.}\ \bibnamefont {Li}}, \bibinfo
  {author} {\bibfnamefont {W.~G.}\ \bibnamefont {Tobias}}, \bibinfo {author}
  {\bibfnamefont {G.}~\bibnamefont {Valtolina}}, \bibinfo {author}
  {\bibfnamefont {G.}~\bibnamefont {Qu{\'e}m{\'e}ner}},\ and\ \bibinfo {author}
  {\bibfnamefont {J.}~\bibnamefont {Ye}},\ }\bibfield  {title} {\bibinfo
  {title} {Resonant collisional shielding of reactive molecules using electric
  fields},\ }\href {https://doi.org/10.1126/science.abe7370} {\bibfield
  {journal} {\bibinfo  {journal} {Science}\ }\textbf {\bibinfo {volume}
  {370}},\ \bibinfo {pages} {1324} (\bibinfo {year} {2020})}\BibitemShut
  {NoStop}%
\bibitem [{\citenamefont {K\"ohler}\ \emph {et~al.}(2006)\citenamefont
  {K\"ohler}, \citenamefont {G\'oral},\ and\ \citenamefont {Julienne}}]{Koh06}%
  \BibitemOpen
  \bibfield  {author} {\bibinfo {author} {\bibfnamefont {T.}~\bibnamefont
  {K\"ohler}}, \bibinfo {author} {\bibfnamefont {K.}~\bibnamefont {G\'oral}},\
  and\ \bibinfo {author} {\bibfnamefont {P.~S.}\ \bibnamefont {Julienne}},\
  }\bibfield  {title} {\bibinfo {title} {Production of cold molecules via
  magnetically tunable {Feshbach} resonances},\ }\href
  {https://doi.org/10.1103/RevModPhys.78.1311} {\bibfield  {journal} {\bibinfo
  {journal} {Rev. Mod. Phys.}\ }\textbf {\bibinfo {volume} {78}},\ \bibinfo
  {pages} {1311} (\bibinfo {year} {2006})}\BibitemShut {NoStop}%
\bibitem [{\citenamefont {Bohn}\ \emph {et~al.}(2017)\citenamefont {Bohn},
  \citenamefont {Rey},\ and\ \citenamefont {Ye}}]{Boh17}%
  \BibitemOpen
  \bibfield  {author} {\bibinfo {author} {\bibfnamefont {J.~L.}\ \bibnamefont
  {Bohn}}, \bibinfo {author} {\bibfnamefont {A.~M.}\ \bibnamefont {Rey}},\ and\
  \bibinfo {author} {\bibfnamefont {J.}~\bibnamefont {Ye}},\ }\bibfield
  {title} {\bibinfo {title} {Cold molecules: Progress in quantum engineering of
  chemistry and quantum matter},\ }\href
  {https://doi.org/10.1126/science.aam6299} {\bibfield  {journal} {\bibinfo
  {journal} {Science}\ }\textbf {\bibinfo {volume} {357}},\ \bibinfo {pages}
  {1002} (\bibinfo {year} {2017})}\BibitemShut {NoStop}%
\bibitem [{\citenamefont {De~Marco}\ \emph {et~al.}(2019)\citenamefont
  {De~Marco}, \citenamefont {Valtolina}, \citenamefont {Matsuda}, \citenamefont
  {Tobias}, \citenamefont {Covey},\ and\ \citenamefont {Ye}}]{DeM19a}%
  \BibitemOpen
  \bibfield  {author} {\bibinfo {author} {\bibfnamefont {L.}~\bibnamefont
  {De~Marco}}, \bibinfo {author} {\bibfnamefont {G.}~\bibnamefont {Valtolina}},
  \bibinfo {author} {\bibfnamefont {K.}~\bibnamefont {Matsuda}}, \bibinfo
  {author} {\bibfnamefont {W.~G.}\ \bibnamefont {Tobias}}, \bibinfo {author}
  {\bibfnamefont {J.~P.}\ \bibnamefont {Covey}},\ and\ \bibinfo {author}
  {\bibfnamefont {J.}~\bibnamefont {Ye}},\ }\bibfield  {title} {\bibinfo
  {title} {A degenerate {Fermi} gas of polar molecules},\ }\href
  {https://doi.org/10.1126/science.aau7230} {\bibfield  {journal} {\bibinfo
  {journal} {Science}\ }\textbf {\bibinfo {volume} {363}},\ \bibinfo {pages}
  {853} (\bibinfo {year} {2019})}\BibitemShut {NoStop}%
\bibitem [{\citenamefont {Schoene}\ \emph {et~al.}(2010)\citenamefont
  {Schoene}, \citenamefont {Thorn},\ and\ \citenamefont {Steck}}]{Sch10}%
  \BibitemOpen
  \bibfield  {author} {\bibinfo {author} {\bibfnamefont {E.~A.}\ \bibnamefont
  {Schoene}}, \bibinfo {author} {\bibfnamefont {J.~J.}\ \bibnamefont {Thorn}},\
  and\ \bibinfo {author} {\bibfnamefont {D.~A.}\ \bibnamefont {Steck}},\
  }\bibfield  {title} {\bibinfo {title} {Cooling atoms with a moving one-way
  barrier},\ }\href {https://doi.org/10.1103/PhysRevA.82.023419} {\bibfield
  {journal} {\bibinfo  {journal} {Phys. Rev. A}\ }\textbf {\bibinfo {volume}
  {82}},\ \bibinfo {pages} {023419} (\bibinfo {year} {2010})}\BibitemShut
  {NoStop}%
\bibitem [{\citenamefont {Vuleti\ifmmode~\acute{c}\else \'{c}\fi{}}\ and\
  \citenamefont {Chu}(2000)}]{Vul00}%
  \BibitemOpen
  \bibfield  {author} {\bibinfo {author} {\bibfnamefont {V.}~\bibnamefont
  {Vuleti\ifmmode~\acute{c}\else \'{c}\fi{}}}\ and\ \bibinfo {author}
  {\bibfnamefont {S.}~\bibnamefont {Chu}},\ }\bibfield  {title} {\bibinfo
  {title} {Laser cooling of atoms, ions, or molecules by coherent scattering},\
  }\href {https://doi.org/10.1103/PhysRevLett.84.3787} {\bibfield  {journal}
  {\bibinfo  {journal} {Phys. Rev. Lett.}\ }\textbf {\bibinfo {volume} {84}},\
  \bibinfo {pages} {3787} (\bibinfo {year} {2000})}\BibitemShut {NoStop}%
\bibitem [{\citenamefont {Maunz}\ \emph {et~al.}(2004)\citenamefont {Maunz},
  \citenamefont {Puppe}, \citenamefont {Schuster}, \citenamefont {Syassen},
  \citenamefont {Pinkse},\ and\ \citenamefont {Rempe}}]{Mau04}%
  \BibitemOpen
  \bibfield  {author} {\bibinfo {author} {\bibfnamefont {P.}~\bibnamefont
  {Maunz}}, \bibinfo {author} {\bibfnamefont {T.}~\bibnamefont {Puppe}},
  \bibinfo {author} {\bibfnamefont {I.}~\bibnamefont {Schuster}}, \bibinfo
  {author} {\bibfnamefont {N.}~\bibnamefont {Syassen}}, \bibinfo {author}
  {\bibfnamefont {P.~W.~H.}\ \bibnamefont {Pinkse}},\ and\ \bibinfo {author}
  {\bibfnamefont {G.}~\bibnamefont {Rempe}},\ }\bibfield  {title} {\bibinfo
  {title} {Cavity cooling of a single atom},\ }\href
  {https://doi.org/10.1038/nature02387} {\bibfield  {journal} {\bibinfo
  {journal} {Nature}\ }\textbf {\bibinfo {volume} {428}},\ \bibinfo {pages}
  {50} (\bibinfo {year} {2004})}\BibitemShut {NoStop}%
\bibitem [{\citenamefont {Hosseini}\ \emph {et~al.}(2017)\citenamefont
  {Hosseini}, \citenamefont {Duan}, \citenamefont {Beck}, \citenamefont
  {Chen},\ and\ \citenamefont {Vuleti\ifmmode~\acute{c}\else
  \'{c}\fi{}}}]{Hos17}%
  \BibitemOpen
  \bibfield  {author} {\bibinfo {author} {\bibfnamefont {M.}~\bibnamefont
  {Hosseini}}, \bibinfo {author} {\bibfnamefont {Y.}~\bibnamefont {Duan}},
  \bibinfo {author} {\bibfnamefont {K.~M.}\ \bibnamefont {Beck}}, \bibinfo
  {author} {\bibfnamefont {Y.-T.}\ \bibnamefont {Chen}},\ and\ \bibinfo
  {author} {\bibfnamefont {V.}~\bibnamefont {Vuleti\ifmmode~\acute{c}\else
  \'{c}\fi{}}},\ }\bibfield  {title} {\bibinfo {title} {Cavity cooling of many
  atoms},\ }\href {https://doi.org/10.1103/PhysRevLett.118.183601} {\bibfield
  {journal} {\bibinfo  {journal} {Phys. Rev. Lett.}\ }\textbf {\bibinfo
  {volume} {118}},\ \bibinfo {pages} {183601} (\bibinfo {year}
  {2017})}\BibitemShut {NoStop}%
\bibitem [{\citenamefont {Deli{\'c}}\ \emph {et~al.}(2020)\citenamefont
  {Deli{\'c}}, \citenamefont {Reisenbauer}, \citenamefont {Dare}, \citenamefont
  {Grass}, \citenamefont {Vuleti{\'c}}, \citenamefont {Kiesel},\ and\
  \citenamefont {Aspelmeyer}}]{Del20}%
  \BibitemOpen
  \bibfield  {author} {\bibinfo {author} {\bibfnamefont {U.}~\bibnamefont
  {Deli{\'c}}}, \bibinfo {author} {\bibfnamefont {M.}~\bibnamefont
  {Reisenbauer}}, \bibinfo {author} {\bibfnamefont {K.}~\bibnamefont {Dare}},
  \bibinfo {author} {\bibfnamefont {D.}~\bibnamefont {Grass}}, \bibinfo
  {author} {\bibfnamefont {V.}~\bibnamefont {Vuleti{\'c}}}, \bibinfo {author}
  {\bibfnamefont {N.}~\bibnamefont {Kiesel}},\ and\ \bibinfo {author}
  {\bibfnamefont {M.}~\bibnamefont {Aspelmeyer}},\ }\bibfield  {title}
  {\bibinfo {title} {Cooling of a levitated nanoparticle to the motional
  quantum ground state},\ }\href {https://doi.org/10.1126/science.aba3993}
  {\bibfield  {journal} {\bibinfo  {journal} {Science}\ }\textbf {\bibinfo
  {volume} {367}},\ \bibinfo {pages} {892} (\bibinfo {year}
  {2020})}\BibitemShut {NoStop}%
\bibitem [{\citenamefont {Grimm}\ \emph {et~al.}(1990)\citenamefont {Grimm},
  \citenamefont {Ovchinnikov}, \citenamefont {Sidorov},\ and\ \citenamefont
  {Letokhov}}]{Gri90}%
  \BibitemOpen
  \bibfield  {author} {\bibinfo {author} {\bibfnamefont {R.}~\bibnamefont
  {Grimm}}, \bibinfo {author} {\bibfnamefont {Y.~B.}\ \bibnamefont
  {Ovchinnikov}}, \bibinfo {author} {\bibfnamefont {A.~I.}\ \bibnamefont
  {Sidorov}},\ and\ \bibinfo {author} {\bibfnamefont {V.~S.}\ \bibnamefont
  {Letokhov}},\ }\bibfield  {title} {\bibinfo {title} {Observation of a strong
  rectified dipole force in a bichromatic standing light wave},\ }\href
  {https://doi.org/10.1103/PhysRevLett.65.1415} {\bibfield  {journal} {\bibinfo
   {journal} {Phys. Rev. Lett.}\ }\textbf {\bibinfo {volume} {65}},\ \bibinfo
  {pages} {1415} (\bibinfo {year} {1990})}\BibitemShut {NoStop}%
\bibitem [{\citenamefont {S\"oding}\ \emph {et~al.}(1997)\citenamefont
  {S\"oding}, \citenamefont {Grimm}, \citenamefont {Ovchinnikov}, \citenamefont
  {Bouyer},\ and\ \citenamefont {Salomon}}]{Sod97}%
  \BibitemOpen
  \bibfield  {author} {\bibinfo {author} {\bibfnamefont {J.}~\bibnamefont
  {S\"oding}}, \bibinfo {author} {\bibfnamefont {R.}~\bibnamefont {Grimm}},
  \bibinfo {author} {\bibfnamefont {Y.~B.}\ \bibnamefont {Ovchinnikov}},
  \bibinfo {author} {\bibfnamefont {P.}~\bibnamefont {Bouyer}},\ and\ \bibinfo
  {author} {\bibfnamefont {C.}~\bibnamefont {Salomon}},\ }\bibfield  {title}
  {\bibinfo {title} {Short-distance atomic beam deceleration with a stimulated
  light force},\ }\href {https://doi.org/10.1103/PhysRevLett.78.1420}
  {\bibfield  {journal} {\bibinfo  {journal} {Phys. Rev. Lett.}\ }\textbf
  {\bibinfo {volume} {78}},\ \bibinfo {pages} {1420} (\bibinfo {year}
  {1997})}\BibitemShut {NoStop}%
\bibitem [{\citenamefont {Corder}\ \emph {et~al.}(2015)\citenamefont {Corder},
  \citenamefont {Arnold},\ and\ \citenamefont {Metcalf}}]{Cor15}%
  \BibitemOpen
  \bibfield  {author} {\bibinfo {author} {\bibfnamefont {C.}~\bibnamefont
  {Corder}}, \bibinfo {author} {\bibfnamefont {B.}~\bibnamefont {Arnold}},\
  and\ \bibinfo {author} {\bibfnamefont {H.}~\bibnamefont {Metcalf}},\
  }\bibfield  {title} {\bibinfo {title} {Laser cooling without spontaneous
  emission},\ }\href {https://doi.org/10.1103/PhysRevLett.114.043002}
  {\bibfield  {journal} {\bibinfo  {journal} {Phys. Rev. Lett.}\ }\textbf
  {\bibinfo {volume} {114}},\ \bibinfo {pages} {043002} (\bibinfo {year}
  {2015})}\BibitemShut {NoStop}%
\bibitem [{\citenamefont {Schmidt-Kaler}\ \emph {et~al.}(2001)\citenamefont
  {Schmidt-Kaler}, \citenamefont {Eschner}, \citenamefont {Morigi},
  \citenamefont {Roos}, \citenamefont {Leibfried}, \citenamefont {Mundt},\ and\
  \citenamefont {Blatt}}]{Sch01}%
  \BibitemOpen
  \bibfield  {author} {\bibinfo {author} {\bibfnamefont {F.}~\bibnamefont
  {Schmidt-Kaler}}, \bibinfo {author} {\bibfnamefont {J.}~\bibnamefont
  {Eschner}}, \bibinfo {author} {\bibfnamefont {G.}~\bibnamefont {Morigi}},
  \bibinfo {author} {\bibfnamefont {C.~F.}\ \bibnamefont {Roos}}, \bibinfo
  {author} {\bibfnamefont {D.}~\bibnamefont {Leibfried}}, \bibinfo {author}
  {\bibfnamefont {A.}~\bibnamefont {Mundt}},\ and\ \bibinfo {author}
  {\bibfnamefont {R.}~\bibnamefont {Blatt}},\ }\bibfield  {title} {\bibinfo
  {title} {Laser cooling with electromagnetically induced transparency:
  application to trapped samples of ions or neutral atoms},\ }\href
  {https://doi.org/10.1007/s003400100721} {\bibfield  {journal} {\bibinfo
  {journal} {Appl. Phys. B}\ }\textbf {\bibinfo {volume} {73}},\ \bibinfo
  {pages} {807} (\bibinfo {year} {2001})}\BibitemShut {NoStop}%
\bibitem [{\citenamefont {Eschner}\ \emph {et~al.}(2003)\citenamefont
  {Eschner}, \citenamefont {Morigi}, \citenamefont {Schmidt-Kaler},\ and\
  \citenamefont {Blatt}}]{Esc03}%
  \BibitemOpen
  \bibfield  {author} {\bibinfo {author} {\bibfnamefont {J.}~\bibnamefont
  {Eschner}}, \bibinfo {author} {\bibfnamefont {G.}~\bibnamefont {Morigi}},
  \bibinfo {author} {\bibfnamefont {F.}~\bibnamefont {Schmidt-Kaler}},\ and\
  \bibinfo {author} {\bibfnamefont {R.}~\bibnamefont {Blatt}},\ }\bibfield
  {title} {\bibinfo {title} {Laser cooling of trapped ions},\ }\href
  {https://doi.org/10.1364/JOSAB.20.001003} {\bibfield  {journal} {\bibinfo
  {journal} {J. Opt. Soc. Am. B}\ }\textbf {\bibinfo {volume} {20}},\ \bibinfo
  {pages} {1003} (\bibinfo {year} {2003})}\BibitemShut {NoStop}%
\bibitem [{\citenamefont {Aspelmeyer}\ \emph {et~al.}(2014)\citenamefont
  {Aspelmeyer}, \citenamefont {Kippenberg},\ and\ \citenamefont
  {Marquardt}}]{Asp14}%
  \BibitemOpen
  \bibfield  {author} {\bibinfo {author} {\bibfnamefont {M.}~\bibnamefont
  {Aspelmeyer}}, \bibinfo {author} {\bibfnamefont {T.~J.}\ \bibnamefont
  {Kippenberg}},\ and\ \bibinfo {author} {\bibfnamefont {F.}~\bibnamefont
  {Marquardt}},\ }\bibfield  {title} {\bibinfo {title} {Cavity optomechanics},\
  }\href {https://doi.org/10.1103/RevModPhys.86.1391} {\bibfield  {journal}
  {\bibinfo  {journal} {Rev. Mod. Phys.}\ }\textbf {\bibinfo {volume} {86}},\
  \bibinfo {pages} {1391} (\bibinfo {year} {2014})}\BibitemShut {NoStop}%
\bibitem [{\citenamefont {Rossi}\ \emph {et~al.}(2018)\citenamefont {Rossi},
  \citenamefont {Mason}, \citenamefont {Chen}, \citenamefont {Tsaturyan},\ and\
  \citenamefont {Schliesser}}]{Ros18b}%
  \BibitemOpen
  \bibfield  {author} {\bibinfo {author} {\bibfnamefont {M.}~\bibnamefont
  {Rossi}}, \bibinfo {author} {\bibfnamefont {D.}~\bibnamefont {Mason}},
  \bibinfo {author} {\bibfnamefont {J.}~\bibnamefont {Chen}}, \bibinfo {author}
  {\bibfnamefont {Y.}~\bibnamefont {Tsaturyan}},\ and\ \bibinfo {author}
  {\bibfnamefont {A.}~\bibnamefont {Schliesser}},\ }\bibfield  {title}
  {\bibinfo {title} {Measurement-based quantum control of mechanical motion},\
  }\href {https://doi.org/10.1038/s41586-018-0643-8} {\bibfield  {journal}
  {\bibinfo  {journal} {Nature}\ }\textbf {\bibinfo {volume} {563}},\ \bibinfo
  {pages} {53} (\bibinfo {year} {2018})}\BibitemShut {NoStop}%
\bibitem [{\citenamefont {Steck}(2021)}]{Ste03}%
  \BibitemOpen
  \bibfield  {author} {\bibinfo {author} {\bibfnamefont {D.}~\bibnamefont
  {Steck}},\ }\href {https://steck.us/alkalidata/rubidium87numbers.pdf}
  {\bibinfo {title} {Rubidium 87 {D} line data}} (\bibinfo {year}
  {2021})\BibitemShut {NoStop}%
\bibitem [{\citenamefont {Youn}\ \emph {et~al.}(2010)\citenamefont {Youn},
  \citenamefont {Lu}, \citenamefont {Ray},\ and\ \citenamefont {Lev}}]{You10}%
  \BibitemOpen
  \bibfield  {author} {\bibinfo {author} {\bibfnamefont {S.~H.}\ \bibnamefont
  {Youn}}, \bibinfo {author} {\bibfnamefont {M.}~\bibnamefont {Lu}}, \bibinfo
  {author} {\bibfnamefont {U.}~\bibnamefont {Ray}},\ and\ \bibinfo {author}
  {\bibfnamefont {B.~L.}\ \bibnamefont {Lev}},\ }\bibfield  {title} {\bibinfo
  {title} {Dysprosium magneto-optical traps},\ }\href
  {https://doi.org/10.1103/PhysRevA.82.043425} {\bibfield  {journal} {\bibinfo
  {journal} {Phys. Rev. A}\ }\textbf {\bibinfo {volume} {82}},\ \bibinfo
  {pages} {043425} (\bibinfo {year} {2010})}\BibitemShut {NoStop}%
\end{thebibliography}
\end{document}